\documentclass[11pt,a4paper]{article}
\pdfoutput=1
\usepackage{jheppub}

\usepackage{amsmath}
\usepackage{verbatim}
\usepackage{amssymb}
\usepackage{inputenc, array}
\usepackage{textcomp}
\usepackage{bm}
\usepackage{physics}
\usepackage{simpler-wick}
\usepackage{appendix}
\usepackage{xcolor}

\newcommand{\e}{\epsilon}
\newcommand{\be}[1]{\begin{equation}\label{#1} }
\newcommand{\ee}{\end{equation}}
\newcommand{\bea}[1]{\begin{eqnarray}\label{#1} }
\newcommand{\eea}{\end{eqnarray}}
\newcommand{\p}{\partial}
\newcommand{\refb}[1]{(\ref{#1})}

\renewcommand{\L}{{\mathcal{L}}}

\newcommand{\bL}{\bar{{\mathcal{L}}}}

\newcommand{\h}{{\bar h}}

\renewcommand{\>}{\rangle}

\newcommand{\txb}{\textcolor{blue}}

\newcommand{\D}{\Delta}

\renewcommand{\a}{\alpha}

\renewcommand{\b}{\beta}
\renewcommand{\t}{\tau}
\newcommand{\s}{\sigma}

\allowdisplaybreaks
\title{Non-Lorentzian Ka{\v{c}}-Moody Algebras}

\author[a, b]{Arjun Bagchi,} \author[a]{Ritankar Chatterjee,} \author[a, c]{Rishabh Kaushik,} \author[a]{Amartya Saha,} \author[a, c]{and Debmalya Sarkar.}
\author{\\}

\affiliation[a]{Indian Institute of Technology Kanpur, Kanpur 208016, India.\\} 

\affiliation[b]{Centre de Physique Theorique, Ecole Polytechnique de Paris, 91128 Palaiseau Cedex, France.\\}

\affiliation[c]{International Centre for Theoretical Sciences (ICTS-TIFR), Bengaluru 560089, India. \\}

\emailAdd{(abagchi, ritankar, amartyas)@iit.ac.in, rishabh.kaushik@icts.res.in, sarkardebmalya01@gmail.com}

\preprint{}
\abstract{We investigate two dimensional (2d) quantum field theories which exhibit Non-Lorentzian Ka{\v{c}}-Moody (NLKM) algebras  as their underlying symmetry. Our investigations encompass both 2d Galilean (speed of light $c\to\infty$) and Carrollian ($c\to0$) CFTs with additional number of infinite non-Abelian currents, stemming from an isomorphism between the two algebras. We alternate between an intrinsic and a limiting analysis. Our
NLKM algebra is constructed first through a contraction and then derived from an intrinsically Carrollian perspective. We then go on to use the symmetries to derive a Non-Lorentzian (NL) Sugawara construction and ultimately write down the NL equivalent of the Knizhnik Zamolodchikov equations. All of these are also derived from contractions, thus providing a robust cross-check of our analyses.}
\begin{document}
\maketitle
\flushbottom
\section{Introduction}
Relativistic conformal field theory (CFT) is one of the most potent tools of modern theoretical physics, with applications ranging from statistical mechanics of phase transitions to quantum gravity through holography and string theory. Especially powerful are methods of two dimensional (2d) CFTs \cite{Belavin:1984vu} where symmetries enhance to two copies of the infinite dimensional Virasoro algebra. The ideas and methods of 2d CFTs are of particular importance to the success of string theory, where this arises as residual symmetry on the string worldsheet after the fixing of conformal gauge \cite{Friedan:1985ge}.

\medskip

Non-abelian current algebras arise on the string worldsheet when one considers strings moving on arbitrary group manifolds \cite{Gepner:1986wi}, generalising the abelian versions which arise for strings propagating on flat backgrounds. These Ka{\v{c}}-Moody algebras give rise to the worldsheet 2d CFT by the Sugawara construction. The construction of strings on arbitrary backgrounds is thus intimately linked to these Ka{\v{c}}-Moody algebras. 

\medskip

Ka{\v{c}}-Moody (KM) algebras also arise when we think of 2d CFTs augmented by additional symmetries \cite{Zamolodchikov:1985wn}. For example, a CFT with additional $U(1)$ global symmetry arises in a number of places, including the study of black holes in AdS$_3$ with charged $U(1)$ hair that are solutions to Einstein-Chern Simons theory \cite{Das:2017vej}. Virasoro with $U(1)$ KM symmetry forms the chiral algebra of a large number of theories including $\mathcal{N}=2$ superconformal field theories and theories with $\mathcal{W}_{1+\infty}$ symmetry. 

\medskip

In this paper, we will be interested in the construction of Non-Lorentzian (NL) versions of Ka{\v{c}}-Moody algebras. Specifically, we are concerned with Galilean and Carrollian CFTs in 2d with additional symmetry. Galilean and Carrollian CFTs are obtained from their relativistic counterparts by a process of contraction where the speed of light is taken to infinity (Galilean theory) or zero (Carrollian theory). In two dimensions, the symmetry algebras turn out to be isomorphic and it is in this 2d case we will focus our attention. We will obtain the algebras of interest by a contraction and then construct various properties of these algebras by methods that have no connections with the limiting procedure and can be thought of as completely intrinsic analyses. We also show that suitable singular limits also reproduce our intrinsic answers. 

\medskip

\subsection*{Possible applications}

We have a variety of applications in mind for our algebraic explorations in this work. 

\medskip

{\em{Holography of flat spacetimes}} 

\smallskip

Following closely related observation in \cite{Bagchi:2010zz}, it was shown in \cite{Duval:2014uva}  that $d$-dimensional Conformal Carrollian algebras are isomorphic to asymptotic symmetry algebras of $(d+1)$ dimensional flat spacetimes discovered first by Bondi, van der Burg, Metzner \cite{Bondi:1962px} and Sachs \cite{Sachs:1962zza} and called BMS algebras. The Carroll CFTs can thus act as putative duals to asypmtotically flat spacetimes \cite{Bagchi:2010zz, Bagchi:2012cy, Bagchi:2016bcd}. Some important evidence for this duality has been provided in e.g. \cite{Bagchi:2012xr, Barnich:2012xq, Barnich:2012aw, Bagchi:2014iea, Bagchi:2015wna, Hartong:2015usd, Jiang:2017ecm}. Of particular importance is the recent result \cite{Bagchi:2022emh} that links 3d Carroll CFTs and scattering amplitudes in 4d asymptotically flat spacetimes. 

\smallskip

Our explorations in this paper are of direct importance in the context of 3d flatspace. Here, the $U(1)$ version of NLKM symmetries are of interest for the study of Flat Space Cosmological solutions \cite{Cornalba:2002fi} with $U(1)$ hair, which are solutions of Einstein-Chern-Simons theory, like the AdS case. This was addressed recently in \cite{Basu:2017aqn, Bagchi:2022xug}. 

\smallskip

The above approach to holography in asymptotically flat spacetimes goes under the name of Carrollian holography. There is an alternate formulation called Celestial holography which posits that there is a 2d (relativistic) CFT that computes $S$-matrix elements in 4d asymptotically flat spacetimes. This has been instrumental in the uncovering of many new results in asymptotic symmetries and scattering amplitudes in 4d. The interested reader is pointed to the excellent reviews \cite{Strominger:2017zoo, Pasterski:2021rjz, Raclariu:2021zjz}. For connections between Celestial and Carrollian holography, we point to \cite{Bagchi:2022emh, Donnay:2022aba, Donnay:2022wvx}. 

\smallskip

Interestingly, of late there have been studies of tree-level massless scattering amplitudes which suggest that the asymptotic symmetries are far richer than the extended BMS group. In \cite{Banerjee:2020zlg, Banerjee:2020vnt}, it was shown that there is an $SL(2)$ current algebra at level zero underlying the symmetries of tree-level graviton amplitudes. More recently, massless scattering amplitudes have revealed an infinite dimensional $w_{1+\infty}$ algebra \cite{Strominger:2021mtt}. 

\smallskip

If we assume that the field theory duals with these additional symmetries would be related to a co-dimension one holographic description of 4d flatspace, and that these theories should live on the whole of the null boundary and not only on the celestial sphere, the structure emerging from the above discussions should only be a part of the whole symmetries, very much like the relation between the two copies of the Virasoro algebra that make up the Celestial CFT and the whole (extended) BMS$_4$ algebra. It is very likely that the algebras of interest would then be the 3d versions of the Carrollian Kac-Moody algebras that we discuss at length in this work. 

\medskip

{\em{Tensionless strings}} 

\smallskip

The tensionless limit of strings, which is analogous to the massless limit of point particles, is an important sector of string theory that remains relatively less explored. In this limit, the string worldsheet becomes null \cite{Schild:1976vq, Isberg:1993av} and the 2d relativistic conformal symmetry that arises on the tensile worldsheet is replaced by 2d Carrollian Conformal symmetry in the tensionless theory \cite{Bagchi:2013bga, Bagchi:2015nca, Bagchi:2020fpr}. The study of tensionless strings on arbitrary group manifolds would naturally incorporate the NLKM algebras we study in this paper. We would, in future, attempt a construction of a Wess-Zumino-Witten model with the NLKM algebras we discuss in this work. 

\smallskip 

It is of interest to mention that in \cite{Lindstrom:2003mg}, it was argued that tensionless strings appeared when the level of the affine algebra corresponding to the WZW model of the strings propagating in a group manifold. We believe that the intrinsic formulation of such strings should involve the NLKM algebras we are studying here. A direction of future work is the connection of these two ideas and the aim would be to show that the NLKM algebras appear when the level of the relativistic affine algebras are dialled to their critical value. 

\medskip

{\em{Other applications}} 

\smallskip

There are, of course, other natural applications. The Galilean version of our story is of relevance for understanding 2d Galilean CFTs with additional symmetry which may appear in real life non-relativistic systems. These would be of interest also in understanding non-relativistic strings in curved Newton-Cartan backgrounds \cite{Harmark:2018cdl}. 

\smallskip 

Interesting enough, Carrollian structures also arise condensed matter systems, e.g. in the physics of flat bands that is of relevance in the context of ``magic" superconductivity in bi-layer graphene and also in fractional quantum Hall systems \cite{Bagchi:2022eui}. It is easy to envision condensed matter systems with additional symmetry and hence our methods in this paper which lay the foundation for systems with Carrollian (and Galilean) affine Lie algebras, should have applicability in a wide number of condensed matter systems. Finally, a $U(1)$ affine algebra was also found to emerge in studies of the BMS scalar field theory in 2d \cite{Hao:2021urq}. It would be of interest to figure out if this is a feature of all free BMS field theories. Our methods outlined in this paper should then be useful even for free BMS field theories.  

\subsection*{Outline of the paper}
In section \ref{section2}, we will start with a brief review of Carrollian and Galilean CFTs in general dimensions, subsequently specializing the discussion to two dimensions followed by a brief introduction to Non Lorentzian Kac Moody (NLKM) algebras via contraction of relativistic Affine Lie Algebras. In section \ref{nlcurr}, we present an intrinsic carrollian derivation to the NLKM Algebra. Then, we formulate the non-Lorentzian version of the Sugawara construction for these algebras in section \ref{section4} and verify its validity by showing its consistency with the OPEs we obtained in section \ref{nlcurr}. In section \ref{section5}, we work out the example of tensionless strings on a flat background geometry which exhibits the $U(1)$ NLKM algebra as the current algebra. Section \ref{section6} contains the derivation of the Non-Lorentzian analog of Knizhnik Zamolodchikov(KZ) equations using the OPE definition of the primaries in section \ref{nlcurr}. And finally in section \ref{section7}, we derive NLKM Algebra, Sugawara Construction and the Non Lorentzian KZ equations through contractions in detail. There are six appendices which collect the details of various calculations that have been skipped in the main text for the ease of readability.

\newpage

\section{Non-Lorentzian Ka{\v{c}}-Moody algebra in 2d }\label{section2}
In this section, we will begin our analysis by building the algebra we wish to study in the remainder of the paper. The NLKM algebra will be defined as a limit from the relativistic Virasoro KM algebra. We will see later how the current part of the algebra can be used to generate the entire algebra by a NL version of the Suwagara construction. We begin by reminding the reader of the Galilean ($c\to\infty$) and the Carrollian ($c\to0$) contractions of relativistic CFTs in generic dimensions and the fact that this leads to isomorphic algebras in $d=2$

\subsection{Carrollian and Galilean CFTs}
The Galilean $(c\to\infty)$ and Carrollian $(c\to0)$ limits of the Poincare algebra leads to two different contractions of the parent relativistic algebra and two different algebras in the limit in general dimensions. 
The Poincare algebra in $D$ dimension $ISO(D-1,1)$ is given by 
\begin{align}\label{chha1}
[P_{\mu},P_{\nu}]=0,  \quad [M_{\mu\nu},P_{\rho}]=-2\eta_{\rho[\mu}P_{\nu]}, \quad [M_{\mu\nu},M_{\rho\sigma}]=4\eta_{[\mu\rho}M_{\sigma]\nu]},
\end{align}
where $\mu={0,1,2...,D-1}$, $P_{\mu}=-\partial_{\mu}$ are translation generators and $M_{\mu\nu}=x_{\mu}\partial_{\nu}-x_{\nu}\partial_{\mu}$ are Lorentz generators. Galilean limit is achieved by taking $c\to\infty$ limit, alternatively by taking the contraction $t\to t$, $x^{i}\to\epsilon x^{i}$, $\epsilon\to 0$ limit where $i\in\{1,2,...,D-1\}$ \cite{Bagchi:2009my, Bagchi:2009ca}. Under this limit we see that
\begin{align}
M_{0i}=t\partial_{i}+x_{i}\partial_{t}\to \frac{1}{\epsilon}t\partial_{i}+\epsilon x_{i}\partial_{t} \quad \implies B_{i}=\lim_{\epsilon\to 0}\epsilon M_{0i}=t\partial_{i}.
\end{align}
The spatial rotation generators $M_{ij}$ remains same under this contraction. Doing this contraction, we end up with the Galilei algebra, where all the non-zero commutators are given by
\begin{align}\label{chha2}
    [M_{ij},M_{kl}]=4\delta_{i[k}M_{l]j]}, \, [M_{ij},P_{k}]=-2\delta_{k[i}P_{j]},\, [M_{ij},B_{k}]=2\delta_{k[i}B_{j]}, \, [B_{i},H]=-P_{i}.
\end{align}
Carrollian limit is achieved by taking $c\to 0$ limit, alternatively by taking the contraction $t\to \epsilon t$, $x^{i}\to x^{i}$, $\epsilon\to 0$. Under this limit, we have 
\begin{align}
     M_{0i} =\epsilon t\partial_{i}+x_{i}\partial_{t}\to \e t\partial_{i}+ \frac{1}{\epsilon}x_{i}\partial_{t} \quad \implies B_{i}=\lim_{\epsilon\to 0}\epsilon M_{0i}=x_{i}\partial_{t},
\end{align}
with $M_{ij}$ intact again. This gives us the Carroll algebra where the non-zero commutators are given by
\begin{align}\label{chha3}
    [M_{ij},M_{kl}]=4\delta_{i[k}M_{l]j]},\, [M_{ij},P_{k}]=-2\delta_{k[i}P_{j]},\, [M_{ij},B_{k}]=-2\delta_{k[i}B_{j]}\,
 [P_{i},B_{j}]=\delta_{ij}H,
\end{align}
where $H=-\partial_{t}$ is the Hamiltonian which has now become a central element. 

\medskip

The tale of the two contractions is also true for the relativistic conformal algebra. In relativistic Conformal symmetry group there are two additional generators
\begin{align}
    D=-x^{\mu}\partial_{\mu}\hspace{5mm}K_{\mu}=-(2x_{\mu}x^{\nu}\partial_{\nu}-x^2\partial_{\mu}),
\end{align}
giving us the following additional commutators along with \eqref{chha1}
\begin{align}
    [D,P_{\mu}]=P_{\mu},\, [D,K_{\mu}]=-K_{\mu},\, [K_{\mu},P_{\nu}]=2(\eta_{\mu\nu}D-M_{\mu\nu}),\,[K_{\rho},L_{\mu\nu}]=2\eta_{\rho[\mu}K_{\nu]}.
\end{align}
Taking the non-relativistic limit this time, we will end up with the following additional generators along with the generators of Galilei algebra 
\begin{align}
  D=-(x_{i}\partial_{i}+t\partial_{t})\hspace{5mm}K=K_{0}=-(2tx_{i}\partial_{i}+t^2\partial_{t})\hspace{5mm}K_{i}=t^2\partial_{i}.
\end{align}
These generators, along with the generators of the Galilei algebra gives us the Galilean Conformal Algebra (GCA) in $D$ dimensions \cite{Bagchi:2009my} where non-zero commutators apart from \eqref{chha2} are given by
\begin{align}
 [K,B_{i}]=K_{i},\hspace{5mm}[K, P_{i}]=2B_{i}\hspace{5mm}[M_{ij},K_{r}]=-2K_{[i}\delta_{j]r}, \nonumber\\
  [H,K_{i}]=-2B_{i}\hspace{5mm}[D,K_{i}]=-2K_{i},\hspace{5mm}[D,P_i]=P_i \nonumber \\
[D,H]=H,\hspace{5mm}[H,K]=-2D,\hspace{5mm}[D,K]=-K. 
 \end{align}
When we take the Carrollian limit of the relativistic Conformal algebra, we get the following new generators apart form the Carroll generators previously encountered: 
\begin{align}
    D=-(x_{i}\partial_{i}+t\partial_{t})\hspace{5mm}K=K_{0}=-x_{i}x_{i}\partial_{t}\hspace{5mm}K_{j}=-2x_{j}(x_{i}\partial_{i}+t\partial_{t})+x_{i}x_{i}\partial_{j}.
\end{align}
These additional generators give us the following non-vanishing commutators along with those given in \eqref{chha3}
\begin{align}\label{chha4}
    [M_{ij},K_{k}]=\delta_{k[j}K_{i]}\hspace{5mm}[B_i,K_j]=\delta_{ij}K,\hspace{5mm}[D,K]=-K\nonumber\\ [K,P_{i}]=-2B_{i}\hspace{5mm}[K_{i},P_{j}]=-2D\delta_{ij}-2M_{ij},\hspace{5mm}[H,K_{i}]=2B_{i}\nonumber\\ [D,H]=-H,\hspace{5mm}[D,P_i]=-P_i\hspace{5mm}[D,K_{i}]=K_i.
\end{align}
\eqref{chha3} and \eqref{chha4} together form Carrollian Conformal Algebra (CCA). We see that in general dimensions the two different contractions give us different algebras that are not isomorphic to each other. 

\medskip

Let us now move to the interesting case of $d=2$. In 2d, the relativistic conformal algebra becomes infinite dimensional and is given by two copies of the Virasoro algebra: 
\begin{align}\label{Vir}
    [\mathcal{L}_{m},\mathcal{L}_{n}]&=(m-n)\mathcal{L}_{m+n}+\frac{c}{12}(m^3-m)\delta_{m+n,0}, \cr
    [{\bar{\mathcal{L}}}_{m},{\bar{\mathcal{L}}}_{n}]&=(m-n){\bar{\mathcal{L}}}_{m+n}+\frac{\bar{c}}{12}(m^3-m)\delta_{m+n,0}, \\
        [{\mathcal{L}}_{m},{\bar{\mathcal{L}}}_{n}]&=0. \nonumber
\end{align}
Here $c, \bar{c}$ are central charges of the Virasoro algebra (not to be confused with the speed of light which we also had called $c$ earlier). The Galilean contraction \cite{Bagchi:2009pe} of the above is given by 
\be{nrlim}
L_{n}= \mathcal{L}_n+\bar{\mathcal{L}}_n, \quad M_{n}=\epsilon(\bar{\mathcal{L}}_{n}-\mathcal{L}_{n}),
\ee
This contraction of the Virasoro algebra leads to 
\begin{subequations} \label{bms}
\begin{align}
    [L_{m},L_{n}]&=(m-n)L_{m+n}+\frac{c_{L}}{12}(m^3-m)\delta_{n+m,0}, \\
    [L_{m},M_{n}]&=(m-n)M_{m+n}+\frac{c_{M}}{12}(m^3-m)\delta_{n+m,0},\\
    [M_{m},M_{n}]&= 0. 
 \end{align}
 \end{subequations}
The way to see that this combination yields the non-relativistic limit, it is instructive to write the generators of the Virasoro algebra in cylindrical cooridinates
\be{}
\mathcal{L}_{n} = e^{in\omega} \p_\omega, \quad {\bar{\mathcal{L}}}_{n} = e^{in{\bar{\omega}}} \p_{\bar{\omega}}, \quad \omega, \bar{\omega} = \t\pm \s, 
\ee
The limit \refb{nrlim} then translates to 
\be{nrcont}
\s \to \e \s, \, \t \to \t, \,\, \e\to 0. 
\ee
which essentially means scaling velocities to be very small compared to 1 and since we are doing all of this in units of speed of light $c=1$, this is indeed the non-relativistic limit $c\to \infty$.

\medskip

On the other hand, the Carrollian contraction of the Virasoro algebra is given by 
\be{urlim}
L_{n}= \mathcal{L}_n - \bar{\mathcal{L}}_{-n}, \quad M_{n}=\epsilon(\mathcal{L}_{n} - \bar{\mathcal{L}}_{-n}).
\ee
Again if we go back to the cylindrical coordinates, this limit translates to 
\be{urcont}
\s \to \s, \, \t \to \e \t, \,\, \e\to 0. 
\ee
The velocities are very large compared to 1 now and this means $v/c \to \infty$, which equivalently translates to $c\to0$. This is thus the Carrollian limit. The surprising thing is that even this contraction yields the same algebra \refb{bms}. In order to avoid confusion with Galilean or Carrollian notions, we will exclusively call this the BMS$_3$ algebra. Galilei and Carroll contractions in $d=2$ yield isomorphic algebras and this is down to the fact that there is only one contracted and one uncontracted direction in each case. The algebra does not differentiate between a contracted spatial and a contracted temporal direction.

\subsection{NL Affine Lie algebras}
We start with the two copies of Virasoro Kac-Moody algebra whose holomorphic part is, 
\begin{align}\label{VirKM}
    [\mathcal{L}_{m},\mathcal{L}_{n}]&=(m-n)\mathcal{L}_{m+n}+\frac{c}{12}(m^3-m)\delta_{m+n,0}, \quad [\mathcal{L}_{m},j^{a}_{n}]=-n j^{a}_{m+n}\nonumber\\
    [j^a_m,j^b_n]&=i\sum_{c=1}^{dim(g)}f^{abc}j^c_{m+n}+mk\delta_{m+n,0}\delta^{ab}.
    \end{align}
There is an equivalent anti-holomorphic part with $\bar{f}_{abc}$, $\bar{c}$ and $\bar{k}$ in place of $f_{abc}$, $c$ and $k$ respectively which are not necessarily equal to their holomorphic counterparts. We are take $f^{abc} \neq \bar{f}^{abc}$ for generality, but the dimensions of the two Lie groups are the same.

\medskip

We will take a contraction of the algebra as follows. We will work with the following linear combinations of the relativistic KM generators: 
\begin{align}\label{chh14}
L_{n}= \mathcal{L}_n+\bar{\mathcal{L}}_n, \quad M_{n}=\epsilon(\bar{\mathcal{L}}_{n}-\mathcal{L}_{n}), \quad J^a_{m}= j^a_{m}+\bar{j}^a_{m}, \quad  K^a_{m}= \epsilon(\bar{j}^a_{m}-j^a_{m}).
\end{align}
We will the consider the limit $\e\to0$. The contracted algebra is given by:
\begin{subequations}\label{gkm}
\begin{align}
    [L_{m},L_{n}]&=(m-n)L_{m+n}+\frac{c_{L}}{12}(m^3-m)\delta_{n+m,0}, \\
    [L_{m},M_{n}]&=(m-n)M_{m+n}+\frac{c_{M}}{12}(m^3-m)\delta_{n+m,0},\\
    [L_m, J^{a}_n]&=-nJ^{a}_{m+n}\ , \ [L_m,K^{a}_n]=-nK^{a}_{m+n} \ , \ [M_m,J^{a}_n]=-nK^{a}_{m+n} \\
        [J^{a}_{m},J^{b}_{n}]&=i\sum_{c=1}^{dim(g)}F^{abc}J^c_{m+n} + i\sum_{c=1}^{dim(g)}G^{abc}K^c_{m+n}+mk_{1}\delta^{ab}\delta_{m+n,0}, \\
[J^{a}_m,K^{b}_n]&=i\sum_{c=1}^{dim(g)}F^{abc}K^c_{m+n}+mk_{2}\delta^{ab}\delta_{m+n,0},
\end{align}
\end{subequations}
with rest of the commutators vanishing. We recognise the first two lines as the familiar BMS$_3$ algebra, equivalently the 2d Galilean or 2d Carrollian Conformal Algebra. Throughout the paper, we will call this sub-algebra the BMS algebra. In the above, the structure constants are related to their relativistic counterparts by
\begin{align}\label{struconst}
F^{abc} = \frac{1}{2}\left(f^{abc} + \bar{f}^{abc}\right)\ , \ G^{abc} = \frac{1}{2\epsilon}\left(\bar{f}^{abc} - f^{abc}\right).
\end{align}
while the central terms are given by 
\be{}\label{centralterms}
c_L=c+\bar{c},\ \ c_{M}=\epsilon(\bar{c}-c), \,  \ k_{1}=\bar{k}+k, \ k_{2}=\epsilon(\bar{k}-k). 
\ee
We will call the algebra \refb{gkm} the 2d {\em Non-Lorentzian Ka{\v{c}}-Moody algebra}.
We can also carry out the contraction ultrarelativistically for which we take the following linear combinations of generators,
\begin{align}\label{carcomb}
    L_n=\mathcal{L}_{n}-\bar{\mathcal{L}}_{-n},  \quad M_n=\epsilon(\mathcal{L}_{n}+\bar{\mathcal{L}}_{-n}), \quad J^a_n=(j^a_n+\bar{j}^a_{-n}), \quad K^a_n=\epsilon(j^a_n-\bar{j}^a_{-n}).
\end{align}
Contracted algebra will be same as \eqref{gkm} with relations analogous to \eqref{struconst} and\eqref{centralterms} taking the following form:
\begin{align}\label{fg-car}
    F^{abc}=\frac{1}{2}\bigg(f^{abc}+\bar{f}^{abc}\bigg) \ , \ G^{abc}=\frac{1}{2\epsilon}\bigg(f^{abc}-\bar{f}^{abc}\bigg)
\end{align}
\begin{align} \label{car-cc}
    c_L=c-\bar{c},\quad c_M=\epsilon(c+\bar{c}), \quad k_1=k-\bar{k},\quad k_2=\epsilon(k+\bar{k}).
\end{align}
This \refb{gkm} will be the algebra of interest for the rest of our paper. 

\medskip

Note that if we start with 2 identical Kac-Moody algebras for the holomorphic and antiholomorphic sections, then after contraction we get 
\be{}
f^{abc} = \bar{f}^{abc} \Rightarrow F^{abc} = f^{abc}, \, G^{abc} =0.
\ee
For most of our analysis, we will use this simplified algebra, but the results can be easily generalised for the general algebra.

\smallskip

We should also clarify something about the new Lie algebra structure and our notation. We started with (the affine version of) two copies of a Lie algebra $g$ of dimension $$n = \text{dim}(g),$$ constructed from the generators $\{j^a, a \in \{1,2,\dots,n\}\}$ and $\{\bar{j}^a, a \in \{1,2,\dots, n\}\}$. Via contraction we obtain a (non-semisimple) Lie algebra $\tilde{g}$ of dimension $$\tilde{n} = {\text{dim}}(\tilde{g}) = 2 \ {\text{dim}}(g)$$ consisting of generators $\{J^a, K^a, a \in \{1,2,\dots,n\}\}$. So in all our notation, the indices $a,b,c$ run from 1 to $n = \text{dim}(g)$ (not $\tilde{n}$), and whenever we write $\text{dim}(g)$ (like in expressions of central charge $c_L$ later in the paper), we mean the dimension of the parent algebra $g$, which is actually half of the dimension of the new algebra $\tilde{g}$. This is done for the tidiness of expressions. 

\medskip

Before we conclude this section, we would like to comment on the choice of contraction of the currents to get to the NLKM algebra. Note that the chosen linear combinations of the Kac Moody generators are motivated by the Galilean/Carrollian contractions analogous to \refb{nrcont} and \refb{urcont}. The generators in cylindrical coordinates look like,
\begin{align}
	j^a_n=j^a\otimes e^{in\omega}, \quad \bar{j}^a_n=\bar{j}^a\otimes e^{in\bar{\omega}}, \quad \omega,\bar{\omega}=\tau\pm \sigma
\end{align}
Galilean limit will correspond to \refb{nrcont} and Carrollian limit corresponds to \refb{urcont}. In Galilean case, we have (upto linear order in $\epsilon$),
\begin{subequations}
\begin{align}
	j^a_n+\bar{j}^a_n=(j^a+\bar{j}^a)\otimes e^{in\tau}+in\sigma \epsilon(j^a-\bar{j}^a)\otimes e^{in\t},\\
	j^a_n-\bar{j}^a_n=(j^a-\bar{j}^a)\otimes e^{in\tau}+in\sigma \epsilon(j^a+\bar{j}^a)\otimes e^{in\t}.
\end{align}
\end{subequations}
We now have two choices, keeping the BMS contraction the same:
\begin{itemize}
	\item $J^a=j^a+\bar{j}^a, \, \, K^a=\epsilon(\bar{j}^a-j^a)\Rightarrow \refb{chh14}$. 
	\item $J^a=\bar{j}^a-j^a, K^a=\epsilon(\bar{j}^a+j^a)\Rightarrow J^a_m=\bar{j}^a_m-j^a_m, \, \, K^a_m=\epsilon(\bar{j}^a_m+j^a_m)$.
\end{itemize}  
Both of these choices lead to relations \refb{centralterms} but \refb{struconst} need to be altered for the second choice to:
\begin{align}
	F^{abc}=\frac{1}{2}(\bar{f}^{abc}-f^{abc}), \ G^{abc}=\frac{1}{2\epsilon}(f^{abc}+\bar{f}^{abc}). 
\end{align}
Similarly, in Carrollian case, we have two choices:  \refb{carcomb} and  
$$J^a_n=j^a_n-j^a_{-n}, \quad K^a_n=\epsilon(j^a_n+j^a_{-n}).$$ 
in which case \refb{car-cc} remains same but we have
\begin{align}
	F^{abc}=\frac{1}{2}(f^{abc}-\bar{f}^{abc}), \quad G^{abc}=\frac{1}{2\epsilon}(f^{abc}+\bar{f}^{abc})
\end{align} 
instead of \refb{fg-car}. Since we will be considering the special case when ${f}^{abc}=\bar{f}^{abc}$, we will stick to the first choice in this paper. For the intrinsic analysis, which only uses the algebra \refb{gkm}, of course these choices do not matter. But when we derive results from the limit, in Sec. 7, it is important to state everything works for the other contraction as well. E.g. Sugawara construction through contraction (Sec 7.3) can be carried out for the second choice if we specialize to the case $\bar{f}^{abc}=-f^{abc}$ in case of Galilean contraction.

\section{An intrinsic Carrollian derivation}\label{nlcurr}
Having derived the non-Lorentzian current algebra through a contraction, we now go on to present an intrinsically Carrollian derivation of the same, where we would not be alluding to a limiting procedure at all. This section heavily borrows from the machinery detailed in  \cite{Saha:2022gjw}, some of the important features of which are described in Appendix \ref{ApCar}. Although we will try and be self consistent so that the section (with the help of the related appendix \ref{ApCar}) stands on its own, for any details that we may have inadvertently skipped in what follows, the reader is referred back to \cite{Saha:2022gjw}.

\subsection{An Infinity of Conserved Quantities}

A 2D Carrollian CFT on the flat Carrollian background $(t,x)$ is invariant under an infinite number of 2D Carrollian conformal (CC) transformations whose infinitesimal versions are given as:
\begin{align}
x^\prime=x+\epsilon^x f(x)\hspace{2.5mm},\hspace{2.5mm}t^\prime=t+\epsilon^xtf^\prime(x)+\epsilon^tg(x)\label{17a}
\end{align}
As a consequence, the EM tensor components classically satisfy the following conditions \cite{Saha:2022gjw}:
\begin{align}
&\partial_{\mu}T^\mu_{\hspace{1.5mm}\nu}=0\hspace{2.5mm};\hspace{2.5mm}T^x_{\hspace{1.5mm}t}=0\hspace{2.5mm};\hspace{2.5mm}T^\mu_{\hspace{1.5mm}\mu}=0\nonumber\\
\Longrightarrow\hspace{2.5mm}&\partial_tT^t_{\hspace{1.5mm}t}=0\hspace{2.5mm};\hspace{2.5mm}\partial_t T^t_{\hspace{1.5mm}x}=\partial_x T^t_{\hspace{1.5mm}t}\label{16a}.
\end{align}
This allows for an infinite number of Noether currents. The two conserved currents corresponding to the symmetry transformation \eqref{17a} are noted below:
\begin{align}
&j^\mu_{\hspace{1.5mm}t}=\left(j^t_{\hspace{1.5mm}t}\hspace{1mm},\hspace{1mm}j^x_{\hspace{1.5mm}t}\right)=\left(g(x)T^t_{\hspace{1.5mm}t}\hspace{1mm},\hspace{1mm}0\right),\nonumber\\
&j^\mu_{\hspace{1.5mm}x}=\left(j^t_{\hspace{1.5mm}x}\hspace{1mm},\hspace{1mm}j^x_{\hspace{1.5mm}x}\right)=\left(f(x)T^t_{\hspace{1.5mm}x}+tf^\prime(x)T^t_{\hspace{1.5mm}t}\hspace{1mm},\hspace{1mm}-f(x)T^t_{\hspace{1.5mm}t}\right).\label{24a}
\end{align}

\medskip

Now, we suppose that there are some other pairs of fields $\{\mathcal{J}^a_x,\mathcal{J}^a_t\}$ in the theory that obey the following conditions analogous to \eqref{16a}:
\begin{align}
\partial_t\mathcal{J}^a_t(t,x)=0\hspace{2.5mm};\hspace{2.5mm}\partial_t \mathcal{J}^a_x(t,x)=\partial_x \mathcal{J}^a_t(t,x)\label{15a}
\end{align}
where $a$ is to be thought of as a `flavor' index (but $t$ and $x$ in subscript are not tensor indices). Using these fields, we can construct an infinite number of conserved quantities:
\begin{subequations}
\begin{align}
&k^{a\mu}=\left(k^{at}\hspace{1mm},\hspace{1mm}k^{ax}\right)=\left(g(x)\mathcal{J}^a_t\hspace{1mm},\hspace{1mm}0\right),\label{5a}\\
&j^{a\mu}=\left(j^{at}\hspace{1mm},\hspace{1mm}j^{ax}\right)=\left(f(x)\mathcal{J}^a_x+tf^\prime(x)
\mathcal{J}^a_t\hspace{1mm},\hspace{1mm}-f(x)\mathcal{J}^a_t\right)\label{6a}.
\end{align} 
\end{subequations}
We shall regard these conserved quantities as the Noether currents associated to some `internal' symmetries of the action.

\medskip

\subsection{Current Ward Identities}
We now consider an infinitesimal internal transformation of a (possibly multi-component) field $\Phi(t,x)$:
\begin{align}
\Phi(t,x)\rightarrow\tilde{\Phi}(t,x)=\Phi(t,x)+\epsilon^a\left(\mathcal{F}_a\cdot\Phi\right)(t,x)\label{2}
\end{align}
where $\mathcal{F}_a\cdot\Phi$ denotes the functional changes of the (multi-component) field $\Phi$ under infinitesimal transformations labeled by $\e^a$. So, the generator $G_a\Phi$ of this transformation is given by:
\begin{align}
-i\epsilon^aG_a\Phi(t,x):=\tilde{\Phi}(t,x)-\Phi(t,x)=\epsilon^a\left(\mathcal{F}_a\cdot\Phi\right)(t,x)
\end{align}

\medskip

We shall now find the Ward identity corresponding to this internal transformation which is assumed to be a symmetry of the 2D Carrollian CFT. For this purpose, we analytically continue the real space variable $x\in\mathbb{R}\cup\{\infty\}$ to the complex plane; thus, the Ward identity reads \cite{Saha:2022gjw}:
\begin{align}
\partial_\mu\langle j^\mu_{\hspace{1.5mm}a}(t,{x})X\rangle\sim -i\sum\limits_{i=1}^n\text{ }\delta(t-{t_i})\left[\frac{-{(\mathcal{F}_a)}_i\cdot\langle X\rangle}{x-x_i}+\sum_{k\geq2}\text{ }\frac{{\langle Y^{(k)}_a\rangle}_i(\mathbf{x_1},...\mathbf{x_n})}{{(x-x_i)}^{k}}\right]
\end{align}
where the yet unknown correlators ${\langle Y^{(k)}_a\rangle}_i$ depend on the transformation properties of the fields in the string-of-fields $X$ and the transformation itself and $\sim$ denotes `modulo terms holomorphic in $x$ inside $[\ldots]$'. We also use the shorthand $\mathbf{x_1} = (t_1, x_1)$. All the correlators are time-ordered. 

\medskip

Let us now assume that the symmetry transformation \eqref{2} is associated to the following conserved current operator:
\begin{align}
k^{a\mu}=\left(\mathcal{J}^a_t\hspace{1mm},\hspace{1mm}0\right)\label{7a}.
\end{align}
The corresponding Ward identity then is:
\begin{align}
&\partial_t\langle \mathcal{J}^a_t(t,{x})X\rangle\sim -i\sum\limits_{i=1}^n\text{ }\delta(t-{t_i})\left[\frac{-{(\mathcal{F}_a)}_i\cdot\langle X\rangle}{x-x_i}+\sum_{k\geq2}\text{ }\frac{{\langle Y^{(k)}_a\rangle}_i(\mathbf{x_1},...\mathbf{x_n})}{{(x-x_i)}^{k}}\right]\nonumber\\
\Rightarrow\hspace{2.5mm}&\langle \mathcal{J}^a_t(t,{x})X\rangle=-i\sum\limits_{i=1}^n\text{ }\theta(t-{t_i})\left[\frac{-{(\mathcal{F}_a)}_i\cdot\langle X\rangle}{x-x_i}+\sum_{k\geq2}\text{ }\frac{{\langle Y^{(k)}_a\rangle}_i(\mathbf{x_1},...\mathbf{x_n})}{{(x-x_i)}^{k}}\right]\label{1a}
\end{align}
where, following \cite{Saha:2022gjw}, the initial condition has been taken to be:
\begin{align}
\lim\limits_{t\rightarrow-\infty}\langle \mathcal{J}^a_t(t,{x})X\rangle=0
\end{align}
and as in 2D relativistic CFT \cite{Zamolodchikov:1985wn}, $\langle \mathcal{J}^a_t(t,{x})X\rangle$ is assumed to be finite whenever $x\neq\{x_i\}$; this condition makes the holomorphic terms inside $[\ldots]$ vanish in this Ward identity.

\medskip

Similarly, if the conserved current: 
\begin{align}
j^{a\mu}=\left(\mathcal{J}^a_x\hspace{1mm},\hspace{1mm}-\mathcal{J}^a_t\right)\label{8a}
\end{align}
is associated to another internal symmetry transformation:
\begin{align}
\Phi(t,x)\rightarrow\tilde{\Phi}(t,x)=\Phi(t,x)+\epsilon^a\left(\mathcal{G}_a\cdot\Phi\right)(t,x)
\end{align}
the corresponding Ward identity is:
\begin{align}
\langle\left(\partial_t\mathcal{J}^a_x(t,{x})-\partial_x\mathcal{J}^a_t(t,x)\right)X\rangle\sim -i\sum\limits_{i=1}^n\text{ }\delta(t-{t_i})\left[\frac{-{(\mathcal{G}_a)}_i\cdot\langle X\rangle}{x-x_i}+\sum_{k\geq2}\text{ }\frac{{\langle Y^{(k)}_a\rangle}_i(\mathbf{x_1},...\mathbf{x_n})}{{(x-x_i)}^{k}}\right]
\end{align}
Assuming the initial condition:
\begin{align}
\lim\limits_{t\rightarrow-\infty}\langle \mathcal{J}^a_x(t,{x})X\rangle=0
\end{align}
and the finite-ness property of $\langle \mathcal{J}^a_x(t,{x})X\rangle$ for $x\neq\{x_i\}$, this Ward identity together with \eqref{1a} lead us to:
\begin{align}
\langle \mathcal{J}^a_x(t,{x})X\rangle=-i\sum\limits_{i=1}^n\text{ }\theta(t-{t_i})\left[\frac{-{(\mathcal{G}_a)}_i\cdot\langle X\rangle}{x-x_i}+\sum_{k\geq2}\text{ }\frac{{\langle Z^{(k)}_a\rangle}_i(\mathbf{x_1},...\mathbf{x_n})}{{(x-x_i)}^{k}}\hspace{25mm}\right.\nonumber\\
\left.-(t-t_i)\left(\frac{-{(\mathcal{F}_a)}_i\cdot\langle X\rangle}{{(x-x_i)}^2}+\sum_{k\geq2}\text{ }k\frac{{\langle Y^{(k)}_a\rangle}_i(\mathbf{x_1},...\mathbf{x_n})}{{(x-x_i)}^{k+1}}\right)\right]\label{2a}
\end{align}
Again, the correlators ${\langle Z^{(k)}_a\rangle}_i$ can not be determined without knowing the explicit internal transformation properties of the fields in $X$.

\medskip

For future references, we note the $i\epsilon$-form \cite{Saha:2022gjw} of the Ward identities \eqref{1a} and \eqref{2a} below with $\Delta \tilde{x}_p:=x-x_p-i\epsilon(t-t_p)$ {\footnote{We hope the reader does not confuse the delta appearing in $\Delta \tilde{x}_p$ with the conformal weight $\Delta$. The $\Delta$ appearing in the difference of coordinates would always appear with a coordinate.}} :
\begin{align}
&i\langle \mathcal{J}^a_t(t,{x})X\rangle=\lim\limits_{\epsilon\rightarrow0^+}\sum\limits_{i=1}^n\left[\frac{-{(\mathcal{F}_a)}_i\cdot\langle X\rangle}{\Delta \tilde{x}_i}+\sum_{k\geq2}\text{ }\frac{{\langle Y^{(k)}_a\rangle}_i}{{(\Delta\tilde{x}_i)}^{k}}\right]\\
&i\langle \mathcal{J}^a_x(t,{x})X\rangle=\lim\limits_{\epsilon\rightarrow0^+}\sum\limits_{i=1}^n\left[\frac{-{(\mathcal{G}_a)}_i\cdot\langle X\rangle}{\Delta\tilde{x}_i}+\sum_{k\geq2}\frac{{\langle Z^{(k)}_a\rangle}_i}{{(\Delta\tilde{x}_i)}^{k}}\right.\nonumber\\
&\hspace{59mm}\left.-(t-t_i)\left(\frac{-{(\mathcal{F}_a)}_i\cdot\langle X\rangle}{{(\Delta\tilde{x}_i)}^2}+\sum_{k\geq2}k\frac{{\langle Y^{(k)}_a\rangle}_i}{{(\Delta\tilde{x}_i)}^{k+1}}\right)\right]
\end{align}

\medskip

Thus, a general (possibly multi-component) 2D Carrollian conformal field $\Phi(t,x)$ has the following OPEs (in the $i\epsilon$-form) with the current-vector components ($\Delta \tilde{x}^\prime:=x^\prime-x-i\epsilon(t^\prime-t)$):
\begin{align}
&i\mathcal{J}^a_t(t^\prime,x^\prime)\Phi(t,x)\sim\lim\limits_{\epsilon\rightarrow0^+}\left[\ldots+\frac{-(\mathcal{F}_a\cdot\Phi)(t,x)}{\Delta \tilde{x}^\prime}\right]\label{3a}\\
&i\mathcal{J}^a_x(t^\prime,x^\prime)\Phi(t,x)\sim\lim\limits_{\epsilon\rightarrow0^+}\left[\ldots+\frac{-(\mathcal{G}_a\cdot\Phi)(t,x)}{\Delta \tilde{x}^\prime}-(t^\prime-t)\left(\ldots+\frac{-(\mathcal{F}_a\cdot\Phi)(t,x)}{{(\Delta \tilde{x}^\prime)}^2}\right)\right]\label{4a}
\end{align}
where $\ldots$ represents higher order poles at $x^\prime=x$ and $\sim$ denotes `modulo terms holomorphic in $x^\prime$ that have vanishing VEVs'.

\medskip

In this work, we shall only consider currents with scaling dimension $\Delta=1$. Comparing the behavior of both sides of the OPEs \eqref{3a} and \eqref{4a} under dilatation, we then infer that the scaling dimensions of the local fields $\mathcal{F}_a\cdot\Phi$ and $\mathcal{G}_a\cdot\Phi$ must be same as that of $\Phi$. Since this is true for any arbitrary local field $\Phi$, we conclude that $\mathcal{F}_a\cdot\Phi$ and $\mathcal{G}_a\cdot\Phi$ must be linear combinations of the components of the multi-component field $\Phi$ that `internally' transforms under a bi-matrix representation of the global current symmetry algebra; all of these components must have an equal scaling dimension. Our goal is to find this global algebra and its infinite extension.

\medskip

We now express $\mathcal{F}_a\cdot\Phi$ and $\mathcal{G}_a\cdot\Phi$ as explicit linear combinations:
\begin{align}
\left(\mathcal{F}_a\cdot\Phi\right)^{i{i^\prime}}=\left(t_K^a\right)^{i}_{\hspace{1mm}j}\Phi^{j{i^\prime}}\hspace{2.5mm};\hspace{2.5mm}\left(\mathcal{G}_a\cdot\Phi\right)^{i{i^\prime}}=\Phi^{i{j^\prime}}\left(t_J^a\right)^{\hspace{1mm}{i^\prime}}_{{j^\prime}}
\end{align}
where $t^a_J$ and $t^a_K$ are just two matrices as of now. Later, we shall relate them to the generators of the internal symmetry algebra.

\medskip

\subsection{Current-primary fields}
In the operator formalism of a QFT, the conserved charge $Q_A$ is the generator of an infinitesimal symmetry transformation on the space of the quantum fields:
\begin{align}
\tilde{\Phi}(\mathbf{x})-{\Phi}(\mathbf{x})=-i\epsilon^A[Q_A\text{ },\text{ }{\Phi}(\mathbf{x})]
\end{align}
In 2D Carroll CFT, the above generator equation for any conserved charge operator $Q_A$ is related to the following contour integral prescription involving an OPE \cite{Saha:2022gjw}\footnote{Section 5 of this reference contains a derivation.}:
\begin{align}
Q_A=\frac{1}{2\pi i}\oint\limits_{C_u} dx\text{ }j^t_{\hspace{1.5mm}A}(t,x)\text{\hspace{5mm}generates\hspace{5mm}}[Q_A\text{ },{\Phi}(t,x)]=\frac{1}{2\pi i}\oint\limits_{x} dx^\prime\text{ }j^t_{\hspace{1.5mm}A}(t^+,x^\prime){\Phi}(t,x)\label{23a}
\end{align}
where $t^+>t$ and the counter-clockwise contour $C_u$ encloses the upper half-plane along with the real line. The contour around $x$ must not enclose any possible singularities of the vector field.

\medskip

The conserved quantum charges $Q^a_t[g]$ and $Q^a_x[f]$ of the respective currents \eqref{5a} and \eqref{6a} are thus given by:
\begin{align}
&Q^a_t[g]=\frac{1}{2\pi i}\oint\limits_{C_u} dx\text{ }g(x)\mathcal{J}^a_t(t,x)\label{28a}\\
&Q^a_x[f]=\frac{1}{2\pi i}\oint\limits_{C_u} dx\left[f(x)\mathcal{J}^a_x(t,x)+tf^{\prime}(x)\mathcal{J}^a_t(t,x)\right]\label{29a}
\end{align}
The conserved charges of all flavors collectively induce the following infinitesimal changes to a generic quantum field, as deduced from the OPEs \eqref{3a} and \eqref{4a}: 
\begin{align}
-i\sum_a\epsilon^a\left[Q^a_t[g^a]\text{ },\text{ }{\Phi}(\mathbf{x})\right]&=-\frac{1}{2\pi}\sum_a\epsilon^a\oint\limits_{x} dx^\prime\text{ }g^a(x^\prime)\mathcal{J}^a_t(t^+,x^\prime){\Phi}(t,x)\nonumber\\
&=\sum_a\epsilon^a\left[g^a(x)\left(t^a_K\cdot{\Phi}\right)(t,x)+(\text{h.d.t.})\right]\label{9a}\\
-i\sum_a\epsilon^a\left[Q^a_x[f^a]\text{ },\text{ }{\Phi}(\mathbf{x})\right]&=-\frac{1}{2\pi}\sum_a\epsilon^a\oint\limits_{x} dx^\prime\text{ }\left[f^a(x^\prime)\mathcal{J}^a_x(t^+,x^\prime)+t^+f^{a\prime}(x^\prime)\mathcal{J}^a_t(t^+,x^\prime)\right]{\Phi}(t,x)\nonumber\\
&=\sum_a\epsilon^a\left[f^a(x)\left({\Phi}\cdot t^a_J\right)(t,x)+tf^{a\prime}(x)\left(t^a_K\cdot{\Phi}\right)(t,x)+(\text{h.d.t.})\right]\label{10a}
\end{align}
where h.d.t. denotes terms necessarily containing derivatives (of order at least 1) of $g^a(x)$ and $f^a(x)$.

\medskip

For the currents \eqref{7a} and \eqref{8a}, we simply have $f(x)=1=g(x)$. For any arbitrary field $\Phi(t,x)$ this immediately leads to:
\begin{subequations}
\begin{align}
&-i\sum_a\epsilon^a\left[Q^a_t[1]\text{ },\text{ }{\Phi}(\mathbf{x})\right]=\sum_a\epsilon^a\left(t^a_K\cdot{\Phi}\right)(t,x)\\
&-i\sum_a\epsilon^a\left[Q^a_x[1]\text{ },\text{ }{\Phi}(\mathbf{x})\right]=\sum_a\epsilon^a\left({\Phi}\cdot t^a_J\right)(t,x)
\end{align}
\end{subequations}
Thus, the finite internal transformation that is generated by the charges $\left\{Q^a_k[1]\right\}$ is:
\begin{align}
\Phi(t,x)\rightarrow\tilde{\Phi}(t,x)=e^{-i\sum_a\epsilon^aQ^a_t[1]}\Phi(t,x)e^{i\sum_a\epsilon^aQ^a_t[1]}=e^{\sum_a\epsilon^at^a_K}\cdot\Phi(t,x)
\end{align}
which is obtained by using the BCH lemma. Similarly, the charges $\left\{Q^a_x[1]\right\}$ generate the following finite transformation:
\begin{align}
\Phi(t,x)\rightarrow\tilde{\Phi}(t,x)=\Phi(t,x)\cdot e^{\sum_a\epsilon^at^a_J}
\end{align}

\medskip

On the other hand, as derived from \eqref{9a}, an arbitrary field finitely transforms under the action of the charges $\left\{Q^a_t[g^a]\right\}$ as:
\begin{align}
\Phi(t,x)\rightarrow\tilde{\Phi}(t,x)=e^{\sum_a\epsilon^ag^a(x)t^a_K}\cdot\Phi(t,x)+\text{ extra terms}\label{11a}
\end{align}
while \eqref{10a} leads to the following finite action of the charges $\left\{Q^a_x[f^a]\right\}$:
\begin{align}
\Phi(t,x)\rightarrow\tilde{\Phi}(t,x)=e^{\sum_a\epsilon^atf^{a\prime}(x)t^a_K}\cdot\Phi(t,x)\cdot e^{\sum_a\epsilon^af^a(x)t^a_J}+\text{ extra terms}\label{12a}
\end{align}

\medskip

In view of the above discussion, we emphasize that while a generic field transforms covariantly under the action of the charges associated to the conserved currents \eqref{7a} and \eqref{8a}, that is not the case for the generic conserved currents \eqref{5a} and \eqref{6a}. A field that transforms covariantly (i.e. for which the extra terms in \eqref{11a} and \eqref{12a} vanish) even under the action of the charges of any currents of the form \eqref{5a} and \eqref{6a} is called a current-primary field. Consequently, there is no h.d.t. in \eqref{9a} and \eqref{10a} appropriate for a current-primary field. This enables us to completely specify the pole structures of the current-primary OPEs for a primary field $\Phi(t,x)$: 
\begin{align}
&\mathcal{J}^a_t(t^\prime,x^\prime)\Phi(t,x)\sim\lim\limits_{\epsilon\rightarrow0^+}i\frac{
t^a_K\cdot\Phi(t,x)}{\Delta \tilde{x}^\prime}\label{30a}\\
&\mathcal{J}^a_x(t^\prime,x^\prime)\Phi(t,x)\sim\lim\limits_{\epsilon\rightarrow0^+}i\left[\frac{\Phi(t,x)\cdot t^a_J}{\Delta \tilde{x}^\prime}-(t^\prime-t)\frac{t^a_K\cdot\Phi(t,x)}{{(\Delta \tilde{x}^\prime)}^2}\right]\label{31a}
\end{align}
that immediately imply the following Ward identities for a string $X$ of primary fields:
\begin{align}
&\langle \mathcal{J}^a_t(t,{x})X\rangle=\lim\limits_{\epsilon\rightarrow0^+}i\sum\limits_{i=1}^n\frac{{(t^a_K)}_i\cdot\langle X\rangle}{\Delta \tilde{x}_i}\\
&\langle \mathcal{J}^a_x(t,{x})X\rangle=\lim\limits_{\epsilon\rightarrow0^+}i\sum\limits_{i=1}^n\left[\frac{\langle X\rangle\cdot{(t^a_J)}_i}{\Delta\tilde{x}_i}-(t-t_i)\frac{{{(t^a_K)}_i}\cdot\langle X\rangle}{{(\Delta\tilde{x}_i)}^2}\right]
\end{align}
where ${(t^a_J)}_i$ and ${(t^a_K)}_i$ denotes transformation-matrices appropriate for the $i$-th primary field in $X$.

\medskip

\subsection{Current-Current OPEs}
To derive the current-current OPEs using the machinery just developed, we shall assume that no field in the theory has negative scaling dimension with the identity being the only field with $\Delta=0$.

\medskip

Under these assumptions, the OPEs between the EM tensor components were derived using only symmetry arguments in \cite{Saha:2022gjw}; the results are:
\begin{align}
    &T^t_{\hspace{1.5mm}t}({t^\prime,x^\prime})T^t_{\hspace{1.5mm}t}(t,x)\sim 0\nonumber\\
&T^t_{\hspace{1.5mm}x}({t^\prime,x^\prime})T^t_{\hspace{1.5mm}t}(t,x)\sim\lim\limits_{\epsilon\rightarrow0^+}-i\left[\frac{-i\frac{c_M}{2}}{{(\Delta\tilde{x}^\prime)}^4}+\frac{2T^t_{\hspace{1.5mm}t}(t,x)}{{(\Delta\tilde{x}^\prime)}^2}+\frac{{\partial_{x}} T^t_{\hspace{1.5mm}t}(t,x)}{\Delta\tilde{x}^\prime}\right]\nonumber\\
&T^t_{\hspace{1.5mm}t}({t^\prime,x^\prime})T^t_{\hspace{1.5mm}x}(t,x)\sim\lim\limits_{\epsilon\rightarrow0^+}-i\left[\frac{-i\frac{c_M}{2}}{{(\Delta\tilde{x}^\prime)}^4}+\frac{2 T^t_{\hspace{1.5mm}t}(t,x)}{{(\Delta\tilde{x}^\prime)}^2}+\frac{{\partial_{t}} T^t_{\hspace{1.5mm}x}(t,x)}{\Delta\tilde{x}^\prime}\right]\label{emgca3}\\
&T^t_{\hspace{1.5mm}x}({t^\prime,x^\prime})T^t_{\hspace{1.5mm}x}(t,x)\sim\lim\limits_{\epsilon\rightarrow0^+} -i\left[\frac{-i\frac{c_L}{2}}{{(\Delta\tilde{x}^\prime)}^4}+\frac{2T^t_{\hspace{1.5mm}x}(t,x)}{{(\Delta\tilde{x}^\prime)}^2}+\frac{{\partial_{x}}T^t_{\hspace{1.5mm}x}(t,x)}{\Delta\tilde{x}^\prime}\right.\nonumber\\
&\left.\hspace{62mm}-({t^\prime}-{t})\left(\frac{-2ic_M}{{(\Delta\tilde{x}^\prime)}^5}+\frac{4 T^t_{\hspace{1.5mm}t}(t,x)}{{(\Delta\tilde{x}^\prime)}^3}+\frac{{\partial_{t}} T^t_{\hspace{1.5mm}x}(t,x)}{{(\Delta\tilde{x}^\prime)}^2}\right)\right].\nonumber
\end{align}
The constants $c_L$ and $c_M$ are the central charges of the 2D Carrollian conformal QFT.

\medskip

We shall use the same technique to find the current-current OPEs below. Keeping in mind that here we are dealing with currents with scaling dimension $\Delta=1$, below we note the most general allowed form of the $\mathcal{J}_t-\mathcal{J}_t$ OPE compatible with the general OPE \eqref{3a}: 
\begin{align}
\mathcal{J}^a_t(t^\prime,x^\prime)\mathcal{J}^b_t(t,x)\sim \lim\limits_{\epsilon\rightarrow0^+}i\left[\frac{A^{a b}}{{\left(\Delta\tilde{x}^\prime\right)}^2}+\frac{\left(t^a_K\cdot \mathcal{J}_t\right)^b(t,x)}{\Delta\tilde{x}^\prime}\right]
\end{align}
where $A^{a b}$ is a field proportional to identity so that it has vanishing scaling dimension. Clearly, $\left(t^a_K\cdot \mathcal{J}_t\right)^b(t,x)$ must have scaling dimension $\Delta=1$. So, in a generic 2D CCFT, $\left(t^a_K\cdot \mathcal{J}_t\right)^b$ must be a linear combination of $\{\mathcal{J}^a_x,\mathcal{J}^a_t\}$. 

\medskip

Correspondingly to the classical conservation equation $\partial_t\mathcal{J}^a_t(t,x)=0$ , in the QFT we should have:
\begin{align}
\mathcal{J}^a_t(t^\prime,x^\prime)\partial_t\mathcal{J}^b_t(t,x)\sim0\hspace{2.5mm}\Rightarrow\hspace{2.5mm}\partial_t\left(t^a_K\cdot \mathcal{J}_t\right)^b(t,x)=0
\end{align}
which means that $\{\mathcal{J}^a_x\}$ can not contribute to the linear combination $\left(t^a_K\cdot \mathcal{J}_t\right)^b$.  

\medskip

Since the currents have scaling dimension $\Delta=1$, they must satisfy the bosonic\footnote{Since, as will be shown later, the currents are 2D CC primary fields with integer scaling dimension, this statement is justified \cite{Saha:2022gjw}.} exchange property. For the $\mathcal{J}^a_t(t,x)$ field, it is:
\begin{align}
\mathcal{J}^a_t(t^\prime,x^\prime)\mathcal{J}^b_t(t,x)=\mathcal{J}^b_t(t,x)\mathcal{J}^a_t(t^\prime,x^\prime)
\end{align}  
This condition implies the following restrictions:
\begin{align}
A^{ab}=A^{ba}\text{\hspace{3.5mm} and \hspace{3.5mm}}\left(t^a_K\cdot \mathcal{J}_t\right)^{b}=-\left(t^b_K\cdot \mathcal{J}_t\right)^{a}
\end{align}

\medskip

Looking at \eqref{4a}, we write the allowed form of a $\mathcal{J}_x-\mathcal{J}_t$ OPE:
\begin{align}
\mathcal{J}^a_x(t^\prime,x^\prime)\mathcal{J}^b_t(t,x)\sim \lim\limits_{\epsilon\rightarrow0^+}i\left[\frac{B^{a b}}{{\left(\Delta\tilde{x}^\prime\right)}^2}+\frac{\left(\mathcal{J}_t\cdot t^a_J\right)^b(t,x)}{\Delta\tilde{x}^\prime}-(t^\prime-t)\left(\frac{2A^{a b}}{{\left(\Delta\tilde{x}^\prime\right)}^3}+\frac{\left(t^a_K\cdot \mathcal{J}_t\right)^{b}(t,x)}{{\left(\Delta\tilde{x}^\prime\right)}^2}\right)\right]\label{13a}
\end{align}
where $B^{ab}$ are some constants. Now, we have the following restrictions:
\begin{align}
\mathcal{J}^a_x(t^\prime,x^\prime)\partial_t\mathcal{J}^b_t(t,x)\sim0\hspace{4mm}\Longrightarrow\hspace{4mm}A^{ab}=0 \text{\hspace{3mm} and \hspace{3mm}} \left(t^a_K\cdot \mathcal{J}_t\right)^{b}=0\text{\hspace{3mm} and \hspace{3mm}}\partial_t\left(\mathcal{J}_t\cdot t^a_J\right)^b(t,x)=0
\end{align}
Thus, again $\{\mathcal{J}^a_x\}$ do not contribute to the linear combination $\left(\mathcal{J}_t\cdot t^a_J\right)^b$.

\medskip

On the other hand, from \eqref{3a}, we get the following $\mathcal{J}_t-\mathcal{J}_x$ OPE:
\begin{align}
\mathcal{J}^b_t(t^\prime,x^\prime)\mathcal{J}^a_x(t,x)\sim \lim\limits_{\epsilon\rightarrow0^+}i\left[\frac{C^{ba}}{{\left(\Delta\tilde{x}^\prime\right)}^2}+\frac{\left(t^b_K\cdot \mathcal{J}_x\right)^{a}(t,x)}{\Delta\tilde{x}^\prime}\right]\label{14a}
\end{align}
with $C^{ab}$ being constants. Using the following bosonic exchange property:
\begin{align*}
\mathcal{J}^b_t(t^\prime,x^\prime)\mathcal{J}^a_x(t,x)=\mathcal{J}^a_x(t,x)\mathcal{J}^b_t(t^\prime,x^\prime)
\end{align*}
to compare the OPE \eqref{13a} with \eqref{14a}, we get the following conditions:
\begin{align}
B^{ab}=C^{ba}\text{\hspace{3.5mm} and \hspace{3.5mm}}\left(\mathcal{J}_t\cdot t^a_J\right)^b=-\left(t^b_K\cdot \mathcal{J}_x\right)^{a}
\end{align}
which implies that $\{\mathcal{J}^a_x\}$ do not appear also in the linear combination $\left(t^b_K\cdot \mathcal{J}_x\right)^{a}$.

\medskip

Finally, we write the $\mathcal{J}_x-\mathcal{J}_x$ OPE in accordance with the general form \eqref{4a}: 
\begin{align}
\mathcal{J}^a_x(t^\prime,x^\prime)\mathcal{J}^b_x(t,x)\sim \lim\limits_{\epsilon\rightarrow0^+}i\left[\frac{D^{a b}}{{\left(\Delta\tilde{x}^\prime\right)}^2}+\frac{\left(\mathcal{J}_x\cdot t^a_J\right)^b(t,x)}{\Delta\tilde{x}^\prime}-(t^\prime-t)\left(\frac{2C^{a b}}{{\left(\Delta\tilde{x}^\prime\right)}^3}+\frac{\left(t^a_K\cdot \mathcal{J}_x\right)^b(t,x)}{{\left(\Delta\tilde{x}^\prime\right)}^2}\right)\right]
\end{align}
from which, the bosonic exchange property:
\begin{align*}
\mathcal{J}^b_x(t^\prime,x^\prime)\mathcal{J}^a_x(t,x)=\mathcal{J}^a_x(t,x)\mathcal{J}^b_x(t^\prime,x^\prime)
\end{align*}
leads to the following conditions:
\begin{align}
&D^{ab}=D^{ba}\hspace{2.5mm};\hspace{2.5mm}C^{ab}=C^{ba}\\
\left(\mathcal{J}_x\cdot t^a_J\right)^b=-\left(\mathcal{J}_x\cdot t^b_J\right)^a\hspace{2.5mm};\hspace{2.5mm}&\left(t^a_K\cdot \mathcal{J}_x\right)^b=-\left(t^b_K\cdot \mathcal{J}_x\right)^a\hspace{2.5mm};\hspace{2.5mm}\partial_t\left(\mathcal{J}_x\cdot t^a_J\right)^b=\partial_x\left(t^a_K\cdot \mathcal{J}_x\right)^b
\end{align}
It can be readily checked that these conditions are compatible with the quantum versions (in the OPE language) of the classical conservation laws \eqref{15a}.

\medskip

We now explicitly write the allowed forms of the linear combinations appearing in the above OPEs:
\begin{align}
&\left(\mathcal{J}_t\cdot t^a_J\right)^b=\left(t^a_K\cdot \mathcal{J}_x\right)^b=F^{abc}\mathcal{J}^c_t&&\text{ with }\hspace{2.5mm}F^{abc}=-F^{bac},\nonumber\\
&\left(\mathcal{J}_x\cdot t^a_J\right)^b=F^{abc}\mathcal{J}^c_x+G^{abc}\mathcal{J}^c_t&&\text{ with }\hspace{2.5mm}G^{abc}=-G^{bac}.
\end{align}

\medskip

Thus, the final forms of the current-current OPEs are:
\begin{align}
&\mathcal{J}^a_t(t^\prime,x^\prime)\mathcal{J}^b_t(t,x)\sim0\nonumber\\
&\mathcal{J}^a_x(t^\prime,x^\prime)\mathcal{J}^b_t(t,x)\sim \lim\limits_{\epsilon\rightarrow0^+}i\left[\frac{C^{ab}}{{\left(\Delta\tilde{x}^\prime\right)}^2}+\frac{F^{abc}\mathcal{J}^c_t(t,x)}{\Delta\tilde{x}^\prime}\right]\nonumber\\
&\mathcal{J}^a_t(t^\prime,x^\prime)\mathcal{J}^b_x(t,x)\sim \lim\limits_{\epsilon\rightarrow0^+}i\left[\frac{C^{ab}}{{\left(\Delta\tilde{x}^\prime\right)}^2}+\frac{F^{abc}\mathcal{J}^c_t(t,x)}{\Delta\tilde{x}^\prime}\right]\label{33a}\\
&\mathcal{J}^a_x(t^\prime,x^\prime)\mathcal{J}^b_x(t,x)\sim \lim\limits_{\epsilon\rightarrow0^+}i\left[\frac{D^{a b}}{{\left(\Delta\tilde{x}^\prime\right)}^2}+\frac{\left(F^{a b c}\mathcal{J}^c_x+G^{abc}\mathcal{J}^c_t\right)(t,x)}{\Delta\tilde{x}^\prime}-(t^\prime-t)\left(\frac{2C^{a b}}{{\left(\Delta\tilde{x}^\prime\right)}^3}+\frac{F^{a b c}\mathcal{J}^c_t(t,x)}{{\left(\Delta\tilde{x}^\prime\right)}^2}\right)\right]\nonumber
\end{align}
These OPEs imply that the currents themselves are not current-primary fields in general.

\medskip

\subsection{Global internal symmetry}
All the correlation functions in the theory must be invariant under the global internal transformations associated to which are the conserved currents \eqref{7a} and \eqref{8a}. This fact will put constraints on $C^{ab}$ and $D^{ab}$, as we will now see.

\medskip

We begin by noting the following 2-point correlators between the currents, from the above OPEs:
\begin{align}
&\left\langle \mathcal{J}^a_t(t^\prime,x^\prime)\mathcal{J}^b_t(t,x)\right\rangle=0\nonumber\\
&\left\langle \mathcal{J}^a_t(t^\prime,x^\prime)\mathcal{J}^b_x(t,x)\right\rangle=\left\langle \mathcal{J}^a_x(t^\prime,x^\prime)\mathcal{J}^b_t(t,x)\right\rangle=\lim\limits_{\epsilon\rightarrow0^+}i\frac{C^{ab}}{{\left(\Delta\tilde{x}^\prime\right)}^2}\\
&\left\langle \mathcal{J}^a_x(t^\prime,x^\prime)\mathcal{J}^b_x(t,x)\right\rangle=\lim\limits_{\epsilon\rightarrow0^+}i\left[\frac{D^{a b}}{{\left(\Delta\tilde{x}^\prime\right)}^2}-(t^\prime-t)\frac{2C^{a b}}{{\left(\Delta\tilde{x}^\prime\right)}^3}\right]\nonumber
\end{align}
since the currents, with $\Delta=1$, must have vanishing VEVs.

\medskip

Next, due to the global internal symmetry, an arbitrary $n$-point correlator in the theory must satisfy:
\begin{align*}
&\sum_{i=1}^n\left\langle{\Phi_1}(t_1,x_1)\ldots\left({\Phi_i}\cdot t^a_J\right)(t_i,x_i)\ldots{\Phi_n}(t_n,x_n)\right\rangle=0\\
&\sum_{i=1}^n\left\langle{\Phi_1}(t_1,x_1)\ldots\left(t^a_K\cdot{\Phi_i}\right)(t_i,x_i)\ldots{\Phi_n}(t_n,x_n)\right\rangle=0
\end{align*}
Thus the 2-point current correlators explicitly satisfy:
\begin{align}
&\left\langle\left(\mathcal{J}_x\cdot t^a_J\right)^b(t_1,x_1)\mathcal{J}^c_x(t_2,x_2)\right\rangle+\left\langle \mathcal{J}^b_x(t_1,x_1)\left(\mathcal{J}_x\cdot t^a_J\right)^c(t_2,x_2)\right\rangle=0\nonumber\\
\Longrightarrow\hspace{2.5mm} &F^{abd}D^{dc}+F^{acd}D^{db}+G^{abd}C^{dc}+G^{acd}C^{db}=0\hspace{5mm}\text{ and }\hspace{5mm}F^{abd}C^{dc}+F^{acd}C^{db}=0\label{18a}
\end{align}

\medskip

The analogues relations obtained from the invariance of the 3-point current correlators are:
\begin{align}
F^{abe}F^{ecf}C^{fd}+F^{cae}F^{ebf}C^{fd}&+F^{ade}F^{bcf}C^{fe}=0\nonumber\\
F^{abe}F^{ecf}D^{fd}+F^{cae}F^{ebf}D^{fd}&+F^{ade}F^{bcf}D^{fe}+\left(F^{abe}G^{ecf}+G^{abe}F^{ecf}\right)C^{fd}\nonumber\\
&+\left(F^{cae}G^{ebf}+G^{cae}F^{ebf}\right)C^{fd}+\left(F^{ade}G^{bcf}+G^{ade}F^{bcf}\right)C^{fe}=0\label{19a}
\end{align}

\medskip

In what follows, we shall see that $\{F^{abc}\}$ and $\{G^{abc}\}$ must also obey the following constraints arising as the Jacobi identity of the infinite-dimensional Lie algebra of the current-modes, which we had previously described in our initial algebraic description from the contraction in \refb{gkm} and also will derive independently later in this section \eqref{34a}:
\begin{align}
&F^{abe}F^{ecd}+F^{cae}F^{ebd}+F^{bce}F^{ead}=0\nonumber\\
&F^{abe}G^{ecd}+G^{abe}F^{ecd}+F^{cae}G^{ebd}+G^{cae}F^{ebd}+F^{bce}G^{ead}+G^{bce}F^{ead}=0\label{20a}
\end{align}
No new constraint for $\{F^{abc}\}$ and $\{G^{abc}\}$ arises from the global internal invariance of $n$-point current correlators for $n\geq4$ .

\medskip

Upon comparison, we notice that \eqref{19a} reduces to \eqref{20a} if we choose:
\begin{align}
D^{ab}=-ik_1\delta^{ab}\hspace{5mm}\text{and}\hspace{5mm}C^{ab}=-ik_2\delta^{ab}\hspace{2.5mm}\text{with}\hspace{2.5mm}k_2\neq0\label{21a}
\end{align}
In that case, $F^{abc}$ and $G^{abc}$ are anti-symmetric in all indices, as seen from \eqref{18a}.

\medskip

\subsection{EM tensor-current OPEs}
We now show under the assumption that no field in the theory has negative scaling dimension with the identity field being the only one with $\Delta=0$ , that the currents $\mathcal{J}^a_x(t,x)$ and $\mathcal{J}^a_t(t,x)$ must transform as a rank-$\frac{1}{2}$ primary multiplet under 2D CC transformations.

\medskip

From \cite{Saha:2022gjw}, we recall the OPEs of a general 2D CC (multi-component) field $\Phi_{(l)}(t,x)$ having scaling dimension $\Delta$, Carrollian boost-charge $\xi$ and boost rank $l$ with the EM tensor components:
\begin{align}
&T^t_{\hspace{1.5mm}x}({t^\prime,x^\prime})\Phi_{(l)}(t,x)\sim\lim\limits_{\epsilon\rightarrow0^+} -i\left[\ldots+\frac{{{\Delta}}\Phi_{(l)}(t,x)}{{(\Delta\tilde{x}^\prime)}^2}+\frac{{\partial_{x}}\Phi_{(l)}(t,x)}{\Delta\tilde{x}^\prime}\right.\nonumber\\
&\left.\hspace{69mm}-({t^\prime}-{t})\left(\ldots+\frac{2\left(\bm{\xi}\cdot\Phi_{(l)}\right)(t,x)}{{(\Delta\tilde{x}^\prime)}^3}+\frac{{\partial_{t}}\Phi_{(l)}(t,x)}{{(\Delta\tilde{x}^\prime)}^2}\right)\right]\nonumber\\
&T^t_{\hspace{1.5mm}t}(t^\prime,x^\prime)\Phi_{(l)}(t,x)\sim\lim\limits_{\epsilon\rightarrow0^+} -i\left[\ldots+\frac{\left(\bm{\xi}\cdot\Phi_{(l)}\right)(t,x)}{{(\Delta\tilde{x}^\prime)}^2}+\frac{{\partial_{t}} \Phi_{(l)}(t,x)}{\Delta\tilde{x}^\prime}\right]\label{Mbappe}
\end{align}
The defining feature of 2D CC primary fields is the vanishing of the higher order poles in the above OPEs that leads to:
\begin{align}
&T^t_{\hspace{1.5mm}x}({t^\prime,x^\prime})\Phi_{(l)}(t,x)\sim\lim\limits_{\epsilon\rightarrow0^+} -i\left[\frac{{{\Delta}}\Phi_{(l)}(t,x)}{{(\Delta\tilde{x}^\prime)}^2}+\frac{{\partial_{x}}\Phi_{(l)}(t,x)}{\Delta\tilde{x}^\prime}\right.\nonumber\\
&\left.\hspace{69mm}-({t^\prime}-{t})\left(\frac{2\left(\bm{\xi}\cdot\Phi_{(l)}\right)(t,x)}{{(\Delta\tilde{x}^\prime)}^3}+\frac{{\partial_{t}}\Phi_{(l)}(t,x)}{{(\Delta\tilde{x}^\prime)}^2}\right)\right]\nonumber\\
&T^t_{\hspace{1.5mm}t}(t^\prime,x^\prime)\Phi_{(l)}(t,x)\sim\lim\limits_{\epsilon\rightarrow0^+} -i\left[\frac{\left(\bm{\xi}\cdot\Phi_{(l)}\right)(t,x)}{{(\Delta\tilde{x}^\prime)}^2}+\frac{{\partial_{t}} \Phi_{(l)}(t,x)}{\Delta\tilde{x}^\prime}\right]\label{emgca4}
\end{align}

\medskip

Now, along with the above assumption, the classical time-independence of the field $\mathcal{J}^a_t$ restricts the general OPE \eqref{Mbappe} to the following form:
\begin{align}
T^t_{\hspace{1.5mm}t}(t^\prime,x^\prime)\mathcal{J}^a_t(t,x)\sim\lim\limits_{\epsilon\rightarrow0^+} -i\left[\frac{A^a_1}{{(\Delta\tilde{x}^\prime)}^3}+\frac{\left(\bm{\xi}\cdot \mathcal{J}^a_t\right)(t,x)}{{(\Delta\tilde{x}^\prime)}^2}\right]
\end{align}
where the field $A^a_1$ is proportional to the identity field. Similarly, we write the most general allowed form of the following OPE from \eqref{Mbappe}: 
\begin{align}
T^t_{\hspace{1.5mm}x}(t^\prime,x^\prime)\mathcal{J}^a_t(t,x)\sim\lim\limits_{\epsilon\rightarrow0^+} -i\left[\frac{A^a_2}{{(\Delta\tilde{x}^\prime)}^3}+\frac{\mathcal{J}^a_t(t,x)}{{(\Delta\tilde{x}^\prime)}^2}+\frac{\partial_x\mathcal{J}^a_t(t,x)}{\Delta\tilde{x}^\prime}-(t^\prime-t)\left\{\frac{3A^a_1}{{(\Delta\tilde{x}^\prime)}^4}+\frac{2\left(\bm{\xi}\cdot \mathcal{J}^a_t\right)(t,x)}{{(\Delta\tilde{x}^\prime)}^3}\right\}\right]
\end{align}
with $A^a_2$ being another constant. But, the quantum counterpart of the conservation law \eqref{15a} leads to:
\begin{align}
&T^t_{\hspace{1.5mm}t}(t^\prime,x^\prime)\partial_t\mathcal{J}^a_t(t,x)\sim0\text{ }\Longrightarrow\text{ }\partial_t\left(\bm{\xi}\cdot\mathcal{J}^a_t\right)(t,x)=0,\nonumber\\
&T^t_{\hspace{1.5mm}x}(t^\prime,x^\prime)\partial_t\mathcal{J}^a_t(t,x)\sim0\text{ }\Longrightarrow\text{ } A^a_1=0\hspace{2.5mm}\text{ and }\hspace{2.5mm}\bm{\xi}\cdot\mathcal{J}^a_t=0.
\end{align}

\medskip

The OPEs for $\mathcal{J}^a_x(t,x)$ may have the most general form given below:
\begin{align}
T^t_{\hspace{1.5mm}t}(t^\prime,x^\prime)\mathcal{J}^a_x(t,x)\sim\lim\limits_{\epsilon\rightarrow0^+} -i\left[\frac{A^a_3}{{(\Delta\tilde{x}^\prime)}^3}+\frac{\left(\bm{\xi}\cdot \mathcal{J}^a_x\right)(t,x)}{{(\Delta\tilde{x}^\prime)}^2}+\frac{\partial_t\mathcal{J}^a_x(t,x)}{\Delta\tilde{x}^\prime}\right]
\end{align}
with the conservation law \eqref{15a} forcing:
\begin{align}
\partial_t\left(\bm{\xi}\cdot \mathcal{J}^a_x\right)(t,x)=0.
\end{align}
The remaining one OPE is given below:
\begin{align}
&T^t_{\hspace{1.5mm}x}(t^\prime,x^\prime)\mathcal{J}^a_x(t,x)\sim\lim\limits_{\epsilon\rightarrow0^+} -i\left[\frac{A^a_4}{{(\Delta\tilde{x}^\prime)}^3}+\frac{\mathcal{J}^a_x(t,x)}{{(\Delta\tilde{x}^\prime)}^2}+\frac{\partial_x\mathcal{J}^a_x(t,x)}{\Delta\tilde{x}^\prime}\right.\\
&\left.\hspace{56.5mm}-(t^\prime-t)\left(\frac{3A^a_3}{{(\Delta\tilde{x}^\prime)}^4}+\frac{2\left(\bm{\xi}\cdot \mathcal{J}^a_x\right)(t,x)}{{(\Delta\tilde{x}^\prime)}^3}+\frac{\partial_t\mathcal{J}^a_x(t,x)}{{(\Delta\tilde{x}^\prime)}^2}\right)\right] \nonumber
\end{align}
where $A^a_3$ and $A^a_4$ are constants. From the quantum version of \eqref{15a}, one then obtains:
\begin{align}
T^t_{\hspace{1.5mm}x}(t^\prime,x^\prime)\left[\partial_t \mathcal{J}^a_x(t,x)-\partial_x \mathcal{J}^a_t(t,x)\right]\sim0\hspace{5mm}
\Longrightarrow\hspace{5mm}\bm{\xi}\cdot \mathcal{J}^a_x=\mathcal{J}^a_t \text{\hspace{3.5mm} and \hspace{3.5mm}} A^a_2=A^a_3
\end{align}
i.e. the currents $\mathcal{J}^a_x(t,x)$ and $\mathcal{J}^a_t(t,x)$ transform under Carrollian boost as a rank-$\frac{1}{2}$ multiplet with boost charge $\xi=1$. 

\medskip

The OPEs for the currents then are:
\begin{align}
&T^t_{\hspace{1.5mm}t}(t^\prime,x^\prime)\mathcal{J}^a_t(t,x)\sim0\nonumber\\
&T^t_{\hspace{1.5mm}x}(t^\prime,x^\prime)\mathcal{J}^a_t(t,x)\sim\lim\limits_{\epsilon\rightarrow0^+} -i\left[\frac{A^a_3}{{(\Delta\tilde{x}^\prime)}^3}+\frac{\mathcal{J}^a_t(t,x)}{{(\Delta\tilde{x}^\prime)}^2}+\frac{\partial_x\mathcal{J}^a_t(t,x)}{\Delta\tilde{x}^\prime}\right]\nonumber\\
&T^t_{\hspace{1.5mm}t}(t^\prime,x^\prime)\mathcal{J}^a_x(t,x)\sim\lim\limits_{\epsilon\rightarrow0^+} -i\left[\frac{A^a_3}{{(\Delta\tilde{x}^\prime)}^3}+\frac{\mathcal{J}^a_t(t,x)}{{(\Delta\tilde{x}^\prime)}^2}+\frac{\partial_t\mathcal{J}^a_x(t,x)}{\Delta\tilde{x}^\prime}\right]\label{22a}\\
&T^t_{\hspace{1.5mm}x}(t^\prime,x^\prime)\mathcal{J}^a_x(t,x)\sim\lim\limits_{\epsilon\rightarrow0^+} -i\left[\frac{A^a_4}{{(\Delta\tilde{x}^\prime)}^3}+\frac{\mathcal{J}^a_x(t,x)}{{(\Delta\tilde{x}^\prime)}^2}+\frac{\partial_x\mathcal{J}^a_x(t,x)}{\Delta\tilde{x}^\prime}\right.\nonumber\\
&\left.\hspace{56.5mm}-(t^\prime-t)\left(\frac{3A^a_3}{{(\Delta\tilde{x}^\prime)}^4}+\frac{2\mathcal{J}^a_t(t,x)}{{(\Delta\tilde{x}^\prime)}^3}+\frac{\partial_t\mathcal{J}^a_x(t,x)}{{(\Delta\tilde{x}^\prime)}^2}\right)\right]\nonumber
\end{align}

\medskip

On the other hand, applying the bosonic exchange property between the currents and the EM tensor components, we obtain:
\begin{align}\label{TJ1}
&\mathcal{J}^a_t(t^\prime,x^\prime)T^t_{\hspace{1.5mm}t}(t,x)\sim0\nonumber\\
&\mathcal{J}^a_t(t^\prime,x^\prime)T^t_{\hspace{1.5mm}x}(t,x)\sim\lim\limits_{\epsilon\rightarrow0^+} -i\left[-\frac{A^a_3}{{(\Delta\tilde{x}^\prime)}^3}+\frac{\mathcal{J}^a_t(t,x)}{{(\Delta\tilde{x}^\prime)}^2}\right]\nonumber\\
&\mathcal{J}^a_x(t^\prime,x^\prime)T^t_{\hspace{1.5mm}t}(t,x)\sim\lim\limits_{\epsilon\rightarrow0^+} -i\left[-\frac{A^a_3}{{(\Delta\tilde{x}^\prime)}^3}+\frac{\mathcal{J}^a_t(t,x)}{{(\Delta\tilde{x}^\prime)}^2}\right]\\
&\mathcal{J}^a_x(t^\prime,x^\prime)T^t_{\hspace{1.5mm}x}(t,x)\sim\lim\limits_{\epsilon\rightarrow0^+} -i\left[-\frac{A^a_4}{{(\Delta\tilde{x}^\prime)}^3}+\frac{\mathcal{J}^a_x(t,x)}{{(\Delta\tilde{x}^\prime)}^2}-(t^\prime-t)\left(-\frac{3A^a_3}{{(\Delta\tilde{x}^\prime)}^4}+\frac{2\mathcal{J}^a_t(t,x)}{{(\Delta\tilde{x}^\prime)}^3}\right)\right]\nonumber
\end{align}
Comparing these with \eqref{3a} and \eqref{4a}, we immediately note the following:
\begin{align}
\left(t^a_K\cdot T^t_{\hspace{1.5mm}t}\right)(t,x)=\left(t^a_K\cdot T^t_{\hspace{1.5mm}x}\right)(t,x)=\left(T^t_{\hspace{1.5mm}t}\cdot t^a_J\right)(t,x)=\left(T^t_{\hspace{1.5mm}x}\cdot t^a_J\right)(t,x)=0
\end{align}
which implies that the EM tensor components actually transform under the singlet representation of the global internal symmetry algebra.

\medskip

Next we look at the consequences of the global internal symmetry on the 2-point correlators between the currents and the EM tensor components to find possible constraints on $A^a_3$ and $A^a_4$. It suffices to consider the following:
\begin{align}
\left\langle\left(\mathcal{J}_x\cdot t^a_J\right)^b(t_1,x_1)T^t_{\hspace{1.5mm}x}(t_2,x_2)\right\rangle+\left\langle \mathcal{J}^b_x(t_1,x_1)\left(T^t_{\hspace{1.5mm}x}\cdot t^a_J\right)(t_2,x_2)\right\rangle=0\hspace{2.5mm}\Longrightarrow\hspace{2.5mm}A^a_3=A^a_4=0
\end{align}
Causing the vanishing of the poles of appropriate orders in the OPEs \eqref{22a}, we finally conclude, comparing with \eqref{emgca4}, that:
\begin{align*}
\text{the currents tranform as a 2D CC primary multiplet of rank-$\frac{1}{2}$ with $\Delta=\xi=1$ .}
\end{align*}

\medskip

Thus, the currents have the following infinitesimal 2D CC transformation property under \eqref{17a}, as is obtained using the prescription \eqref{23a} for the corresponding conserved currents \eqref{24a}:
\begin{align}
&-iG_x\mathcal{J}^a_t(t,x)=-[f(x)\partial_x+f^\prime(x)]\mathcal{J}^a_t(t,x)\hspace{5mm};\hspace{5mm}-iG_t\mathcal{J}^a_t(t,x)=0\nonumber\\
&-iG_t\mathcal{J}^a_x(t,x)=-g(x)\partial_t\mathcal{J}^a_x(t,x)-g^{\prime}(x)\mathcal{J}^a_t(t,x)\label{AS}\\
&-iG_x\mathcal{J}^a_x(t,x)=-[f(x)\partial_x+tf^\prime(x)\partial_t+f^\prime(x)]\mathcal{J}^a_x(t,x)-tf^{\prime\prime}(x)\mathcal{J}^a_t(t,x)\nonumber
\end{align}

\medskip

As an aside, from \eqref{TJ1} we note down below the infinitesimal internal transformation properties of the EM tensor components, generated by the conserved charges of the currents \eqref{5a} and \eqref{6a}:
\begin{align}
&-iG_JT^t_{\hspace{1.5mm}t}(t,x)=-f^{a\prime}(x)\mathcal{J}^a_t(t,x)\hspace{6mm};\hspace{5mm}-iG_KT^t_{\hspace{1.5mm}t}(t,x)=0\label{32a}\\
&-iG_KT^t_{\hspace{1.5mm}x}(t,x)=-g^{a{\prime}}(x)\mathcal{J}^a_t(t,x)\hspace{5mm};\hspace{5mm}-iG_JT^t_{\hspace{1.5mm}x}(t,x)=-f^{a\prime}(x)\mathcal{J}^a_x(t,x)-tf^{a\prime\prime}(x)\mathcal{J}^a_t(t,x)  \nonumber
\end{align}

\medskip

\subsection{The Algebra of Modes}
The EM tensor components have the following mode-expansions:
\begin{align}
    &{T}^t_{\hspace{1.5mm}t}(t,x)=-i\sum_{n\in\mathbb{Z}}x^{-n-2}M_n\hspace{7mm};\hspace{7mm}{T}^t_{\hspace{1.5mm}x}(t,x)=-i\sum_{n\in\mathbb{Z}}x^{-n-2}\text{ }[L_n-(n+2)\frac{t}{x}M_n]\label{emgca2}\\
\Longrightarrow\hspace{2.5mm}&M_n=i\oint\limits_0\frac{dx}{2\pi i}\text{ }x^{n+1}\text{ }{T}^t_{\hspace{1.5mm}t}(t,x)\hspace{4mm};\hspace{4mm}L_n=i\oint\limits_0\frac{dx}{2\pi i}\left[x^{n+1}{T}^t_{\hspace{1.5mm}x}(t,x)+(n+1)x^nt\text{ }{T}^t_{\hspace{1.5mm}t}(t,x)\right]\label{eq:36}
\end{align}
In \cite{Saha:2022gjw}, it was shown from the EM tensor OPEs \eqref{emgca3} that the EM tensor modes indeed generate the centrally extended BMS$_3$ algebra:
\begin{align}
    &\left[M_n\hspace{1mm},\hspace{1mm}M_m\right]=0\nonumber\\
&\left[L_n\hspace{1mm},\hspace{1mm}M_m\right]=(n-m)M_{n+m}+\frac{c_M}{12}(n^3-n)\delta_{n+m,0}\\
&\left[L_n\hspace{1mm},\hspace{1mm}L_m\right]=(n-m)L_{n+m}+\frac{c_L}{12}(n^3-n)\delta_{n+m,0}\nonumber
\end{align}

\medskip

The infinitesimal 2D CC transformation properties of the currents are expressed in the operator language \cite{Saha:2022gjw} using the prescription \eqref{23a} for the charges in \eqref{eq:36}:
\begin{align}
&[L_n\text{ },\mathcal{J}^a_x(t,x)]=[x^{n+1}\partial_x+t(n+1)x^n\partial_t+(n+1)x^n]\mathcal{J}^a_x(t,x)+t(n+1)nx^{n-1}\mathcal{J}^a_t(t,x)\nonumber\\
&[M_n\text{ },\mathcal{J}^a_x(t,x)]=x^{n+1}\partial_t\mathcal{J}^a_x(t,x)+(n+1)x^{n}\mathcal{J}^a_t(t,x)\\
&[L_n\text{ },\mathcal{J}^a_t(t,x)]=x^{n+1}\partial_x\mathcal{J}^a_t(t,x)+(n+1)x^{n}\mathcal{J}^a_t(t,x)\hspace{5mm};\hspace{5mm}[M_n\text{ },\mathcal{J}^a_t(t,x)]=0\nonumber
\end{align} 

\medskip

Next, the classical conservation laws \eqref{15a} imply the following space-time dependence of the fields $\{\mathcal{J}^a_t\}$ and $\{\mathcal{J}^a_x\}$:
\begin{align*}
&\partial_t\mathcal{J}^a_t(t,x)=0\hspace{2.5mm}\Longrightarrow\hspace{2.5mm}\mathcal{J}^a_t(t,x)=\mathcal{J}^a_t(x)\\
&\partial_t\mathcal{J}^a_x(t,x)=\partial_x\mathcal{J}^a_t(x)\hspace{2.5mm}\Longrightarrow\hspace{2.5mm}\mathcal{J}^a_x(t,x)=t\partial_x\mathcal{J}^a_t(x)+R^a(x)
\end{align*}
with $R^a(x)$ being arbitrary functions.

\medskip

Guided by the above functional dependence and using the fact that the pair of the fields $\mathcal{J}^a_x$ and $\mathcal{J}^a_t$ forms a 2D CC primary rank-$\frac{1}{2}$ multiplet with scaling dimension $\Delta=1$, we write down the mode-expansions following \cite{Saha:2022gjw}:
\begin{align}
&\mathcal{J}^a_t(x)=\sum_{n\in\mathbb{Z}}x^{-n-1}K^a_n\hspace{5mm};\hspace{5mm}\mathcal{J}^a_x(t,x)=\sum_{n\in\mathbb{Z}}x^{-n-1}\left[J^a_n-(n+1)\frac{t}{x} K^a_n\right]\label{26a}\\
&K^a_n=\oint\limits_{C_u}\frac{dx}{2\pi i}\text{ }x^{n}\mathcal{J}^a_t(x)\hspace{5.5mm};\hspace{5mm}J^a_n=\oint\limits_{C_u}\frac{dx}{2\pi i}\left[x^{n}\mathcal{J}^a_x(t,x)+nx^{n-1}t\mathcal{J}^a_t(x)\right]\label{27a}
\end{align}
where the counter-clockwise contour $C_u$ encloses the upper half-plane and the real line. 

\medskip

Comparing \eqref{27a} with \eqref{28a} and \eqref{29a}, we immediately see that:
\begin{align*}
&\text{$J^a_n$ is the conserved charge of the current $j^{a\mu}=\left(x^n\mathcal{J}^a_x+tnx^{n-1}
\mathcal{J}^a_t\hspace{1mm},\hspace{1mm}-x^n\mathcal{J}^a_t\right)$}\\
&\text{$K^a_n$ is the conserved charge of the current $k^{a\mu}=\left(x^n\mathcal{J}^a_t\hspace{1mm},\hspace{1mm}0\right)$}
\end{align*}
Thus, using the prescription \eqref{23a} on the OPEs \eqref{30a} and \eqref{31a} we reach the definition of a current primary field ${\Phi}(t,x)$ in the operator formalism (for any $n\in\mathbb{Z}$): 
\begin{subequations}\label{curprimcom}
\begin{align}
& [K^a_n\text{ },{\Phi}(t,x)]=ix^n\left(t^a_K\cdot{\Phi}\right)(t,x), \\
& [J^a_n\text{ },{\Phi}(t,x)]=i\left[x^n\left(\Phi\cdot t^a_J\right)+tnx^{n-1}\left(t^a_K\cdot{\Phi}\right)\right](t,x)
\end{align}
\end{subequations}
For the EM tensor components, the analogues commutation relations are found directly from \eqref{32a}:
\begin{align}
&[K^a_n\text{ },T^t_{\hspace{1.5mm}t}(t,x)]=0,\, \, [J^a_n\text{ },T^t_{\hspace{1.5mm}t}(t,x)]=-inx^{n-1}\mathcal{J}^a_t(t,x), \, \,[K^a_n\text{ },T^t_{\hspace{1.5mm}x}(t,x)]=-inx^{n-1}\mathcal{J}^a_t(t,x), \nonumber\\
&[J^a_n\text{ },T^t_{\hspace{1.5mm}x}(t,x)]=-i\left[nx^{n-1}\mathcal{J}^a_x+tn(n-1)x^{n-2}\mathcal{J}^a_t\right](t,x)
\end{align}
Substituting the current mode-expansion \eqref{26a} and EM tensor mode expansion \eqref{emgca2} in these operator relations, we obtain the cross-commutation relations between the modes of the EM tensor and the currents:
\begin{align}
[L_n\text{ },J^a_m]=-mJ^a_{m+n}\hspace{3mm};\hspace{3mm}[M_n\text{ },J^a_m]=-mK^a_{m+n}=[L_n\text{ },K^a_m]\hspace{3mm};\hspace{3mm}[M_n\text{ },K^a_m]=0
\end{align}

\medskip

Similarly, from the current-current OPEs \eqref{33a}, using the condition \eqref{21a} and the current mode-expansion \eqref{26a}, we reach the Lie algebra of the current modes:
\begin{align}
&[J^a_n\text{ },J^b_m]=iF^{abc}J^c_{n+m}+iG^{abc}K^c_{n+m}+nk_1\delta^{ab}\delta_{n+m,0}\nonumber\\
&[J^a_n\text{ },K^b_m]=iF^{abc}K^c_{n+m}+nk_2\delta^{ab}\delta_{n+m,0}\hspace{5mm};\hspace{5mm}[K^a_n\text{ },K^b_m]=0\label{34a}
\end{align}

\medskip

The global internal symmetry is governed by the subalgebra of the zero-modes:
\begin{align}
[J^a_0\text{ },J^b_0]=iF^{abc}J^c_{0}+iG^{abc}K^c_{0}\hspace{5mm};\hspace{5mm}[J^a_0\text{ },K^b_0]=iF^{abc}K^c_{0}\hspace{5mm};\hspace{5mm}[K^a_0\text{ },K^b_0]=0
\end{align}

This is precisely the algebra we have obtained for the NL currents in \refb{gkm}.\\

\medskip

\newpage

\section{Sugawara Construction}\label{section4}
As is well known, in 2d CFT, the Sugawara construction is employed to construct (the modes of) the energy momentum tensor in terms of (the modes of) the currents. In this section, here we will attempt to construct a NL version of  the Sugawara construction and express $L_m$ and $M_m$ in terms of $J_m$ and $K_m$. We shall see that the $L_m$'s and $M_m$'s constructed in such manner indeed satisfy the BMS algebra. Similar constructions have been studied earlier in \cite{Rasmussen:2019zfu, Ragoucy:2020elp}. 

\subsection{Intrinsic construction from algebra}\label{sectionsuga}
We begin here by assuming we have only the algebra of the NL currents, i.e. 
\begin{subequations}\label{jk}
\begin{align}
[J^{a}_{m},J^{b}_{n}]&=i f^{abc}J^c_{m+n} +mk_{1}\delta^{ab}\delta_{m+n,0}, \\
[J^{a}_m,K^{b}_n]&=i f^{abc}K^c_{m+n}+mk_{2}\delta^{ab}\delta_{m+n,0}, \quad [K^{a}_m, K^{b}_n]=0. 
\end{align}
\end{subequations}
In the above, the sum over the group indices e.g. $f^{abc}J^c_{m+n} = \sum_{c=1}^{dim(\bar{g})} f^{abc}J^c_{m+n}$ is implied. Comparing with \refb{gkm}, as stressed before, we work with the case where $F^{abc}=f^{abc}, G^{abc}=0$. Let us first consider the zero modes of $J$ and $K$. Putting $n=m=0$ in (\ref{jk}), the algebra for the zero modes become
\begin{align}
[J^{a}_{0},J^{b}_{0}]=i f^{abc}J^c_{0} \ ; \ \quad [J^{a}_0,K^{b}_0]=i f^{abc}K^c_{0}\ ; \ \quad [K^{a}_{0}, K^{b}_{0}]=0
\end{align}
Now let us look for quadratic Casimir operators for the above algebra. The possible combinations are
\begin{align}\label{casimir}
    \sum_a J^a_0J^a_0 \ , \ \sum_a J^a_0K^a_0 \ , \ \sum_a K^a_0J^a_0 \ , \ \sum_a K^a_0K^a_0.
\end{align}
Keeping in mind that Casimir operators must commute with all the generators, we can exclude $ \sum_a J^a_0J^a_0 $ since it does not commute with $K^a_0$
\begin{align}
\Big[\sum_a J^a_0J^a_0, K^b_0\Big]=i \sum_{a,c}f^{abc}(J^aK^c+K^cJ^a)\neq 0
\end{align}
All other combinations in (\ref{casimir}) commute with all $J^a_0$'s and $K^a_0$'s. Hence a generic Casimir operator constructed from the zero modes of $J$ and $K$ will be a linear combination of the above three combinations of $J^a_0$'s and $K^a_0$'s. We therefore want to construct the zero level generator $L_0, M_0$ from these combinations.

\medskip

Since $L_{0}$ and $M_{0}$ are expected to be quadratic in terms of $J^{a}_{n}$'s and $K^{a}_n$'s, we can write down the following generic expression for them (The term  $\sum JJ$ is excluded since we have already seen that the term $\sum_a J^a_0J^a_0$ does not contribute to the zero mode part of $L_{0}$ and $M_{0}$)
\begin{subequations}\label{chh16}
\begin{align}
L_{0}&=\alpha\sum_{a,l,n} J^a_lK^a_n+\beta\sum_{a,l,n} K^a_lJ^a_n+\rho\sum_{a,l,n}K^a_lK^a_n \\
M_{0}&=\mu\sum_{a,l,n} J^a_lK^a_n+\nu\sum_{a,l,n} K^a_lJ^a_n+\eta\sum_{a,l,n}K^a_lK^a_n
\end{align}
\end{subequations}
Now putting the conditions that $$[L_0, J^a_n] = -nJ^a_n, [M_0, J^a_n] = -nK^a_n$$ and looking at just the level of current generators on the RHS, we can see that only $(l= -n) $ terms should contribute, so we obtain
\begin{subequations}
\begin{equation}
L_{0}=\alpha\sum_{a,n}J^a_nK^a_{-n}+\beta\sum_{a,n} K^a_nJ^a_{-n}+\rho\sum_{a,n} K^a_nK^a_{-n} = \alpha X_0 + \beta Y_0 + \rho Z_0
\end{equation}
\begin{equation}
M_{0}=\mu\sum_{a,l}J^a_lK^a_{-l}+\nu\sum_{a,l}K^a_lJ^a_{-l}+\eta\sum_{a,l}K^a_lK^a_{-l} = \mu X_0 + \nu Y_0 + \eta Z_0
\end{equation}
\end{subequations}
Now take 
\begin{equation}
    \begin{split}
        Y_0 &= \sum_{a,l} K^a_l J^a_{-l} = \sum_{a,l} J^a_{-l} K^a_l + k_2\frac{\text{dim}g}{2}\sum_{l =-\infty}^{\infty} l = X_0 + c,
    \end{split}
\end{equation}
where c is a (possibly infinite) constant. Since we always have the independence of redefining BMS generators by constant shifts, we can define level 0 BMS generators as
\begin{equation}
L_0 \equiv L'_{0}=\frac{\alpha + \beta}{2} (X_0 +Y_0) + \rho Z_0, \quad M_0 \equiv M'_{0}= \frac{\mu +\nu}{2}(X_0 + Y_0) + \eta Z_0.
\end{equation}
The normal ordering of the operators $L_0, M_0$ is achieved by individual normal ordering of $X_0, Y_0, Z_0$. We generalise the definition to the other BMS generators as (with normal ordering)
\begin{align}\label{chh94}
   L_m&=\sum_{a=1}^{\text{dim}g}\left[ \frac{\alpha + \beta}{2}\left\{ \sum_{l\leq-1}(J^a_lK^a_{m-l}+K^a_lJ^a_{m-l})+\sum_{l>-1}(J^a_{m-l}K^a_l+K^a_{m-l}J^a_l)\right\}+ \rho \sum_{l}K^a_l K^a_{m-l} \right] \nonumber\\
  M_m&=\sum_{a=1}^{\text{dim}g}\left[ \frac{\mu + \nu}{2}\left\{ \sum_{l\leq-1}(J^a_lK^a_{m-l}+K^a_lJ^a_{m-l})+\sum_{l>-1}(J^a_{m-l}K^a_l+K^a_{m-l}J^a_l)\right\}+ \eta \sum_{l}K^a_l K^a_{m-l} \right]
\end{align}
Using the form of the BMS generators in (\ref{chh94}), and substituting in the BMS-current cross commutators, i.e.
\begin{equation}
    [L_m, J^a_n] = -nJ^a_{m+n} \ ; \ [L_m, K^a_n] = -nK^a_{m+n} \
    ; \ [M_m, J^a_n] = -nK^a_{m+n}, 
\end{equation}
we obtain the following values for the coefficients
\begin{equation}
        \alpha + \beta = \frac{1}{k_2} \ ; \ \rho = -\frac{k_1 + 2C_g}{2k_2^2}; \quad
    \mu + \nu = 0 \ ; \ \eta = \frac{1}{2k_2}.
\end{equation}
So we obtain the final form for our NL Sugawara construction
\begin{align}\label{chh4}
   L_m&=\frac{1}{2k_2}\sum_{a=1}^{dim(g)}\left\{ \sum_{l\leq-1}(J^a_lK^a_{m-l}+K^a_lJ^a_{m-l})+\sum_{l>-1}(J^a_{m-l}K^a_l+K^a_{m-l}J^a_l)-\frac{(k_1+2C_g)}{k_2}\sum_{l}K^a_lK^a_{m-l} \right \}\nonumber\\
  M_m&=\frac{1}{2k_2}\sum_{a=1}^{dim(g)}\sum_{l}K^a_lK^a_{m-l}
\end{align}
As a check of the validity of the analysis, we compute the algebra of the $L$ and $M$ generators. These 
satisfy the following algebra (see Appendix \ref{AppA} for detailed calculations)
\begin{align}\label{chh1}
[L_m,L_n]=(m-n)L_{m+n}+\frac{dim(g)}{6}(m^3-m)\delta_{m+n,0} \ ; \ [L_m,M_n]=(m-n)M_{m+n}
\end{align}
Hence, we see that by doing the NL Sugawara construction of NL currents, we can find BMS algebra with 
\be{}
c_L=2 \text{dim}(g), \quad c_M=0.
\ee

\medskip

\paragraph{Non-zero $c_M$:} We can obtain a non-zero $c_M$ with a slight modification to the Sugawara construction presented above. In case of Virasoro algebra with additional symmetries, we can define new Virasoro generators \cite{Caroca:2017onr} as 
\begin{equation}\label{chh9}
\tilde{\mathcal{L}}_n=\mathcal{L}^{S}_n+in\theta^aj^{a}_n+\frac{1}{2}k\theta^2\delta_{n,0}
\end{equation}
where $\mathcal{L}^{S}_n$ is the Virasoro generators obtained from Sugawara construction and $\theta=\theta^{a}t^{a}$ is a vector belonging to the Lie algebra with generators $t^{a}$. It can be showed that (see Appendix \ref{amod_Suga}) if $\mathcal{L}_n$ satisfy Virasoro algebra with central charge $c$ then $\tilde{\mathcal{L}}_n$ will satisfy Virasoro algebra with shifted central charge $\tilde{c}$ where
\begin{equation}\label{chh10}
\tilde{c}=c+12k\theta^2.
\end{equation}
In case of NL Kac-Moody algebra, inspired by this, we introduce the following redefinitions
\begin{align}\label{chh11}
    \tilde{L}_n&=L^S_n+in\theta^aJ^{a}_n+\frac{1}{2}k_1\theta^2\delta_{n,0}\nonumber\\
    \tilde{M}_n&=M^S_n+in\theta^aK^{a}_n+\frac{1}{2}k_2\theta^2\delta_{n,0}
\end{align}
where $L^S_n$ and $M^S_n$ are given by (\ref{chh94}). This redefinition will give us the following algebra (see Appendix \ref{amod_Suga})
\begin{align}\label{chh12}
     [\tilde{L}_m,\tilde{L}_n]&=(m-n)\tilde{L}_{m+n}+\frac{c_{L}}{12}(m^3-m)\delta_{n+m,0} \nonumber \\
    [\tilde{L}_m,\tilde{M}_{n}]&=(m-n)\tilde{M}_{m+n}+\frac{c_{M}}{12}(m^3-m)\delta_{n+m,0}
\end{align}
where the central charges are given by
\begin{align}\label{chh13}
    c_L&=2dim(g)+12k_{1}\theta^2\nonumber \\
    c_M&=12k_{2}\theta^2
\end{align}
Hence we can obtain the fully centrally extended BMS algebra.

\subsection{Consistency with OPEs}
We will recast the NL Sugawara construction in terms of the NL EM tensor that we introduced in Sec.~\ref{nlcurr} instead of the generators of the BMS: $\{L_n, M_n\}$. $\{L_n, M_n\}$ are of course modes of the NL EM tensor. Hence the calculations in this subsection provide a sanity check for our analysis making sure the various formulations are consistent with each other. 

\medskip

The NL Sugawara construction gave us (\ref{chh4}). We now normal order the products to rewrite this as
\begin{equation}\label{Suga_n}
    \begin{split}
        &L_n = \frac{1}{2k_2} \sum_{l,a} \Bigl( :J^a_l K^a_{n-l}: + :K^a_l J^a_{n-l}: -\frac{k_1 + 2C_g}{k_2}:K^a_lK^a_{n-l}: \Bigr)\\
        &M_n = \frac{1}{2k_2} \sum_{l, a} :K^a_lK^a_{n-l}:
    \end{split}
\end{equation}
Here, $:A_l :$ is a shorthand for normal ordered products. Now substituting (\ref{Suga_n}) in \refb{emgca2} and rearranging the terms, we get
\txb{uv to xt}
\begin{equation}\label{Suga_f}
    \begin{split}
        &T_u(u, v) = \frac{1}{2k_2}\sum_a \left[ \bigl[(\mathcal{J}^a_u \mathcal{J}^a_v)(u,v) + (\mathcal{J}^a_v \mathcal{J}^a_u)(u,v)\bigr] - \frac{k_1 + 2C_g}{k_2}(\mathcal{J}^a_v \mathcal{J}^a_v)(v)\right]\\
        &T_v(v) = \frac{1}{2k_2} \sum_a (\mathcal{J}^a_v \mathcal{J}^a_v)(v)
    \end{split}
\end{equation}
Here $(\dots)$ means normal ordered products of fields, which can be expressed in terms of contour integrals. Here we are using $u, v$ coordinates instead of $x, t$ of Sec.~\ref{nlcurr} because this construction applies to both Galilean and Carrollian CFTs, because of the similarity between the 2 theories mentioned in the introduction. So by substituting $u \to x, v \to t$ or $u \to t, v \to x $, we can obtain the Galilean or Carrollian theory expressions, respectively.\\
So (\ref{Suga_f}) gives the field expression for the NL Sugawara construction. For the modified Sugawara construction as defined in (\ref{chh11}), the EM tensor fields turn out to be (by doing the mode expansion using the new generators)
\begin{equation}\label{mod_suga_2}
    \begin{split}
        &\tilde{T}_u(u,v) = T_u(u,v) -i\theta^a\partial_v\mathcal{J}^a_u(u, v) - i\theta^a\frac{\mathcal{J}^a_u(u,v)}{v} - i\theta^a\frac{u\mathcal{J}^a_v(v)}{v^2} +\frac{1}{2}\frac{k_1\theta^2}{v^2} + \frac{uk_2\theta^2}{v^3}\\  
        &\tilde{T}_v(v) = T_v(v) - i\theta^a \partial_v\mathcal{J}^a_v(v) - i \theta^a\frac{\mathcal{J}^a_v(v)}{v} +\frac{1}{2}\frac{k_2 \theta^2}{v^2}
    \end{split}
\end{equation}
where, $T_u(u, v), T_v(u, v)$ refers to the quantities in (\ref{Suga_f}), i.e. the normal NL Sugawara construction expressions.

\paragraph{$T-\mathcal{J}$ OPE:} Now, for a consistency check of our analysis so far, we will rederive the OPEs from the above definitions of the NL Sugawara construction. We begin with the T-J OPEs. Taking the definitions as in (\ref{Suga_f}) and contracting with the currents, we get the following expressions (See Appendix \ref{aTJ} for detailed calculation):
\begin{equation}
    \begin{split}
        &T_v(u_1, v_1) \mathcal{J}^b_v(u_2, v_2) \sim \, \text{regular}\\
        &T_u(u_1, v_1) \mathcal{J}^b_v(u_2, v_2) \sim \frac{\mathcal{J}^b_v(u_2, v_2)}{v_{12}^2} + \frac{\partial_v \mathcal{J}^b_v(u_2, v_2)}{v_{12}} + \dots\\
        &T_v(u_1, v_1) \mathcal{J}^a_u(u_2, v_2) \sim \frac{\mathcal{J}^a_v(u_2, v_2)}{v_{12}^2} + \frac{\partial_v \mathcal{J}^a_v(u_2, v_2)}{v_{12}} + \dots\\
        &T_u(u_1, v_1) \mathcal{J}^a_u(u_2, v_2) \sim \frac{\mathcal{J}^a_u(u_2, v_2)}{v_{12}^2} + \frac{\partial_v \mathcal{J}^a_u(u_2, v_2)}{v_{12}} + \frac{u_{12}}{v_{12}^2}\partial_u \mathcal{J}^a_u(u_2, v_2)\\
        &\qquad\qquad\qquad\qquad\qquad\qquad+ \frac{2u_{12}}{v_{12}}\left( \frac{\mathcal{J}^a_v(u_2, v_2)}{v_{12}^2} + \frac{\partial_v \mathcal{J}^a_v(u_2, v_2)}{v_{12}} \right) + \dots
    \end{split}
\end{equation} 
Which are equivalent to the $[L_m, J^a_n]$ type commutation relations given in (\ref{gkm}). Clearly these relation satisfy the OPE relations (\ref{TJ1}) that the currents and the EM tensors of a theory are supposed to satisfy.

\paragraph{$T-T$ OPE:} Next we use the definition of the Sugawara construction to compute the T-T type OPEs. Doing the calculations (see Appendix \ref{aTT} for detailed Calculations), we get, 
\begin{align}\label{TT}
T_u(u_1, v_1)T_u(u_2, v_2) \sim  &\frac{2 T_u(u_2, v_2)}{v_{12}^2} + \frac{\partial_v T_u(u_2, v_2)}{v_{12}} + \frac{u_{12}}{v_{12}^2}\partial_u T_u(u_2, v_2) \nonumber\\
        &+ \frac{2u_{12}}{v_{12}}\Bigl( \frac{2T_v(u_2, v_2)}{v_{12}^2} + \frac{\partial_v T_v(u_2, v_2)}{v_{12}} \Bigr) + \frac{dim(g)}{v_{12}^4} + \dots  \nonumber \\
        T_u(u_1, v_1) T_v(u_2, v_2) \sim &\frac{2T_v(u_2, v_2)}{v_{12}^2} + \frac{\partial_v T_v(u_2, v_2)}{v_{12}} + \dots\\
        T_v(u_1, v_1)T_v(u_2, v_2) \sim & \, \text{regular}  \nonumber
\end{align}
These results (\ref{TT}) match with the OPEs of the energy momentum tensors of a 2d Carrollian or Galilean CFT, as given in (\ref{emgca3}). So, the NL Sugawara construction really gives the EM tensors of a 2d NL CFT (proved both at algebra level and now at field level). Also from the OPE expressions, we can verify the earlier results of $c_L = 2dim(g), c_M = 0$ obtained earlier. Again if we start with the modified Sugawara construction (\ref{mod_suga_2}), we can obtain same OPE relations as above, but with modified central charges $c_L = 2dim(g) + 12k_1 \theta^2, c_M = 12k_2\theta^2$. To see this, let's look at, for example, the $\tilde{T}_u(u_1,v_1) \tilde{T}_v(v_2)$ OPE and focus our attention to a specific term
\begin{equation}
\tilde{T}_u(u_1, v_1) \tilde{T}_v(v_2) \sim -\theta^a\theta^b \partial_{v}\mathcal{J}^a_u(u_1, v_1) \partial_{v} \mathcal{J}^a_v(v_2) + \dots \sim -\theta^2 \partial_{v_1}\partial_{v_2} \frac{k_2}{v_{12}^2} +\dots\sim \frac{6k_2\theta^2}{v_{12}^4} + \dots \nonumber
\end{equation}
Matching the above expression with (\ref{emgca3}), we can see $c_M = 12 k_2 \theta^2$. Similarly we can obtain the other shifted central charge.

\newpage

\section{Tensionless String as NLKM}\label{section5}

The tensionless limit of string theory is the limit that is diametrically opposite to the usual point particle limit where known supergravity appears from strings. This limit that explores the very strong gravity regime and, when quantized, the very highly quantum and highly stringy regime of string theory. In this section, we explore how this is connected to the NLKM structures we have discussed in this paper. We will see that the currents satisfying the $U(1)$ NLKM algebra come up intrinsically when we are looking at tensionless strings propagating in flat spacetimes. 

\medskip

As is very well known, the action for a tensile relativistic string propagating in a flat $d$-dimensional spacetime is given by the Polyakov action: 
\be{}
S_{\text{P}} = T \int d^2\xi \, \sqrt{\gamma} \gamma^{\a\b} \partial_\alpha X^\mu \partial_\beta X^\nu \eta_{\mu\nu}. 
\ee
In order to take the tensionless limit, it is helpful to work with the phase space action of string. One can then systematically take the limit \cite{} and the resulting action can be cast in a Polyakov like form, which we will call the ILST action after the authors:  
\begin{equation}\label{ilst}
    S_{\text{ILST}} = \int d^2\xi \, V^\alpha V^\beta \partial_\alpha X^\mu \partial_\beta X^\nu \eta_{\mu\nu}. 
\end{equation}
Here $X^\mu$ are the coordinates in the background flat space $\eta_{\mu\nu}$ which are scalar fields on the worldsheet parametrized by $\xi^a=\sigma, \tau$. The worldsheet metric $\gamma^{\a \b}$ degenerates in the limit and the following replacement is made:
\be{}
T\sqrt{\gamma} \gamma^{\a\b} \to V^\a V^\b,
\ee
where $V^\alpha$ is a vector density. The action is invariant under worldsheet diffeomorphisms and hence one needs to fix a gauge. It is helpful to go to the equivalent of the conformal gauge 
\be{cong}
V^\alpha = (v, 0).
\ee
There is some symmetry left over even after this gauge fixing. In the usual tensile string, the residual symmetry gives two copies of the Virasoro algebra and the appearance of a 2d CFT on the worldsheet is the central reason why we understand string theory as well as we do. Now in the tensionless case, the residual gauge symmetry that appears on the worldsheet is the BMS$_3$ algebra. This now dictates the theory of tensionless strings. The reason behind the appearance of the BMS algebra is that the tensionless string in flat spacetimes is actually a null string. This is the string equivalent of a massless point particle which is constrained to travel on a null geodesic of the ambient spacetime. The string sweeps out a worldsheet which is null and since there are no mass terms, the resulting action has to have conformal Carrollian symmetry in $d=2$ or equivalently BMS$_3$. 
 
\medskip
 
Now, let us connect to our discussion in this paper. In the favourable conformal gauge \refb{cong}, the equations of motion and constraints arising from \refb{ilst} take the form  
\begin{equation}\label{nreom}
  {\text{EOM:}} \quad  \ddot{X}^\mu =0 \ ; \ \quad {\text{Constraints:}} \quad \dot{X}^2 = 0 \ , \ \dot{X}.X' = 0.
\end{equation}
A convenient form of the solution is given by 
\begin{equation}\label{nrsoln}
    X^\mu(\sigma, \tau) = x^\mu + \sqrt{2c'}\bigl( J^\mu_0 \sigma + K^\mu_0\tau + i\sum_{n \neq 0} \frac{1}{n}(J^\mu_n - in\tau K^\mu_n)e^{-in\sigma} \bigr)
\end{equation}
Here $\sigma, \tau$ are the coordinates on the cylinder, which are related to the planar Non-Lorentzian coordinates as 
\begin{equation}
    u = e^{-i\sigma} \ ; \ v = -\tau e^{-i\sigma}
\end{equation}
The periodic condition $X^\mu(\sigma + 2\pi, \tau) = X^\mu(\sigma, \tau)$ forces us to have $J^\mu_0=0$. From (\ref{nrsoln}), we obtain
\begin{equation}\label{nrcurrent}
    \begin{split}
        J^\mu_\tau = \partial_\sigma X^\mu =  \sqrt{2c'}\sum_{n} (J^\mu_n - in\tau K^\mu_n)e^{-in\sigma}, \quad 
        J^\mu_\sigma = -\partial_\tau X^\mu = - \sqrt{2c'}\sum_{n}K^\mu_n e^{-in\sigma} 
    \end{split}
\end{equation}
Clearly the currents in (\ref{nrcurrent}) satisfy similar relations as \eqref{15a}, i.e. 
\begin{equation}
    \partial_\tau J^\mu_\sigma = 0 \ ; \ \partial_\tau J^\mu_\tau + \partial_\sigma J^\mu_\sigma =0.
\end{equation}
If we canonically quantise the system by demanding $[X^\mu(\tau, \sigma), \Pi^\nu(\tau, \sigma')] = i\delta(\sigma - \sigma')\eta^{\mu\nu}$, we obtain the following relations for the current modes
\begin{equation}\label{nralg}
    [J^\mu_m, J^\nu_n] = [K^\mu_m, K^\nu_n] = 0 \ ; \ [J^\mu_m, K^\nu_n]= 2m\delta_{m+n,0}\eta^{\mu\nu}
\end{equation}
Clearly, the current modes for the same spacetime index $\mu$ follow a special case of $U(1)$ NL Kac-Moody algebra (look at (\ref{gkm}) for comparison), with $k^{(\mu\nu)}_1=0, k^{(\mu\nu)}_2=2\eta^{\mu\nu}$.

\medskip

Next if we look into the (classical) energy momentum tensor of this theory, their mode expansion coefficients are given by
\begin{equation}\label{nrgca}
    L_n = \frac{1}{2} \sum_{m} J^\mu_{m}K^\nu_{n-m}\eta_{\mu\nu} \ ; \ M_n = \frac{1}{4} \sum_{m} K^\mu_{m}K^\nu_{n-m}\eta_{\mu\nu}.
\end{equation}
This matches classically with the relation (\ref{chh4}) for the NL Sugawara construction with 
\be{}
k_1=0, \quad k_2=2.
\ee 
Note that $C_g=0$ for $U(1)$ algebra. The generators in (\ref{nrgca}) satisfy the (centreless) BMS$_3$ algebra, which is expected as the action (\ref{ilst}) has BMS symmetry as a residual gauge symmetry.

\medskip

Thus we can see that the Non-Lorentzian $U(1)$ Kac-Moody algebra and the associated NL Sugawara construction come up in the theory of tensionless or null strings propagating on a flat background geometry. An outstanding question is what happens when we look at the propagation of null strings on arbitrary curved manifolds. These can be viewed as group manifolds and there will be the NL equivalent of a Wess-Zumino-Witten model appearing for tensionless strings in this context. We expect that the non-abelian NLKM algebras explored in this work would naturally appear on these NL WZW models. This is work in progress and we hope to report on this in the near future. 

\newpage

\section{Non-Lorentzian Knizhnik Zamolodchikov Equations}\label{section6}
In usual Lorentzian Kac-Moody algebras, the Knzhnik-Zamolodchikov equations are the linear differential equations that are satisfied by the correlation functions of these CFTs endowed with additional affine Lie symmetry. In this section, we write down the non-Lorentzian analogues of these KZ equations based on our construction of the NLKM algebra and its representations in this paper. 

\medskip

In what follows, we will outline the steps to get to the NL Knizhnik Zamolodchikov equations. The details of the calculation are presented in a separate appendix \ref{aKZ_field}. 

\medskip

We begin with the OPE definition of BMS primary field (\ref{emgca4}) and obtain
\begin{equation}\label{GCAp3}
    \begin{split}
        &\partial_u\Phi(u', v') = -\oint_{v'}\frac{dv}{2\pi i}\oint_{u'}\frac{du}{2 \pi i}(u-u')^{-1} T_v(u, v) \Phi(u', v') = - \oint_{v'}\frac{dv}{2\pi i} T_v(v) \Phi(u', v')  \\
        &\partial_v\Phi(u',v') = \oint_{v'}\frac{dv}{2\pi i}\oint_{u'}\frac{du}{2 \pi i}(u-u')^{-1} T_u(u, v) \Phi(u', v') = \oint_{v'}\frac{dv}{2\pi i} T_u(u', v) \Phi(u', v')
    \end{split}
\end{equation}
Next, we take the correlation function of a string of primary fields as shown below
\bea{}
\langle\partial_u\Phi(u, v) \Phi_1(u_1, v_1)\dots\Phi_n(u_n, v_n) \rangle && = \langle \partial_u\Phi(u, v) X(\{u_i, v_i\}) \rangle = \oint_{v} \frac{dv'}{2\pi i} \langle T_v(u', v')\Phi(u,v) X(\{u_i, v_i\}) \rangle \cr
    && = \oint_{v} \frac{dv'}{2\pi i} \langle \frac{1}{2k_2}\Bigl((\mathcal{J}^a_v\mathcal{J}^a_v)(u', v')\Bigr)\Phi(u,v) X(\{u_i, v_i\}) \rangle \nonumber
\eea
and use relations (\ref{emgca4},  
\ref{GCAp3}) to finally get the following expression (details in Ap. \ref{aKZ_field})
\begin{equation}
    \begin{split}
        &\langle \partial_u\Phi(u, v) X(\{u_i, v_i\}) \rangle= -\frac{1}{k_2}\Bigl(\sum_{j}\frac{t^a_{R, K} \otimes t^a_{R_j, K}}{(v - v_j)} \Bigr)\langle\Phi(u, v)X(\{u_i, v_i\}) \rangle
    \end{split}
\end{equation}
which can be written as 
\begin{equation}\label{nlkz1}
    \Bigl(\partial_{u_i} + \frac{1}{k_2}\sum_{j \neq i}\frac{t^a_{R_i, K} \otimes t^a_{R_j, K}}{(v_i - v_j)} \Bigr)\langle\Phi_1(u_1, v_1)\dots \Phi_n(u_n, v_n) \rangle =0.
\end{equation}
This is one of the Non-Lorentzian Knizhnik Zamolodchikov equations.

\medskip

Similarly we can start with  
\begin{equation}
    \begin{split}
        &\langle\partial_v\Phi(u, v) \Phi_1(u_1, v_1)\dots\Phi_n(u_n, v_n) \rangle = \langle \partial_v\Phi(u, v) X(\{u_i, v_i\}) \rangle = \oint_{v} \frac{dv'}{2\pi i} \langle T_u(u', v')\Phi(u,v) X(\{u_i, v_i\}) \rangle\\
        &= \oint_{v} \frac{dv'}{2\pi i} \langle \frac{1}{2k_2}\Bigl( (\mathcal{J}^a_v\mathcal{J}^a_u)(u', v') + (\mathcal{J}^a_u\mathcal{J}^a_v)(u', v') - \frac{k_1+ 2C_g}{k_2} (\mathcal{J}^a_v\mathcal{J}^a_v)(u', v')\Bigr)\Phi(u,v) X(\{u_i, v_i\}) \rangle
    \end{split}
\end{equation}
From here, we can proceed in the same way as  Appendix \ref{aKZ_field} to obtain the other  equation
\begin{equation}\label{nlkz2}
\begin{split}
        {\Bigl(\partial_{v_i} - \frac{1}{k_2}\sum_{j \neq i} \frac{1}{(v_i - v_j)} \sum_{a= 1}^{dim g} \Bigl[\bigl( t^a_{R_i, J} \otimes t^a_{R_j, K} + t^a_{R_i, K} \otimes t^a_{R_j, J}\bigr)} & + {\bigl(\frac{u_i - u_j}{v_i - v_j} - \frac{k_1+ 2C_g}{k_2} \bigr)t^a_{R_i, K} \otimes t^a_{R_j, K}\Bigr] \Bigr)}\\
        & \hspace{-2cm} {\langle \Phi_{R_1}(u_1, v_1)\Phi_{R_2}(u_2, v_2)...\Phi_{R_n}(u_n, v_n)  \rangle = 0}
    \end{split}
\end{equation}
This above equation \eqref{nlkz2} is the second of the Non-Lorentzian Knizhnik Zamolodchikov equations. The solutions to these two equations \refb{nlkz1}, \refb{nlkz2} would give the correlation functions of the underlying NLKM theory. For usual relativistic theories, the KZ equations are difficult to solve in general, but one can do it for the four-point functions yielding a closed solution in terms of hypergeometric functions. It would be instructive to check whether something similar happens here. If one can find the solutions to the NLKM four point functions, it would also be a nice exercise to check whether these can be arrived at as a limit of the relativistic answers. This process of attempting to generate the non-Lorentzian answers from the relativistic ones is something we outline in detail in the section that follows.

\section{NLKM from Contractions}\label{section7}
In this section, we rederive various results we have obtained earlier in the paper through a systematic limit on the algebraic structures obtained in the relativistic set-up. Before moving to recovering the answers, we spend a bit of time understanding which limit is appropriate for our purposes.

\smallskip

\subsection{A brief detour to representations of BMS}
We said in the introduction that there were two distinct contractions that land us up on the BMS$_3$ algebra starting from two copies of the Virasoro algebra. One of them was a Carrollian or ultra-relativistic limit \refb{urlim}, where there was a mixing of positive and negative Virasoro modes creating the BMS generators, while the other was a Galilean or non-relativistic \refb{nrlim}, where no mixing took place. This mixing of modes is critical for the understanding of the representations in the limit. 

\medskip

We begin with the Galilean contraction. In the parent relativistic CFT, the theory is best described in terms of the highest weight representation. The states of the theory are labelled by the zero modes:
\be{}
\L_0 |h, \h\> = h |h, \h\>, \quad \bL_0 |h, \h\> = \h |h, \h\>
\ee
There is a class of states called primary states which are annihilated by all positive modes:
\be{}
\L_n |h, \h\>_p = 0, \quad \bL_n |h, \h\>_p = 0, \quad \forall n>0.
\ee
The Virasoro modules are built on these primary states by acting with raising operators $\L_{-n}$. Now, looking back at the Galilean contraction \refb{nrlim}, we see that 
\be{}
\L_n =\frac{1}{2}\left(L_n + \frac{1}{\e} M_n\right), \quad \bL_n = \frac{1}{2}\left(L_n - \frac{1}{\e} M_n\right)
\ee
The 2d CFT primary conditions then boil down to:
\be{}
L_0 |\D, \xi\> = \D |\D, \xi\>, \, M_0 |\D, \xi\> = \xi |\D, \xi\>; \quad L_n |\D, \xi\>_p = 0, \, M_n |\D, \xi\>_p = 0, \, \forall n>0.
\ee
In the above, the assumption is that the state $|h, \h\>$ goes to the state $|\D, \xi\>$ in the limit. So we see that highest weight states map to highest weight states in the Galilean contraction. This analysis holds in a similar way when we consider the full NLKM algebra. We will be using this for the analysis in the rest of the section. 

\medskip

But before we get there, let us point out why we are not using the Carroll limit \refb{urlim} for this purpose. In the Carroll contraction, we can read off 
\be{}
\L_n =\frac{1}{2}\left(L_n + \frac{1}{\e} M_n\right), \quad \bL_{n} = \frac{1}{2}\left(-L_{-n} + \frac{1}{\e} M_{-n}\right)
\ee
The 2d CFT primary conditions in the Carroll limit become:
\be{}
M_0 |M, s\> = M |M, s\>, \quad L_0 |M, s\> = s |M, s\>, \quad M_n |M, s\> = 0,  \quad \forall n \neq 0.
\ee
In the above, the state $|h, \h\>$ in the Carroll limit becomes the state $|M, s\>$. This is clearly not a highest weight state. The set of these states form what is called the induced representation. The story is again similar for the full NLKM. In the analysis that follows, we will not focus on the induced representations. It would be of interest to consider them in future work. 

We should stress however, that the answers we have obtained in the intrinsic Carrollian form earlier are for highest weight representations. The Galilean limit would be an effective way of reproducing these answers with a flip of space and time directions at the end of the analysis. 

\subsection{Contraction of the affine parameters}
In this section we will establish the mode expansion of the NLKM currents \eqref{27a} from another approach, by doing a contraction of the original loop extended algebra. Starting from the finite Lie algebra 
\begin{equation} 
[j^a, j^b] = if^{abc}j^c \ ; \ [\bar{j}^a, \bar{j}^b] = if^{abc}\bar{j}^c
\end{equation} 
we can define loop extended generators 
\begin{align}
    j^a_n=j^a\otimes z^n \ ; \ \bar{j}^a_n=\bar{j}^a \otimes \bar{z}^n,
\end{align}
which satisfy 
\begin{equation}
    [j^a_n, j^b_m] = i f^{abc} j^c \otimes z^{n+m} = i f^{abc} j^c_{n+m}.
\end{equation}
Loop extended algebra obtained above admits a central extension to give us the current algebra in \eqref{Vir}.
Now, we can do a contraction of the affine parameter as $z=t+\epsilon x$ and $\bar{z}=t-\epsilon x$ in the Galilean limit. We then obtain (keeping upto first order in $\epsilon$),
\begin{align}\label{epsexp}
    j^a_n&=j^a\otimes(t^n+n\epsilon xt^{n-1}+...) \ ; \
    \bar{j}^a_n=\bar{j}^a\otimes (t^n-n\epsilon xt^{n-1}+...)
\end{align}
Now we can define new contracted generators of the finite algebra as follows:
\begin{align}
    J^a=j^a+\bar{j}^a \ ; \ K^a=\epsilon(\bar{j}^a-j^a)
\end{align}
Above definition can be extended to general modes of $J$ and $K$ using (\ref{epsexp}),
\begin{subequations}\label{genmode}
\begin{align}
    J^a_n&=\lim_{\epsilon\rightarrow 0}(j^a_n+ \bar{j}^a_n)  =J^a\otimes t^n-n K^a\otimes xt^{n-1}\\
    K^a_n&=\lim_{\epsilon\rightarrow 0}\epsilon(\bar{j}^a_n-j^a_n) =K^a\otimes t^n
\end{align}
\end{subequations}
In this way, we get the structure of the power series expansion of $J$ and $K$ in terms of $x$ and $t$, which is the Galilean analog of \eqref{27a}, or identical upto the flip of temporal and spatial directions. 

\medskip

If we define primary field $\phi(x,t)$ as a representation of the finite algebra, i.e. in terms of the action of zero modes of the generators $J$ and $K$, 
\begin{equation}
    \begin{split}
        [J^a_0,\phi(x,t)]&\equiv[J^a,\phi(x,t)]=t^a_J\phi(x,t)\\
        [K^a_0,\phi(x,t)]&\equiv[K^a,\phi(x,t)]=t^a_K\phi(x,t)
    \end{split}
\end{equation}
We can get the action of general modes using (\ref{genmode}),
\begin{equation}\label{GKMpt}
    \begin{split}
        [J^a_n,\phi(x,t)]&=t^a_J\phi(x,t)\otimes t^n-nt^a_K\phi(x,t)\otimes xt^{n-1}\equiv t^nt^a_J \phi(x,t)-nxt^{n-1}t^a_K\phi(x,t)\\
        [K^a_n,\phi(x,t)]&=t^a_K\phi(x,t)\otimes t^n\equiv t^n t^a_K \phi(x,t)
    \end{split}
\end{equation}
This definition of a primary field agrees with our earlier result \eqref{curprimcom}.

\medskip

This action of $J^a_n$ and $K^a_n$ on primary fields also appears when we draw motivation from \cite{Bagchi:2009ca}. We can define the primary field as,
\begin{equation}\label{algdefprim}
    \begin{split}
    [J^a_0,\phi(0,0)]=t^a_J\phi(0,0) \ &; \ [K^a_0,\phi(0,0)]=t^a_k\phi(0,0),\\
     [J^a_n,\phi(0,0)]=0 \ &; \ [K^a_n,\phi(0,0)]=0  \  \ \  \ \forall \ n >0.
     \end{split}
\end{equation}
For $n\geq 0$ and $U=e^{tL_{-1}-xM_{-1}}$,
\begin{align}\label{jphicalc}
        [J^a_n,\phi(x,t)]&=[J^a_n,U\phi(0)U^{-1}]=U[U^{-1}J^a_nU,\phi(0)]U^{-1}
\end{align}
Consider,
\begin{align}
        U^{-1}J^a_nU&=e^{-tL_{-1}+xM_{-1}}J^a_n e^{tL_{-1}-xM_{-1}}\nonumber\\
                &=\sum_{k=0}^n \frac{t^k}{k!}\frac{n!}{(n-k)!}J^a_{n-k}-nx\sum_{k=0}^{n-1}\frac{t^k}{k!}\frac{(n-1)!}{(n-k-1)!}K^a_{n-k-1}
\end{align}
where we have used the Baker-Campbell-Hausdorff(BCH) formula twice. Putting this back in (\ref{jphicalc}), we finally get,
\begin{equation}\label{GKMpa}
    \begin{split}
        [J^a_n,\phi(x,t)]&=U\left(\sum_{k=0}^n \frac{t^k}{k!}\frac{n!}{(n-k)!}[J^a_{n-k},\phi(0)]-nx\sum_{k=0}^{n-1}\frac{t^k}{k!}\frac{(n-1)!}{(n-k-1)!}[K^a_{n-k-1},\phi(0)]\right)U^{-1}\\
     &=(t^nt^a_J-nxt^{n-1}t^a_K)\phi(x,t)
    \end{split}
\end{equation}
where only the terms corresponding to $k=n$ and $k=n-1$ survived respectively in the first and the second sum because of ($\ref{algdefprim}$). 
Similarly for $K^a_n$, we get,
\begin{equation}\label{GKMpb}
    \begin{split}
        [K^a_n,\phi(x,t)]=U\left( \sum_{k=0}^n\frac{t^k}{k!}\frac{n!}{(n-k)!}[K^a_{n-k},\phi(0)]\right)U^{-1} =t^nt^a_K\phi(x,t)
    \end{split}
\end{equation}
These results verify (\ref{GKMpt}) again from a different perspective. 

\subsection{Contracting the Sugawara construction}
In section \ref{sectionsuga}, we constructed the BMS generators by taking quadratic products of the NL currents and ended up with \eqref{chh4}. In this subsection, we shall reproduce the same from the Galilean limit of Sugawara construction for CFT. We start from the following expression for virasoro modes $\mathcal{L}_m$ and $\bar{\mathcal{L}}_m$ in relativistic sugawara construction.
\begin{subequations}\label{relsuga}
\begin{align}
    \mathcal{L}_m&=\gamma\sum_{a=1}^{dim(g)}(\sum_{l\leq-1}j^a_lj^a_{m-l}+\sum_{l>-1}j^a_{m-l}j^a_l) \\
\bar{\mathcal{L}}_m&=\bar{\gamma}\sum_{a=1}^{dim(g)}(\sum_{l\leq-1}\bar{j}^a_l\bar{j}^a_{m-l}+\sum_{l>-1}\bar{j}^a_{m-l}\bar{j}^a_l) \\
 \text{ where } \gamma=\frac{1}{2(k+C_g)},&  \bar{\gamma}=\frac{1}{2(\bar{k}+C_g)}\text{ and } C_g=-\frac{1}{2dim(g)}\sum_{b,c}f^{bac}f^{bcd}
\end{align}
\end{subequations}
We can now take the Galilean limit by using the inverted version of relations (\ref{chh14}) and get the following form of Virasoro Generators by collecting the terms with same order in $\epsilon$,
\begin{align}
    \mathcal{L}_m=\frac{\gamma}{4}(A_m-\frac{B_m}{\epsilon}+\frac{C_m}{\epsilon^2}) \ ; \    \bar{\mathcal{L}}_m=\frac{\bar{\gamma}}{4}(A_m+\frac{B_m}{\epsilon}+\frac{C_m}{\epsilon^2})
\end{align}
where, 
\begin{subequations}\label{chh3}
\begin{align}
A_m&=\sum_{a=1}^{dim(g)}\sum_{l\leq-1}J^a_lJ^a_{m-l}+\sum_{l>-1}J^a_{m-l}J^a_l \\
B_m&=\sum_{a=1}^{dim(g)}\{\sum_{l\leq-1}(J^a_lK^a_{m-l}+K^a_lJ^a_{m-l})+\sum_{l>-1}(J^a_{m-l}K^a_l+K^a_{m-l}J^a_l)\}\\
C_m&=\sum_{a=1}^{dim(g)}\sum_{l}K^a_lK^a_{m-l}
\end{align}
\end{subequations}
where we have employed the commutativity of $K$'s in order to write  $\frac{1}{\epsilon^2}$ term($C_m$) as a sum from $l=-\infty$ to $\infty$.
We can use \refb{nrlim} and the above relations to obtain: 
\begin{subequations}\label{chh2}
\begin{align}
    L_m&=\frac{1}{4}\Bigl((\gamma+\bar{\gamma})A_m-\frac{(\gamma-\bar{\gamma})}{\epsilon}B_m+\frac{(\gamma+\bar{\gamma})}{\epsilon^2}C_m\Bigr)\\
    M_m&=-\frac{1}{4}\Bigl(\epsilon(\gamma-\bar{\gamma})A_m-(\gamma+\bar{\gamma})B_m+\frac{(\gamma-\bar{\gamma})}{\epsilon}C_m\Bigr)
\end{align}
\end{subequations}
Inverting the relations (\ref{chh14}), we can write $k$ and $\bar{k}$ in terms of $k_1$ and $k_2$,
\begin{align}\label{chh7}
    k=\frac{1}{2}(k_1-\frac{k_2}{\epsilon})\ ;& \ \bar{k}=\frac{1}{2}(k_1+\frac{k_2}{\epsilon})
\end{align}
Using the definitions of $\gamma$ and $\bar{\gamma}$ in (\ref{relsuga}) and using \eqref{chh7}, we can write,
\begin{align}\label{gamrel}
    \lim_{\epsilon\to 0}\frac{\gamma+\bar{\gamma}}{\epsilon^2}=-\frac{2(k_1+2C_g)}{k_2^2} \text{ and } \lim_{\epsilon\to 0}\frac{\gamma-\bar{\gamma}}{\epsilon}=-\frac{2}{k_2}
\end{align}
Hence, in limit $\epsilon\rightarrow 0$ \eqref{chh2} can be written as,
\begin{align}
    L_m=\frac{1}{2k_2}\bigg(B_m-\frac{(k_1+2C_g)}{k_2}C_m\bigg)\text{ and }M_m=\frac{1}{2k_2}C_m
\end{align}
which agrees with \eqref{chh4}.

\medskip

As we have already seen in the previous section that \eqref{chh4} satisfies BMS algebra with central charges $c_L=2dim(g)$ and $c_M=0$. This fact can also be verified using the following definitions of central charges in the relativistic Sugawara construction,
\begin{align}\label{chh8}
    c=\frac{kdim(g)}{k+C_g} \ ; \ \bar{c}=\frac{\bar{k}dim(g)}{\bar{k}+C_g}
\end{align}
Using \eqref{chh8} and following similar steps as before using the (\ref{chh7}), we can get the following,
\begin{align}
c_L=\lim_{\epsilon \to 0} (c+\bar{c})
&=\lim_{\epsilon \to 0}\bigg\{ dim(g) \bigg(\frac{\frac{1}{2}(k_1-\frac{k_2}{\epsilon})}{\frac{1}{2}(k_1-\frac{k_2}{\epsilon})+C_g}+\frac{\frac{1}{2}(k_1+\frac{k_2}{\epsilon})}{\frac{1}{2}(k_1+\frac{k_2}{\epsilon})+C_g}\bigg)\bigg\}\nonumber\\
&\Rightarrow c_L=2dim(g)
\end{align}
Similarly, we can get,
\begin{align}
c_M=\lim_{\epsilon \to 0} \epsilon(\bar{c}-c)=0
\end{align}
These are the same values of central charges appeared in the $L$, $M$ commutation relations as we got in \eqref{chh1}. 

\medskip

The modification we have done in order to get non-zero $c_M$ too can be retrieved from contraction. For this, we can start from \eqref{chh9} and its conjugate. Now if we take the limit in \eqref{chh14}, we shall again retrieve the redefined $L_n$s and $M_n$s as we have seen in \eqref{chh11}, and the commutators will be same as \eqref{chh12} with central charges \eqref{chh13}.

\subsection{NLKZ equations from Contraction}
Finally, we show how to obtain the non-Lorentzian Knizhnik Zamolodchikov equations from a limit of the ones for a 2d CFT with additional symmetry. Some of the details of the analysis are contained in Appendix \ref{AppG}. We start with the original Knizhnik Zamolodchikov equations:
\begin{subequations}
\begin{align}
     \left(\partial_{w_i}-2\gamma\sum_{j\neq i}\frac{\sum_a (t^a_{R_i})^{r_i}_{s_i}(t^a_{R_j})^{r_j}_{s_j}}{w_i-w_j}\right)\langle...\phi^{s_i}_{R_i}(w_j)...\phi^{s_j}_{R_j}(w_j)...\rangle=0 \label{kz1}\\
      \left(\partial_{\bar{w}_i}-2\bar{\gamma}\sum_{j\neq i}\frac{\sum_a (\bar{t}^a_{\bar{R}_i})^{\bar{r}_i}_{\bar{s}_i}(\bar{t}^a_{\bar{R}_j})^{\bar{r}_j}_{\bar{s}_j}}{\bar{w}_i-\bar{w}_j}\right)\langle...\bar{\phi}^{\bar{s}_i}_{\bar{R}_i}(\bar{w}_j)...\bar{\phi}^{\bar{s}_j}_{\bar{R}_j}(\bar{w}_j)...\rangle=0 \label{kz2} 
\end{align}
\end{subequations}
where
\begin{align}
(t^a_{R_i})^{r_i}_{s_i}(t^a_{R_j})^{r_j}_{s_j}\langle...\phi^{s_i}_{R_i}(w_j)...\phi^{s_j}_{R_j}(w_j)...\rangle=((t^a_{R_i}\otimes t^a_{R_j})\langle...\phi_{R_i}(w_j)...\phi_{R_j}(w_j)...\rangle)^{r_i,r_j}
\end{align}
Using the separability of the primary fields, $\Phi_{R,\bar{R}}^{r,\bar{r}}(w,\bar{w})=\phi_{R}^r(w)\otimes' \bar{\phi}_{\bar{R}}^{\bar{r}}(\bar{w})$, we can write,
\begin{align}
    \langle\Phi_{R_1,\bar{R}_1}^{r_1,\bar{r}_1}(w_1,\bar{w}_1)...\Phi_{R_N,\bar{R}_N}^{r_N,\bar{r}_N}(w_N,\bar{w}_N)\rangle=\langle\phi_{R_1}^{r_1}(w_1)...\phi_{R_N}^{r_N}(w_N)\rangle\langle\bar{\phi}_{\bar{R}_1}^{\bar{r}_1}(\bar{w}_1)...\bar{\phi}_{\bar{R}_N}^{\bar{r}_N}(\bar{w}_N)\rangle
\end{align}
where the primed tensor product($\otimes'$) ensures independent action of operators on chiral and anti-chiral sectors whereas the unprimed tensor product($\otimes$) ensures independent action on $i$th and $j$th insertion in the $n$-point function.\\

Using the following linear combinations in limit $\epsilon\rightarrow 0$,
\begin{align}
(\ref{kz1}) \times \langle\bar{\phi}_{\bar{R}_1}^{\bar{r}_1}(\bar{w}_1)...\bar{\phi}_{\bar{R}_N}^{\bar{r}_N}(\bar{w}_N)\rangle+(\ref{kz2})\times\langle\phi_{R_1}^{r_1}(w_1)...\phi_{R_N}^{r_N}(w_N)\rangle=0\nonumber\\
 \epsilon\{(\ref{kz1}) \times\langle\bar{\phi}_{\bar{R}_1}^{\bar{r}_1}(\bar{w}_1)...\bar{\phi}_{\bar{R}_N}^{\bar{r}_N}(\bar{w}_N)\rangle-(\ref{kz2})\times\langle\phi_{R_1}^{r_1}(w_1)...\phi_{R_N}^{r_N}(w_N)\rangle\}=0
\end{align}
We get (shown in details in Appendix \ref{AppG}),
\begin{equation}
    \begin{split}
        & \bigg[\partial_{t_i}-\frac{1}{k_2}\sum_{j\neq i}\bigg\{\frac{\sum_a (t^a_{R_i,J}\otimes t^a_{R_j,K}+t^a_{R_i,K}\otimes t^a_{R_j,J})}{t_{ij}}+\bigg(\frac{x_{ij}}{t_{ij}^2}-\frac{(k_1+2C_g)}{k_2t_{ij}}\bigg)\sum_a (t^a_{R_i,K}\otimes t^a_{R_j,K})\bigg\}\bigg]\\
        &\hspace{250pt}\langle...\Phi_{R_i,\bar{R}_i}(x_i,t_i)...\Phi_{R_j,\bar{R_j}}(x_j,t_j)...\rangle=0\\ 
        &\bigg[ \partial_{x_i}+\frac{1}{k_2}\sum_{j\neq i} \frac{\sum_a (t^a_{R_i,K}\otimes t^a_{R_j,K})}{t_{ij}} \bigg]\langle...\Phi_{R_i,\bar{R}_i}(x_i,t_i)...\Phi_{R_j,\bar{R}_j}(x_j,t_j)...\rangle=0
    \end{split}
\end{equation}
where, $t^a_{R_i,J}=t^a_{R_i}\otimes'\bar{I}+I\otimes' \bar{t}^a_{\bar{R}_i}$ and $t^a_{R_i,J}=\epsilon(I\otimes' \bar{t}^a_{\bar{R}_i}-t^a_{R_i}\otimes'\bar{I})$ in limit $\epsilon \rightarrow 0$. The above equations are same as what we got form intrinsic analysis i.e. (\ref{nlkz1}) and (\ref{nlkz2}), with $t\to v$ and $x\to u$ as its supposed to for a Galilean contracted result.

\newpage
\section{Conclusions}

\subsection{Summary}
In this paper we have explored aspects of Non-Lorentzian CFTs with additional Lie Algebraic symmetries. First we have reproduced the Non-Lorentzian Kač-Moody algebra by taking singular limit from the Virasoro Kač-moody algebra. 

\medskip

After this we attempt to construct the same from intrinsic viewpoint without any knowledge of the parent algebra. We see that 2d Carrollian Conformal symmetry allows an infinite number of Noether currents. We then take a few pairs of conserved currents (introducing flavour indices to distinguish them) satisfying the conditions for EM tensor components in 2d Carrollian Conformal symmetry. After this we derive the Ward identities associated with those conserved currents. We also derive OPEs of a general 2d Carrollian Conformal field with the current. After this we find the conserved charge operators associated with these currents. The transformation generated by these charges on a generic field is also derived. Here we first encounter the transformation matrices which we encounter in relativistic CFT with Kac-Moody algebra. We introduce current primary fields which turn out to be analogous to the Virasoro Kač-Moody primary fields. After this the current current OPEs are derived. While doing so, the structure constants emerge from the action of the transformation matrices on the currents. After this, global internal symmetry is applied on two point and three point correlation and it turns out that the structure constants we have encountered while calculating the current current OPEs satisfy Jacobi identity. After this we derive the OPE between the Energy Momentum Tensor components and the current components. Using all these OPEs we derive the algebra of the current modes and the Energy Momentum tensor modes. The algebra transpire to be identical to the algebra obtained from limits of Virasoro Kač-Moody algebra.

\medskip

Later in the paper, we attempt to construct the EM tensor modes (which forms the BMS algebra) from the current modes through Sugawara construction. Here we see that the Sugawara construction only takes us to the BMS algebra with one of the central charges to be zero. We needed another modification to the Sugawara construction in order to get a fully centrally extended BMS algebra. Using the expression of the EM tensor modes in terms of the current modes, we calculate the OPEs of EM tensor fields with themselves and with currents and see that the OPEs thus derived matches with the OPEs in the earlier section.

\medskip

After this we have a brief look at the tensionless string. When we look at the mode expansion of the coordinates (treating them as scalar fields), we see that the modes satisfy a special case of Non Lorentzian $U(1)$ Kac-Moody algebra. We also look at the expression of modes of the classical energy momentum tensor in terms of these $U(1)$ modes and see that classically this matches with the expression we have earlier derived for Non-Lorentzian Sugawara construction. Hence we see that $U(1)$ NLKM currents are intrinsically present in the tensionless string.

\medskip

In section 6, we take a correlation function of a string of BMS current primary fields. Using the OPE definition of the BMS primary field, we arrive at the Non-Lorentzian version of the Knizhnik Zamolodchikov equations. Finally, in section 7, we derive all the earlier results by taking limits from the parent algebra.

\subsection{Discussions and future directions}
We mentioned in the introduction that our results in this paper lay the groundwork for a large number of applications, most importantly to the construction of a holographic dictionary for asymptotic flat spacetimes and also the understanding of tensionless strings. We have built the underlying algebraic structures in this paper which would be of importance to the quantum field theories that are at the heart of these problems. 

\medskip

One of the most important and immediate next steps is to construct a Non-Lorentzian Wess-Zumino-Witten model that realises these symmetries. This would be central to the understanding of tensionless null strings moving on arbitrarily curved manifolds. The work on this is currently underway. 

\medskip

The tensionless limit of string theory on arbitrary backgrounds is intimately related to this and as mentioned in the introduction, we wish to revisit the construction of  \cite{Lindstrom:2003mg} in the light of our findings in this paper. We believe that this limit would lead to null tensionless strings in AdS. This is to be contrasted with tensionless strings in AdS considered in e.g. \cite{Gaberdiel:2018rqv} and subsequent work in this direction, which are tensionless but not null. A better understanding of the differences and perhaps similarities between the two approaches would be important. 

\medskip

A generalisation of the methods outlined in this paper would be carried out for 3d Carrollian and 3d Galilean theories. The structures are expected to remain similar for the 3d Galilean theories, as the infinite dimensional structure of the algebra without the extra currents remains intact when generalised to higher dimensions. But for Carrollian theories, owing to the fundamental difference between BMS$_3$ and BMS$_4$, where one copy of the Virasoro in the 3d case gets enhanced to two Virasoros in the 4d case and supertranslations develop two legs instead of one, the construction of the quantum field theories with additional non-abelian currents would be more involved. 

\bigskip 

\subsection*{Acknowledgements}
We thank Aritra Banerjee, Rudranil Basu and Niels Obers for interesting discussions and comments on an initial version of the manuscript. 

\smallskip

The work of AB is partially supported by a Swarnajayanti fellowship (SB/SJF/2019-20/08) from the Science and Engineering Research Board (SERB) India, the SERB grant (CRG/2020/002035), and a visiting professorship at \'{E}cole Polytechnique Paris. AB also acknowledges the warm hospitality of the Niels Bohr Institute, Copenhagen during later stages of this work. RC is supported by the CSIR grant File No: 09/092(0991)/2018-EMR-I. AS is financially supported by a PMRF fellowship, MHRD, India. RK acknowledges the support of the Department of Atomic Energy, Government of India, under project number RTI4001. DS would like to thank
ICTS, Bengaluru for hospitality during the course of this project.

\bigskip \bigskip

\newpage

\section*{APPENDICES}
\appendix

\section{Carroll Multiplets}\label{ApCar}
In this appendix, we review the construction of Carrollian boost multiplets in two dimensions.

\medskip

In two space-time dimensions, the Carrollian boost (CB) transformation is defined as: $x\rightarrow x^\prime=x\text{ , }t\rightarrow t^\prime=t+vx${ }; or equivalently, as:
\begin{align}
\begin{pmatrix}
x\\
t
\end{pmatrix}\longrightarrow \begin{pmatrix}
x^\prime\\
t^\prime
\end{pmatrix}= \left[\exp{\begin{pmatrix}
0 & 0\\
v & 0
\end{pmatrix}}\right]\begin{pmatrix}
x\\
t
\end{pmatrix}\text{ }\Longleftrightarrow\text{ } x^\mu\rightarrow {x^{\prime}}^\mu={\left[e^{v\mathbf{B_{(2)}}}\right]}^{\mu}_{\hspace{2mm}\nu}\text{ }x^\nu
\end{align}
with 
\begin{align}
\mathbf{B_{(2)}}:=\begin{pmatrix}
0 & 0\\
1 & 0
\end{pmatrix}
\end{align}
being the 2D representation of the CB generator $\mathbf{B}$ that is clearly not diagonalizable.

\medskip

Taking a cue from the Lorentz covariance of Lorentz tensors, it was postulated in \cite{} that a rank-$n$ Carrollian Cartesian tensor field $\Phi$ with `boost-charge' $\xi$ transforms under the Carrollian boost as:
\begin{align}
&{\Phi}^{\mu_1...\mu_n}(t,x)\longrightarrow{\tilde{\Phi}}^{\mu_1...\mu_n}(t^\prime,x^\prime)={\left[e^{-\xi v\mathbf{B_{(2)}}}\right]}^{\mu_1}_{\hspace{3mm}\nu_1}...{\left[e^{-\xi v\mathbf{B_{(2)}}}\right]}^{\mu_n}_{\hspace{3mm}\nu_n}{\Phi}^{\nu_1...\nu_n}(t,x)\nonumber\\
\Longleftrightarrow\qquad &{\mathbf{\Phi}}(t,x)\longrightarrow{{\mathbf{\tilde{\Phi}}}}(t^\prime,x^\prime)=\left[\bigotimes_{i=1}^n e^{-\xi v\mathbf{B_{(2)}}}\right]{\mathbf{\Phi}}(t,x)=e^{-\xi v\bigoplus\limits_{i=1}^n\mathbf{B_{(2)}}}{\mathbf{\Phi}}(t,x)\label{x:1}
\end{align}
where $\mu_i, \nu_i$ are Carrollian space-time indices and for matrices, the left index denotes row while the right one denotes column; repeated indices are summed over and, in \eqref{x:1}, indices are suppressed. It is to be noted that the up/down appearance of a tensor-index does not matter; only the left/right ordering is important.

\medskip

Clearly, the Carrollian Cartesian tensors defined above are decomposible. So, we now construct indecomposible Carrollian multiplets from these tensors. We begin by recognizing that:
\begin{align}
\mathbf{B_{(2)}}=\mathbf{J}_{(l=\frac{1}{2})}^\mathbf{-}
\end{align}
which is the lowering ladder operator in the $\text{SU}(2)$ spin-$\frac{1}{2}$ representation. Thus, $\bigoplus\limits_{i=1}^n\mathbf{B_{(2)}}$ in \eqref{x:1} can be decomposed into indecomposable representations of $\mathbf{J}^\mathbf{-} $ using the technique of `addition of $n$ spin-$\frac{1}{2}$ angular momenta' in quantum mechanics, such that:
\begin{align}
\mathbf{B_{(d)}}\equiv\mathbf{J}_{(l=\frac{d-1}{2})}^\mathbf{-}
\end{align} 
It is evident that the representations $\mathbf{B_{(d)}}$ of the classical CB generator are indecomposable since their only generalized eigenvalue is 0 and it has geometric multiplicity 1. But, these representations are reducible for $d\geq2$. 

\medskip

A multi-component field transforming under the $d$-dimensional representation of CB, $\mathbf{B_{(d)}}$, will be called a Carrollian multiplet of rank $\frac{d-1}{2}$ with $d$ components, denoted by 
\begin{align*}
\Phi_{(l=\frac{d-1}{2})}^m \text{\hspace{2.5mm} with \hspace{2.5mm}} m=\frac{1-d}{2},\frac{3-d}{2},...,\frac{d-1}{2}
\end{align*}
By treating the $\mu=t$ index as spin-$\frac{1}{2}$ up-state and the $\mu=x$ index as spin-$\frac{1}{2}$ down-state, components $\Phi_{(l)}^m$ of a Carrollian multiplet arise precisely as such linear combinations (with proper Clebsch-Gordon coefficients) of the components of a Cartesian tensor of an allowed rank $n$ that would appear while expanding the $|l,m\rangle$ states in an allowed $|s_1,s_2,...,s_n\rangle$ basis (where $|s_i\rangle$ are $\mathbf{J}_{(\frac{1}{2})}^\mathbf{z}$ eigenstates). So, as a linear combination of the components of a rank-$n$ Cartesian tensor, one can obtain multipltes of ranks: $0,1,2,...,\frac{n}{2}$ for even $n$ and $\frac{1}{2},\frac{3}{2},\frac{5}{2},...,\frac{n}{2}$ for odd $n$. As an example, we see how Carrollian multiplets of ranks $\frac{1}{2}$ and $\frac{3}{2}$ are constructed from a rank-3 Cartesian tensor:
\begin{align*}
&\Phi_{(\frac{3}{2})}^{\frac{3}{2}}(t,x):=\Phi^{ttt}(t,x)\text{ }\text{ }\qquad\text{ }\text{ }\Phi_{(\frac{3}{2})}^{\frac{1}{2}}(t,x):=\frac{1}{\sqrt{3}}\left[\Phi^{ttx}+\Phi^{txt}+\Phi^{xtt}\right](t,x)\nonumber\\
&\Phi_{(\frac{3}{2})}^{-\frac{1}{2}}(t,x):=\frac{1}{\sqrt{3}}\left[\Phi^{txx}+\Phi^{xtx}+\Phi^{xxt}\right](t,x)\text{ }\text{ }\qquad\text{ }\text{ }\Phi_{(\frac{3}{2})}^{-\frac{3}{2}}(t,x):=\Phi^{xxx}(t,x)\\
&\Phi_{(\frac{1}{2})}^{\frac{1}{2}}(t,x):=\frac{1}{\sqrt{a^2+b^2+c^2}}\left[a\Phi^{ttx}+b\Phi^{txt}+c\Phi^{xtt}\right](t,x) \text{ }\text{ }\qquad\text{with $a+b+c=0$}\nonumber\\
&\Phi_{(\frac{1}{2})}^{-\frac{1}{2}}(t,x):=\frac{1}{\sqrt{a^2+b^2+c^2}}\left[a\Phi^{xxt}+b\Phi^{xtx}+c\Phi^{txx}\right](t,x)\nonumber
\end{align*}
(As the tuple $(a,b,c)$ in $\mathbb{R}^3$ lies on the plane $a+b+c=0$ which is spanned by two basis vectors, two linearly independent rank-$\frac{1}{2}$ multiplets arise.)

\medskip

A rank-$l$ Carrollian multiplet with boost-charge $\xi$ thus transforms under the $2l+1$ dimensional representation of the CB as:
\begin{align}
\Phi_{(l)}^m(t,x)\longrightarrow\tilde{\Phi}_{(l)}^m(t^\prime,x^\prime)={\left[e^{-\xi v\mathbf{J}_{(l)}^{\mathbf{-}}}\right]}^m_{\hspace{2mm} m^\prime}{\Phi}_{(l)}^{m^\prime}(t,x)\label{x:2}
\end{align}

\medskip

After constructing the Carrollian multiplets from the Cartesian tensors as demonstrated above, the components of the multiplets can always be redefined such that:
\begin{align*}
\text{in \eqref{x:2}, }\mathbf{J}_{(l)}^{\mathbf{-}}\text{ is replaced by }\mathbf{M}_{(l)}:=\text{sub-diag }{(1,1,...,1)}_{2l+1}\text{ .}
\end{align*}
Hence, instead of the actual $\mathbf{J}_{(l)}^{\mathbf{-}}$ matrix, only the indecomposable Jordan-block structure is important for defining the CB transformation property of the Carrollian multiplets.

\medskip

We conclude this appendix with the following observation. Since the finite dimensional indecomposable representations of $\mathbf{B}$ are not symmetric (or Hermitian), one can start with:
\begin{align*}
\begin{pmatrix}
t\\
x
\end{pmatrix}\longrightarrow \begin{pmatrix}
t^\prime\\
x^\prime
\end{pmatrix}= \left[\exp{\begin{pmatrix}
0 & v\\
0 & 0
\end{pmatrix}}\right]\begin{pmatrix}
t\\
x
\end{pmatrix}\text{ }\Longleftrightarrow\text{ } x^\mu\rightarrow {x^{\prime}}^\mu={\left[e^{v\mathbf{B^\prime_{(2)}}}\right]}^{\mu}_{\hspace{2mm}\nu}\text{ }x^\nu
\end{align*}
where 
\begin{align*}
\mathbf{B^\prime_{(2)}}:=\begin{pmatrix}
0 & 1\\
0 & 0
\end{pmatrix}
\end{align*}
and follow the preceding argument to construct
\begin{align*}
\mathbf{B^\prime_{(d)}}\equiv\mathbf{J}_{(l=\frac{d-1}{2})}^\mathbf{+}
\end{align*}
which is the $\text{SU}(2)$ raising ladder operator. But, as $\mathbf{J}_{(l)}^\mathbf{+}={(\mathbf{J}_{(l)}^\mathbf{-})}^{\textbf{T}}$, the raising and lowering operators' representation matrices are related to each other by the similarity transformation: 
\begin{align*}
S=\text{anti-diag }{(1,1,...,1)}_{2l+1}
\end{align*}
and consequently, $\mathbf{B}$ and $\mathbf{B^\prime}$ furnish two equivalent representations of the CB generator.

\bigskip \bigskip

\section{Calculation of Sugawara Construction Commutators}\label{AppA}

In this appendix, we provide the details of the calculation of commutators of the NLKM algebra from the Sugawara construction. 

\subsection*{Calculating $[M_m, J^b_n]$}
\begin{align}
[M_m, J^b_n]&=\frac{1}{2k_2}\sum_{a}\sum_{l}[K^a_lK^a_{m-l},J^b_n] \qquad\qquad\qquad\qquad\qquad\qquad\qquad\text{                         (Using (\ref{chh4}))}\nonumber\\
&=\frac{1}{2k_2}\sum_{a}\sum_{l}\Bigl([K^a_l,J^b_n]K^a_{m-l}+K^a_l[K^a_{m-l},J^b_n]\Bigr)\nonumber\\
&=\frac{1}{2k_2}\sum_{a}\sum_{l}\bigg\{(-i\sum_{c} f^{bac}K^c_{n+l}-nk_2\delta_{n+l,0}\delta^{ab})K^a_{m-l}\nonumber\\
&\qquad+K^a_l(-i\sum_{c} f^{bac}K^c_{m+n-l}-nk_2\delta_{m+n-l,0}\delta^{ab})\bigg\} \text{ \qquad\qquad(Using (\ref{gkm}))}\nonumber\\
&=\frac{i}{2k_2}\sum_{a,c}\sum_{l}f^{abc}\Bigl( K^c_{n+l}K^a_{m-l}+K^a_lK^c_{m+n-l}\Bigr)-nK^b_{m+n} \qquad(\because -f^{bac}=f^{abc})\nonumber
\end{align}
where we have omitted the limits in summation over $a$ and $c$ and it is understood that it runs over $1$ to $dim(g)$. Now, $\sum_{l}K^c_{n+l}K^a_{m-l}=\sum_{l}K^c_{m+n-l}K^a_{l}$ simply by translating $l$. Hence,
\begin{align}\label{mjcom}
    [M_m,J^b_n]=\frac{i}{k_2}\sum_{a,c}\sum_{l}f^{abc}K^a_lK^c_{m+n-l}-nK^b_{m+n}
\end{align}
Now, again using antisymmetry property of $f^{abc}$
\begin{align}
    \sum_{a,c}\sum_{l}f^{abc}K^a_lK^c_{m+n-l}&=\sum_{a,c}\sum_{l}f^{abc}K^a_{m+n-l}K^c_{l}\nonumber\\
    &=\sum_{a,c}\sum_{l}f^{cba}K^c_{m+n-l}K^a_{l}\nonumber\\
    &=-\sum_{a,c}\sum_{l}f^{abc}K^a_{l}K^c_{m+n-l}\nonumber\\
    \Rightarrow \sum_{a,c}\sum_{l}f^{abc}K^a_lK^c_{m+n-l}=0
\end{align}
Hence, \eqref{mjcom} can be written as, 
\begin{align}
    [M_m,J^b_n]=-nK^b_{m+n}
\end{align}
\subsection*{Calculating $[L_m,K^b_n]$}
Again, using (\ref{chh4}) and (\ref{gkm}), we can get the following,
\begin{equation}
    \begin{split}
        [L_m, K^b_n]&=\frac{1}{2k_2}\sum_{a}\bigg[ \sum_{l\leq-1}([J^a_l ,K^b_n]K^a_{m-l}+K^a_l[ J^a_{m-l},K^b_n])\\
        &\hspace{50 pt}+\sum_{l>-1}([J^a_{m-l},K^b_n]K^a_l+K^a_{m-l}[J^a_l,K^b_n])\bigg]\\
        &=\frac{1}{2k_2}\sum_{a}\bigg[ \sum_{l\leq-1}\bigg\{i\sum_c f^{abc}(K^c_{l+n}K^a_{m-l}+K^a_lK^c_{m+n-l})+k_2lK^a_{m-l}\delta_{l+n,0}\delta^{ab}\\
        &\hspace{10 pt}+k_2(m-l)K^a_l\delta_{m+n-l,0}\delta^{ab}\bigg\} 
        \quad +\sum_{l>-1}\bigg\{i\sum_c f^{abc}(K^c_{m+n-l}K^a_l+K^a_{m-l}K^c_{l+n})\\ &\hspace{130 pt}+k_2(m-l)K^a_l\delta_{m+n-l,0}\delta^{ab}+k_2lK^a_{m-l}\delta_{l+n,0}\delta^{ab}\bigg\} \bigg]\\
        &=\frac{i}{k_2}\sum_{a,c}\sum_lf^{abc}(K^c_{m+n-l}K^a_l+K^a_{m-l}K^c_{l+n})\\
        &\qquad \qquad \qquad \qquad  +\frac{1}{2k_2}\sum_a\sum_l\bigg\{k_2(m-l)K^a_l\delta_{m+n-l,0}\delta^{ab}+k_2lK^a_{m-l}\delta_{l+n,0}\delta^{ab}\bigg\}
    \end{split}
\end{equation}

First term vanishes again because of the antisymmetry of $f$ and commutativity of $K$'s. Therefore,
\begin{align}\label{A.2}
    [L_m, K^b_n]=-nK^{b}_{m+n}
\end{align}
\subsection*{Calculating $[L_m,J^b_n]$}
Following similar steps as before, we get,
\begin{align}
    [L_m,J^b_n]=&\frac{1}{2k_2}\sum_{a}\bigg\{ \sum_{l\leq-1}([J^a_lK^a_{m-l},J^b_n]+[K^a_lJ^a_{m-l},J^b_n])+\sum_{l>-1}([J^a_{m-l}K^a_l,J^b_n]\nonumber \\
    &\qquad\qquad\qquad\qquad \qquad+[K^a_{m-l}J^a_l,J^b_n])\bigg\}-\frac{(k_1+2C_g)}{k_2}[M_m,J^b_n]
\end{align}
First term can be simplified as,
 \begin{align}
\sum_a\sum_{l\leq-1}&([J^a_lK^a_{m-l},J^b_n]+[K^a_lJ^a_{m-l},J^b_n])\nonumber\\ 
&=\sum_a  \sum_{l\leq-1}\bigg\{J^a_l[K^a_{m-l},J^b_n]+[J^a_l,J^b_n]K^a_{m-l}+K^a_l[J^a_{m-l},J^b_n]+[K^a_l,J^b_n]J^a_{m-l}\bigg\}\nonumber\\
&=\sum_a\sum_{l\leq-1}\bigg\{i\sum_c f^{abc}J^a_lK^c_{m+n-l}+i\sum_c f^{abc}J^c_{l+n}K^a_{m-l}+i\sum_c f^{abc}K^a_{l}J^c_{m+n-l}\nonumber\\
&\quad+i\sum_c f^{abc}K^c_{l+n}J^a_{m-l}-k_2nJ^a_l\delta_{m+n-l,0}\delta^{ab}
+k_1 l K^a_{m-l}\delta_{l+n,0}\delta^{ab}\nonumber\\
&\quad+k_1 (m-l) K^a_{l}\delta_{m+n-l,0}\delta^{ab}-k_2nJ^a_{m-l}\delta_{l+n,0}\delta^{ab}\bigg\}\nonumber \\
&=\sum_{l\leq-1}\bigg\{i\sum_{a,c} f^{abc}J^a_lK^c_{m+n-l}+i\sum_{a,c} f^{abc}J^c_{l+n}K^a_{m-l}+i\sum_{a,c} f^{abc}K^a_{l}J^c_{m+n-l}\nonumber\\
&\quad+i\sum_{a,c} f^{abc}K^c_{l+n}J^a_{m-l}\bigg\}-n\sum_{l\leq-1}(k_2J^b_{m+n}+k_1 K^b_{m+n})(\delta_{m+n-l,0}
+\delta_{l+n,0})
\end{align}
Similarly, the second term looks like,
\begin{align}
    \sum_a\sum_{l>-1}&([K^a_{m-l}J^a_l,J^b_n]+[J^a_{m-l}K^a_l,J^b_n])\nonumber\\
    &=\sum_{l>-1}\bigg\{i\sum_{a,c} f^{abc}K^a_{m-l}J^c_{l+n}+i\sum_{a,c} f^{abc}K^c_{m+n-l}J^a_l+i\sum_{a,c} f^{abc}J^a_{m-l}K^c_{l+n}\nonumber\\
    &\quad+i\sum_{a,c} f^{abc}J^c_{m+n-l}K^a_{l}\bigg\}-n\sum_{l>-1}(k_2J^b_{m+n}+k_1 K^b_{m+n})(\delta_{m+n-l,0}
+\delta_{l+n,0})
\end{align}
Hence, we have,
\begin{align}
    [L_m,J^b_n]
     &=\frac{1}{2k_2}\bigg[i\sum_{a,c}f^{abc}\bigg\{\sum_{0\leq l\leq n-1}J^c_{l}K^a_{m+n-l}-\sum_{0\leq l\leq n-1}K^a_lJ^c_{m+n-l}-\sum_{0\leq l\leq n-1}K^a_{m+n-l}J^c_l\nonumber\\
     &\quad+\sum_{0\leq l\leq n-1}J^c_{m+n-l}K^a_l\bigg\}\bigg]-nJ^b_{m+n}+\frac{2C_g}{k_2}nK^b_{m+n}\nonumber\\
     &=\frac{1}{2k_2}\bigg[i\sum_{a,c}f^{abc}\bigg\{\sum_{0\leq l\leq n-1}[J^c_{l},K^a_{m+n-l}]+\sum_{0\leq l\leq n-1}[J^c_{m+n-l},K^a_l]\bigg\}\bigg]\nonumber\\
     &\quad-nJ^b_{m+n}+\frac{2C_g}{k_2}nK^b_{m+n}\nonumber\\
     &=\frac{1}{2k_2}\bigg[-2n\sum_{a,c,d}f^{abc}f^{cad}K^d_{m+n}+2ik_2\sum_{a,c}\sum_{0\leq l\leq n-1}f^{abc}\delta_{m+n,0}\delta^{ab}\bigg]\nonumber\\
     &\quad-nJ^b_{m+n}+\frac{2C_g}{k_2}nK^b_{m+n}\nonumber\\
      &=-nJ^b_{m+n}+\frac{n}{k_2}(2C_gK^b_{m+n}-\sum_{a,c,d}f^{abc}f^{cad}K^d_{m+n})\nonumber\\
      &=-nJ^b_{m+n}\nonumber
\end{align}
Hence, we have, 
\begin{align}\label{A.3}
    [L_m,J^b_n]=-nJ^b_{m+n}
\end{align}
\subsection*{Calculating $[L_m,M_n]$ }
\begin{align}
    [L_m,M_n]&=\frac{1}{2k_2}\sum_{a}\sum_{l}\bigg([L_m,K^a_l]K^a_{n-l}+K^a_l[L_m,K^a_{n-l}]\bigg) \nonumber\\
    &=\frac{1}{2k_2}\sum_{a}\sum_{l}\bigg\{-lK^a_{m+l}K^a_{n-l}-(n-l)K^a_lK^a_{m+n-l}\bigg\}\qquad \qquad \qquad \text{ (Using (\ref{A.2}))}\nonumber\\
    &=\frac{1}{2k_2}\sum_{a}\sum_{l}\{-(l-m)K^a_{l}K^a_{n+m-l}-(n-l)K^a_lK^a_{m+n-l}\}\nonumber\\
    &=(m-n)\frac{1}{2k_2}\sum_{a}\sum_{l}K^a_{l}K^a_{n+m-l}\nonumber
\end{align}
where , we have done the re-labelling $l\rightarrow l-m$ in the second step. Hence, we get,
\begin{align}\label{A.4}
    [L_m,M_n]=(m-n)M_{m+n}
\end{align}

\subsection*{Calculating $[L_m,L_n]$}
It can be carried out in similar fashion by writing one of the $L$'s in terms of $J$'s and $K$'s using (\ref{chh4}) and then using (\ref{A.2}), (\ref{A.3}) and (\ref{A.4}), we can do the following,
\begin{align}
    [L_m,L_n]=&\frac{1}{2k_2}\sum_a\bigg\{ \sum_{l\leq-1}([L_m,J^a_l]K^a_{n-l}+J^a_l[L_m,K^a_{n-l}]+[L_m,K^a_l]J^a_{n-l}+K^a_l[L_m,J^a_{n-l}])\nonumber\\
    &+\sum_{l>-1}([L_m,J^a_{n-l}]K^a_l+J^a_{n-l}[L_m,K^a_l]+[L_m,K^a_{n-l}]J^a_l+K^a_{n-l}[L_m,J^a_l])\bigg\}\nonumber\\
    &-\frac{(k_1+2C_g)}{k_2}[L_m,\frac{1}{2k_2}\sum_{l}K^a_lK^a_{n-l}]\nonumber\\
    &=\frac{1}{2k_2}\sum_a\bigg\{ \sum_{l\leq-1}(-lJ^a_{m+l}K^a_{n-l}+lJ^a_lK^a_{m+n-l}-lK^a_{m+l}J^a_{n-l}+lK^a_lJ^a_{m+n-l})\nonumber\\
    &+\sum_{l>-1}(lJ^a_{m+n-l}K^a_l-lJ^a_{n-l}K^a_{m+l}+lK^a_{m+n-l}J^a_l-lK^a_{n-l}J^a_{m+l})\bigg\}\nonumber\\
    &-m\frac{(k_1+2C_g)}{k_2}\frac{1}{2k_2}\sum_a\sum_lK^a_lK^a_{m+n-l}-nL_{m+n}
    \end{align}
    Changing the index $l\rightarrow (l-m)$ in the negative terms in the curly brackets and then simplifying, we can get,
    \begin{align}
     [L_m,L_n] &=\frac{1}{2k_2}\sum_a\bigg\{ \sum_{l\leq-1}m(J^a_{l}K^a_{n+m-l}+K^a_{l}J^a_{m+n-l})
     +\sum_{l> -1}m(J^a_{n+m-l}K^a_{l}+K^a_{n+m-l}J^a_{l})\nonumber\\
     & + \sum_{l=0}^{m-1}(m-l)([J^a_{l},K^a_{n+m-l}]-[J^a_{n+m-l},K^a_{l}])\bigg\}\nonumber\\
     &-m\frac{(k_1+2C_g)}{k_2}\frac{1}{2k_2}\sum_a\sum_lK^a_lK^a_{m+n-l}-nL_{m+n}\nonumber\\
     &=\frac{1}{2k_2}\sum_a\bigg\{ \sum_{l=0}^{m-1}(m-l)(k_2l\delta_{m+n,0}-k_2(n+m-l)\delta_{m+n,0})\bigg\}+(m-n)L_{m+n}
     \end{align}
     which upon further simplification gives,
     \begin{align}\label{A.5}
    [L_m,L_n]=(m-n)L_{m+n}+\frac{dim(g)}{6}m(m^2-1)\delta_{m+n,0}
     \end{align}
Other commutation relations of type $[M_m,K^b_n]$ and $[M_m,M_n]$ vanish trivially because of the vanishing $[K^a_m,K^b_n]$ commutator.

\section{Modified Sugawara Construction}\label{amod_Suga}
In this appendix, we give details of the modified Sugawara construction. First we start with the modified construction in the relativistic case and then explain the construction in the non-Lorentzian case. 

\medskip

We begin by calculating $[\tilde{\mathcal{L}}_m,\tilde{\mathcal{L}}_n]$ with $\tilde{\mathcal{L}}_m$ given in (\ref{chh9})
\begin{align}
    [\tilde{\mathcal{L}}_m,\tilde{\mathcal{L}}_n]&=[\mathcal{L}^{S}_m+im\theta^aj^{a}_m+\frac{1}{2}k\theta^2\delta_{m,0},\mathcal{L}^{S}_n+in\theta^bj^{b}_n+\frac{1}{2}k\theta^2\delta_{n,0}]\nonumber\\
    &=[\mathcal{L}^{S}_m,\mathcal{L}^{S}_n]+in\theta^b[\mathcal{L}^{S}_m,j^b_n]+im\theta^a[j^a_m,\mathcal{L}^{S}_n]-mn\theta^a\theta^b[j^a_m,j^b_n]\nonumber\\
    &=(m-n)\mathcal{L}^{S}_{m+n}-in^2\theta^bj^{b}_{m+n}+im^2\theta^a j^{a}_{m+n}-mn\theta^a\theta^b\Big(if^{abc}j^{c}_{m+n}+mk\delta_{m+n}\delta_{ab}\Big)\nonumber\\
    &=(m-n)\tilde{\mathcal{L}}_{m+n}-i(m^2-n^2)\theta^aj^a_{m+n}-\frac{1}{2}k\theta^2(m-n)\delta_{m+n,0}+i(m^2-n^2)\theta^aj^a_{m+n}\nonumber\\
    &\hspace{40mm}-imn\theta^a\theta^bf^{abc}j^{c}_{m+n}+m^3k\theta^2\delta_{m+n,0}+\frac{c}{12}(m^3-m)\delta_{m+n,0}\nonumber\\
    &=(m-n)\tilde{\mathcal{L}}_{m+n}+\Big(\frac{c}{12}+k\theta^2\Big)(m^3-m)\delta_{m+n,0}\nonumber\\
    &=(m-n)\tilde{\mathcal{L}}_{m+n}+\frac{\tilde{c}}{12}(m^3-m)\delta_{m+n,0}
\end{align}
Here $\tilde{c}=c+12k\theta^2$. In the fifth line we have used the fact that $\theta^a\theta^b f^{abc}$ vanishes due to antisymmetry of $f^{abc}$ over indices $a$ and $b$, also the fact that $n\delta_{m+n,0}=-m\delta_{m+n,0}$.\\
\\ Now defining $\tilde{L}$ and $\tilde{M}$ as in (\ref{chh11}) we would like to calculate $[\tilde{L}_m,\tilde{L}_n]$ and $[\tilde{L}_m,\tilde{M}_n]$. The calculation of $[\tilde{L}_m,\tilde{L}_n]$ will be exactly similar to that of $[\tilde{\mathcal{L}}_m,\tilde{\mathcal{L}}_n]$, while that of $[\tilde{L}_m,\tilde{M}_n]$ will be as following
\begin{align}
   [\tilde{L}_m,\tilde{M}_n]&=[L^{S}_m+im\theta^aJ^{a}_m+\frac{1}{2}k_2\theta^2\delta_{n,0},M^{S}_n+in\theta^bK^{b}_n+\frac{1}{2}k_2\theta^2\delta_{n,0}] \nonumber\\
   &=[L^{S}_m,M^{S}_n]+im\theta^a[L^{S}_m,K^a_n]+in\theta^b[J^a_m,M^{S}_n]-mn\theta^a\theta^b[J^b_m,K^a_n]\nonumber\\ 
   &=(m-n)M^{S}_{m+n}-in^2\theta^bK^{b}_{m+n}+im^2\theta^a K^{a}_{m+n}-mn\theta^a\theta^b\Big(if^{abc}K^{c}_{m+n}+mk_2\delta_{m+n}\delta_{ab}\Big)\nonumber\\ 
    &=(m-n)M^{S}_{m+n}-i(m^2-n^2)\theta^aK^a_{m+n}-\frac{1}{2}k_2\theta^2(m-n)\delta_{m+n,0}+i(m^2-n^2)\theta^aK^a_{m+n}\nonumber\\ 
    &\hspace{82.5mm}-imn\theta^a\theta^bf^{abc}K^{c}_{m+n}+m^3k_2\theta^2\delta_{m+n,0}\nonumber\\ 
    &=(m-n)M_{m+n}+k_2\theta^2(m^3-m)\delta_{m+n,0}
\end{align}
Hence, just by doing a slight modification to the Sugawara construction we end up with fully centrally extended BMS algebra with central charges given in (\ref{chh13}).

\section{Details of OPE calculations}
\subsection*{Calculation of T-J OPE}\label{aTJ}
First consider
\begin{equation}\label{tj1}
    \begin{split}
        &\wick{ \c1{\mathcal{J}^a_v(u_1, v_1)} \sum_b\Bigl( \c1{(\mathcal{J}^b_u \mathcal{J}^b_v)(u_2, v_2)} \Bigr)}\\
        &= \oint_{v_2} \frac{dv'}{v'-v_2} \oint_{u_2} \frac{du'}{u'-u_2} \sum_b\left[ \wick{\c1{\mathcal{J}^a_v(u_1, v_1)} \c1{\mathcal{J}^b_u(u', v')} \mathcal{J}^b_v(u_2, v_2) + \mathcal{J}^b_u(u', v')\c2{\mathcal{J}^a_v(u_1, v_1)} \c2{\mathcal{J}^b_v(u_2, v_2)}} \right]\\
        &= \oint_{v_2} \frac{dv'}{v'-v_2} \oint_{u_2} \frac{du'}{u'-u_2} \sum_b\left[ if^{abc} \frac{\mathcal{J}^c_v(u', v') \mathcal{J}^b_v(u_2, v_2)}{(v_1 - v')} + k_2 \delta^{ab}\frac{\mathcal{J}^b_v(u_2, v_2)}{(v_1 - v')^2} + \mathcal{J}^b_u(u', v') \times 0\right]\\
        &= \oint_{v_2} \frac{dv'}{v'-v_2} \oint_{u_2} \frac{du'}{u'-u_2} \left[ \sum_b if^{abc}\frac{(\mathcal{J}^c_v \mathcal{J}^b_v)(u_2, v_2)}{(v_1 - v')} + k_2 \frac{\mathcal{J}^a_v(u_2, v_2)}{(v_1 - v')^2}\right]\\
        &= \sum_b if^{abc}\frac{(\mathcal{J}^c_v \mathcal{J}^b_v)(u_2, v_2)}{v_{12}} + k_2\frac{\mathcal{J}^a_v(u_2, v_2)}{v_{12}^2}
    \end{split}
\end{equation}
Similarly we get 
\begin{equation}\label{TJ2}
    \begin{split}
        &\wick{ \c1{\mathcal{J}^a_v(u_1, v_1)} \sum_b\Bigl( \c1{(\mathcal{J}^b_v \mathcal{J}^b_u)(u_2, v_2)} \Bigr)}\\
        &= \oint_{v_2} \frac{dv'}{v'-v_2} \oint_{u_2} \frac{du'}{u'-u_2} \left[ \sum_b if^{abc}\frac{(\mathcal{J}^b_v \mathcal{J}^c_v)(u_2, v_2)}{v_{12}} + k_2 \frac{\mathcal{J}^a_v(u', v')}{v_{12}^2}\right]\\
        &= \sum_b iF^{abc}\frac{(\mathcal{J}^b_v \mathcal{J}^c_v)(u_2, v_2)}{v_{12}} + k_2\frac{\mathcal{J}^a_v(u_2, v_2)}{v_{12}^2}
    \end{split}
\end{equation}
Summing up (\ref{TJ1}) and (\ref{TJ2}), we see that the first terms in the expressions cancel each other due to the antisymmetry of the structure constant, so we get
\begin{equation}\label{TJ3}
    \begin{split}
        &\wick{ \c1{\mathcal{J}^a_v(u_1,  v_1)} \sum_b\Bigl( \c1{(\mathcal{J}^b_u \mathcal{J}^b_v)(u_2, v_2) + (\mathcal{J}^b_v \mathcal{J}^b_u)(u_2, v_2)} \Bigr)}\\
        &= 2k_2\frac{\mathcal{J}^a_v(u_2, v_2)}{v_{12}^2}\\
        &= 2k_2\left( \frac{\mathcal{J}^a_v(u_1,  v_1)}{v_{21}^2} + \frac{\partial_v \mathcal{J}^a_v(u_1,  v_1)}{v_{21}}\right)
    \end{split}
\end{equation}
Also it can be trivially shown that 
\begin{equation}\label{TJ4}
    \begin{split}
        &\wick{\c1{\mathcal{J}^a_v(u_1, v_1)} \sum_b \Bigl( \c1{(\mathcal{J}^a_v \mathcal{J}^a_v)(u_2, v_2)}\Bigr)} = 0
    \end{split}
\end{equation}
From (\ref{TJ3}) and (\ref{TJ4}), we can determine
\begin{equation}
    \begin{split}
        &T_v(u_1, v_1) \mathcal{J}^b_v(u_2, v_2) \sim \, regular\\
        &T_u(u_1, v_1) \mathcal{J}^b_v(u_2, v_2) \sim \frac{\mathcal{J}^b_v(u_2, v_2)}{v_{12}^2} + \frac{\partial_v \mathcal{J}^b_v(u_2, v_2)}{v_{12}} + \dots
    \end{split}
\end{equation}
Similarly we get
\begin{equation}
    \begin{split}
        &T_v(u_1, v_1) \mathcal{J}^a_u(u_2, v_2) \sim \frac{\mathcal{J}^a_v(u_2, v_2)}{v_{12}^2} + \frac{\partial_v \mathcal{J}^a_v(u_2, v_2)}{v_{12}} + \dots\\
        &T_u(u_1, v_1) \mathcal{J}^a_u(u_2, v_2) \sim \frac{\mathcal{J}^a_u(u_2, v_2)}{v_{12}^2} + \frac{\partial_v \mathcal{J}^a_u(u_2, v_2)}{v_{12}} + \frac{u_{12}}{v_{12}^2}\partial_u \mathcal{J}^a_u(u_2, v_2)\\
        &\qquad\qquad\qquad\qquad\qquad\qquad+ \frac{2u_{12}}{v_{12}}\left( \frac{\mathcal{J}^a_v(u_1, v_1)}{v_{12}^2} + \frac{\partial_v \mathcal{J}^a_v(u_2, v_2)}{v_{12}^2} \right) + \dots
    \end{split}
\end{equation} 
\subsection*{Calculation of T-T OPE}\label{aTT}
First consider
\begin{equation}
    \begin{split}
        &\wick{ \c1{T_u(u_1, v_1)} \c1{(\mathcal{J}^a_v \mathcal{J}^a_v)(u_2, v_2)}}\\
        &= \oint_{v_2} \frac{dv'}{v'-v_2} \oint_{u_2} \frac{du'}{u'-u_2} \Bigl(\wick{\c1{T_u(u_1, v_1)} \c1{\mathcal{J}^a_v(u, v)} \mathcal{J}^a_v(u_2, v_2) +  \mathcal{J}^a_v(u, v)\c2{T_u(u_1, v_1)} \c2{v_2, u_2}} \Bigr)\\
        &=\oint_{v_2} \frac{dv'}{v'-v_2} \oint_{u_2} \frac{du'}{u'-u_2} \Bigl( \frac{\mathcal{J}^a_v(u, v)\mathcal{J}^a_v(u_2, v_2)}{(v_1 - v')^2} + \frac{\partial_{v'} \mathcal{J}^a_v(u, v) \mathcal{J}^a_v(u_2, v_2)}{(v_1-v')}\\
        &\qquad\qquad\qquad\qquad+ \frac{\mathcal{J}^a_v(u, v) \mathcal{J}^a_v(u_2, v_2)}{v_{12}^2} +\frac{\mathcal{J}^a_v(u, v)\partial_{v_2}\mathcal{J}^a_v(u_2, v_2) }{v_{12}} \Bigr)\\
        &= 2\frac{(\mathcal{J}^a_v \mathcal{J}^a_v)(u_2, v_2)}{v_{12}^2} + \oint_{v_2} \frac{dv'}{v'-v_2} \oint_{u_2} \frac{du'}{u'-u_2} \Bigl( \frac{\partial_{v'}[(\mathcal{J}^a_v \mathcal{J}^a_v)(u_2, v_2) + (\partial_v \mathcal{J}^a_v \mathcal{J}^a_v)(u_2, v_2) + \dots]}{(v_1 - v')}\\
        &\qquad\qquad\qquad\qquad\qquad\qquad\qquad\qquad+ \frac{\partial_{v_2}[(\mathcal{J}^a_v \mathcal{J}^a_v)(u_2, v_2) + (\partial_v \mathcal{J}^a_v \mathcal{J}^a_v)(u_2, v_2)+ \dots]}{v_{12}} \Bigr)\\
        &= 2\frac{(\mathcal{J}^a_v \mathcal{J}^a_v)(u_2, v_2)}{v_{12}^2} + \frac{\partial_v (\mathcal{J}^a_v \mathcal{J}^a_v)(u_2, v_2)}{v_{12}} 
    \end{split}
\end{equation}
So we obtain
\begin{equation}\label{TT1}
    T_u(u_1, v_1) T_v(u_2, v_2) \sim 2\frac{(\mathcal{J}^a_v \mathcal{J}^a_v)(u_2, v_2)}{v_{12}^2} + \frac{\partial_v (\mathcal{J}^a_v \mathcal{J}^a_v)(u_2, v_2)}{v_{12}} + \dots
\end{equation}
Similarly, we can check
\begin{equation}\label{TT2}
    T_v(u_1, v_1)T_v(u_2, v_2) \sim \, regular,
\end{equation}
and
\begin{equation}\label{TT3}
    \begin{split}
        T_u(u_1, v_1)T_u(u_2, v_2) \sim  2\frac{T_u(u_2, v_2)}{v_{12}^2} +& \frac{\partial_v T_u(u_2, v_2)}{v_{12}} + \frac{u_{12}}{v_{12}^2}\partial_u T_u(u_2, v_2)\\
        &+ \frac{2u_{12}}{v_{12}}\Bigl( 2\frac{T_v(u_2, v_2)}{v_{12}^2} + \frac{\partial_v T_v(u_2, v_2)}{v_{12}} \Bigr) + \frac{dim(g)}{v_{12}^4} + \dots
    \end{split}
\end{equation}

\section{K-Z equation in field theory approach}\label{aKZ_field}
We start with
\begin{equation}
    \begin{split}
        &\langle\partial_u\Phi(u,v) \Phi_1(u_1, v_1)\dots\Phi_n(u_n, v_n) \rangle = \langle \partial_u\Phi(u,v) X \rangle\\
        &= -\oint_{v} \frac{dv'}{2\pi i} \langle T_v(u', v')\Phi(u,v) X \rangle\\
        &= -\oint_{v} \frac{dv'}{2\pi i} \frac{1}{2k_2}\langle(\mathcal{J}^a_v\mathcal{J}^a_v)(u', v')\Phi(u,v) X \rangle\\
        &= -\oint_{v} \frac{dv'}{2\pi i} \frac{1}{2k_2} \oint_{v'} \frac{dv''}{2\pi i} \frac{1}{(v''-v')} \langle\Bigl( \wick{\c1{\mathcal{J}^a_v(u'', v'')} \c1{\Phi(u,v)} \mathcal{J}^a_v(u', v') + \mathcal{J}^a_v(u'', v'') \c2{\mathcal{J}^a_v(u', v')} \c2{\Phi(u,v)}}\Bigr) X \rangle\\
        &\qquad\qquad- \oint_{v} \frac{dv'}{2\pi i} \frac{1}{2k_2} \langle (\mathcal{J}^a_v\mathcal{J}^a_v\Phi)(u,v)X \rangle + \dots\\
        &= -\oint_{v} \frac{dv'}{2\pi i} \frac{1}{2k_2} \oint_{v'} \frac{dv''}{2\pi i} \frac{1}{(v''-v')} \langle \Bigl( \wick{\frac{t^a_k}{(v''-v)} \Phi(u,v)\mathcal{J}^a_v(u', v') + \frac{t^a_K}{(v'-v)} \mathcal{J}^a_v(u'', v'')\Phi(u,v) }\Bigr) X \rangle + 0\\
        &= I_1 + I_2
    \end{split}
\end{equation}
(Where, second line is (\ref{GCAp3}), third line is the Non-Lorentzian Sugawara construction, fourth line uses definition of normal ordering and the fact that OPE = contractions + normal ordered product + \dots)\\
Now, 
\begin{equation}
    \begin{split}
        I_1 &= -\frac{t^a_k}{2k_2}\oint_{v} \frac{dv'}{2\pi i}  \oint_{v'} \frac{dv''}{2\pi i} \frac{1}{(v'' - v')(v'' - v)} \langle \Phi(u,v)\mathcal{J}^a_v(u',v')X \rangle\\
        &= -\frac{t^a_k}{2k_2}\oint_{v} \frac{dv'}{2\pi i} \frac{1}{(v' - v)} \langle \Phi(u,v)\mathcal{J}^a_v(u',v')X \rangle\\
        &= \frac{t^a_k}{2k_2}\sum_j\oint_{v_j} \frac{dv'}{2\pi i} \frac{1}{(v' - v)}\langle \Phi(u,v) \Phi_1(u_1,v_1)...\bigl(\mathcal{J}^a_v(u',v')\Phi_j(u_j ,v_j)\bigr)\dots \Phi_n(u_n,v_n) \rangle\\
        &= \frac{1}{2k_2}\sum_j\oint_{v_j} \frac{dv'}{2\pi i} \frac{t^a_K \otimes t^a_{K,j}}{(v' - v)(v' - v_j)}\langle \Phi(u,v) \Phi_1(u_1,v_1)...\Phi_j(u_j ,v_j)\dots \Phi_n(u_n,v_n) \rangle\\
        &= \frac{1}{2k_2}\sum_j \frac{t^a_K \otimes t^a_{K,j}}{(v_j - v)}\langle \Phi(u,v) \Phi_1(u_1,v_1)...\Phi_j(u_j ,v_j)\dots \Phi_n(u_n,v_n) \rangle
    \end{split}
\end{equation}
(third line involves a change in contour)\\
and similarly, 
\begin{equation}
    \begin{split}
        I_2 &= -\frac{t^a_k}{2k_2}\oint_{v} \frac{dv'}{2\pi i}  \oint_{v'} \frac{dv''}{2\pi i} \frac{1}{(v'' - v')(v' - v)} \langle \mathcal{J}^a_v(v'', x'')\Phi(u,v)X \rangle\\
        &= \frac{1}{2k_2}\sum_j \frac{t^a_K \otimes t^a_{K,j}}{(v_j - v)}\langle \Phi(u,v) \Phi_1(u_1,v_1)...\Phi_j(u_j ,v_j)\dots \Phi_n(u_n,v_n) \rangle\\
        &= I_1
    \end{split}
\end{equation}
Using the above two results,we get one of the Carrollian K-Z equations
\begin{equation}
    \Bigl(\partial_{u_i} + \frac{1}{k_2}\sum_{j \neq i}\frac{t^a_{R_i, K} \otimes t^a_{R_j, K}}{v_{ij}} \Bigr)\langle\Phi_1(u_1,v_1)\dots \Phi_n(u_n,v_n) \rangle =0
\end{equation}
The other equation can be obtained similarly.

\section{NL KZ equation as a limit}\label{AppG}
Taking the following linear combination of the equations (\ref{kz1}) and (\ref{kz2}),
\begin{align}\label{f1}
(\ref{kz1}) \times \langle\bar{\phi}_{\bar{R}_1}^{\bar{r}_1}(\bar{w}_1)...\bar{\phi}_{\bar{R}_N}^{\bar{r}_N}(\bar{w}_N)\rangle+((\ref{kz2}))\times\langle\phi_{R_1}^{r_1}(w_1)...\phi_{R_N}^{r_N}(w_N)\rangle=0
\end{align}
gives us,
\begin{align}
      \left( \partial_{w_i} +\partial_{\bar{w}_i}-2\gamma\sum_{j\neq i}\frac{\sum_a (t^a_{R_i}\otimes' \bar{I})^{r_i,\bar{r}_i}_{s_i,\bar{s}_i}(t^a_{R_j}\otimes' \bar{I})^{r_j,\bar{r}_j}_{s_j,\bar{s}_j}}{w_i-w_j}-2\bar{\gamma}\sum_{j\neq i}\frac{\sum_a (I\otimes'\bar{t}^a_{\bar{R}_i})^{r_i,\bar{r}_i}_{s_i,\bar{s}_i} (I\otimes'\bar{t}^a_{\bar{R}_j})^{r_j,\bar{r}_j}_{s_j,\bar{s}_j}}{\bar{w}_i-\bar{w}_j} \right)\nonumber\\
     \langle...\Phi_{R_i,\bar{R}_i}^{s_i,\bar{s}_i}(w_i,\bar{w}_i)...\Phi_{R_j,\bar{R}_j}^{s_j,\bar{s}_j}(w_j,\bar{w}_j)...\rangle=0
\end{align}
We have $(w_i,\bar{w}_i)=t\pm \epsilon x\Rightarrow  \partial_{w_i}=\frac{1}{2}(\partial_{t_i}+\frac{\partial_{x_i}}{\epsilon}) ;  \partial_{\bar{w}_i}=\frac{1}{2}(\partial_{t_i}-\frac{\partial_{x_i}}{\epsilon})$ such that the Carrollian limit is achieved by taking $\epsilon\rightarrow 0$. 
Using these we get,
\begin{align}
    \Rightarrow \left( \partial_{t_i}-2\sum_{j\neq i}X \right)
     \langle...\Phi_{R_i,\bar{R}_i}^{s_i,\bar{s}_i}(x_i,t_i)...\Phi_{R_j,\bar{R}_j}^{s_j,\bar{s}_j}(x_j,t_j)...\rangle=0
\end{align}
where(with $t_{ij}=t_i-t_j$ and $x_{ij}=x_i-x_j$), we have,
\begin{align}
   X=\left\{\frac{\gamma}{t_{ij}+\epsilon x_{ij}}\sum_a (t^a_{R_i}\otimes' \bar{I})^{r_i,\bar{r}_i}_{s_i,\bar{s}_i}(t^a_{R_j}\otimes' \bar{I})^{r_j,\bar{r}_j}_{s_j,\bar{s}_j}+\frac{\bar{\gamma}}{t_{ij}-\epsilon x_{ij}}\sum_a (I\otimes'\bar{t}^a_{\bar{R}_i})^{r_i,\bar{r}_i}_{s_i,\bar{s}_i} (I\otimes'\bar{t}^a_{\bar{R}_j})^{r_j,\bar{r}_j}_{s_j,\bar{s}_j}\right\}
\end{align}
Inverting relations (1.11), we have,
\begin{align}\label{f4}
    t^a_{R_i}\otimes' \bar{I}=\frac{1}{2}(t^a_{R_i,J}-\frac{t^a_{R_i,K}}{\epsilon}) \ ; \ I\otimes' \bar{t}^a_{\bar{R}_i}=\frac{1}{2}(t^a_{R_i,J}+\frac{t^a_{R_i,K}}{\epsilon})
\end{align}
Hence,
\begin{align}
&X=\frac{1}{4}\sum_a\bigg\{\frac{\gamma}{t_{ij}+\epsilon x_{ij}} \left(t^a_{R_i,J}-\frac{t^a_{R_i,K}}{\epsilon}\right)^{r_i,\bar{r}_i}_{s_i,\bar{s}_i}\left(t^a_{R_j,J}-\frac{t^a_{R_j,K}}{\epsilon}\right)^{r_j,\bar{r}_j}_{s_j,\bar{s}_j}\nonumber\\
&\hspace{55pt}+\frac{\bar{\gamma}}{t_{ij}-\epsilon x_{ij}} \left(t^a_{R_i,J}+\frac{t^a_{R_i,K}}{\epsilon}\right)^{r_i,\bar{r}_i}_{s_i,\bar{s}_i} \left(t^a_{R_j,J}+\frac{t^a_{R_j,K}}{\epsilon}\right)^{r_j,\bar{r}_j}_{s_j,\bar{s}_j}\bigg\}\nonumber\\
&=\frac{1}{4}\sum_a\bigg\{\frac{\gamma}{t_{ij}+\epsilon x_{ij}} \bigg(t^a_{R_i,J}\otimes t^a_{R_j,J}-\frac{1}{\epsilon}(t^a_{R_i,J}\otimes t^a_{R_j,K}+t^a_{R_i,K}\otimes t^a_{R_j,J})\nonumber\\
&\hspace{10 pt}+\frac{1}{\epsilon^2}t^a_{R_i,K}\otimes t^a_{R_j,K}\bigg)^{r_i,\bar{r}_i;r_j,\bar{r}_j}_{s_i,\bar{s}_i;s_j,\bar{s}_j}+\frac{\bar{\gamma}}{t_{ij}-\epsilon x_{ij}}\bigg(t^a_{R_i,J}\otimes t^a_{R_j,J}+\frac{1}{\epsilon}(t^a_{R_i,J}\otimes t^a_{R_j,K}+t^a_{R_i,K}\otimes t^a_{R_j,J})\nonumber\\
&\hspace{250 pt}+\frac{1}{\epsilon^2}t^a_{R_i,K}\otimes t^a_{R_j,K}\bigg)^{r_i,\bar{r}_i;r_j,\bar{r}_j}_{s_i,\bar{s}_i;s_j,\bar{s}_j} \bigg\}
\end{align}
We introduce notation for convenience,
\begin{align}\label{f5}
    \sum_a (t^a_{R_i,A}\otimes t^a_{R_j,B})^{r_i,\bar{r}_i;r_j,\bar{r}_j}_{s_i,\bar{s}_i;s_j,\bar{s}_j}=t^{ij}_{AB}   \ \ \text{(where $A,B=J,K$) }
\end{align}
Therefore,
\begin{align}\label{f3}
X=&\frac{1}{4}\bigg\{\frac{\gamma}{t_{ij}}(1-\epsilon \frac{x_{ij}}{t_{ij}}+\epsilon^2 \frac{x^2_{ij}}{t^2_{ij}}+...) (t^{ij}_{JJ}-\frac{1}{\epsilon}(t^{ij}_{JK}+t^{ij}_{KJ})+\frac{1}{\epsilon^2}t^{ij}_{KK})\nonumber\\
&+\frac{\bar{\gamma}}{t_{ij}}(1+\epsilon \frac{x_{ij}}{t_{ij}}+\epsilon^2 \frac{x^2_{ij}}{t^2_{ij}}+...)(t^{ij}_{JJ}+\frac{1}{\epsilon}(t^{ij}_{JK}+t^{ij}_{KJ})+\frac{1}{\epsilon^2}t^{ij}_{KK})\bigg\}\nonumber\\
\Rightarrow X&=\frac{1}{4t_{ij}}\bigg\{-\frac{(\gamma-\bar{\gamma})}{\epsilon}(t^{ij}_{JK}+t^{ij}_{KJ})+\frac{\gamma+\bar{\gamma}}{\epsilon^2}t^{ij}_{KK})-\frac{x_{ij}}{4t^2_{ij}}\frac{(\gamma-\bar{\gamma})}{\epsilon}t^{ij}_{KK}\bigg\}
\end{align}
In limit $\epsilon\rightarrow 0$ and using (\ref{gamrel}), we get,
\begin{align}
    \Rightarrow X&=\frac{1}{4t_{ij}}\bigg\{\frac{2}{k_2}(t^{ij}_{JK}+t^{ij}_{KJ})-\frac{2(k_1+2C_g)}{k_2^2}t^{ij}_{KK})+\frac{x_{ij}}{4t^2_{ij}}\frac{2}{k_2}t^{ij}_{KK}\bigg\}\nonumber
    \\\Rightarrow X&=\frac{1}{2k_2}\bigg\{\frac{(t^{ij}_{JK}+t^{ij}_{KJ})}{t_{ij}}+(\frac{x_{ij}}{t_{ij}^2}-\frac{(k_1+2C_g)}{k_2t_{ij}})t^{ij}_{KK}\bigg\}
\end{align}
Therefore, (\ref{f1}) can be finally written as( in limit $\epsilon\rightarrow 0$),
\begin{align}
      & \Rightarrow \bigg[ \partial_{t_i}-\sum_{j\neq i}\frac{1}{k_2}\bigg\{\frac{(t^{ij}_{JK}+t^{ij}_{KJ})}{t_{ij}}+\bigg(\frac{x_{ij}}{t_{ij}^2}-\frac{(k_1+2C_g)}{k_2t_{ij}}\bigg)t^{ij}_{KK}\bigg\} \bigg]\nonumber\\
       &\hspace{250 pt}\langle...\Phi_{x_i,t_i}^{s_i,\bar{s}_i}(x_i,t_i)...\Phi_{R_j,\bar{R}_j}^{s_j,\bar{s}_j}(x_j,t_j)...\rangle=0\nonumber
\end{align}
\begin{align}
      \Rightarrow &\bigg[\partial_{t_i}-\frac{1}{k_2}\sum_{j\neq i}\bigg\{\frac{\sum_a (t^a_{R_i,J}\otimes t^a_{R_j,K}+t^a_{R_i,K}\otimes t^a_{R_j,J})^{r_i,\bar{r}_i;r_j,\bar{r}_j}_{s_i,\bar{s}_i;s_j,\bar{s}_j}}{t_{ij}}\nonumber\\
      &+\bigg(\frac{x_{ij}}{t_{ij}^2}-\frac{(k_1+2C_g)}{k_2t_{ij}}\bigg)\sum_a (t^a_{R_i,K}\otimes t^a_{R_j,K})^{r_i,\bar{r}_i;r_j,\bar{r}_j}_{s_i,\bar{s}_i;s_j,\bar{s}_j}\bigg\}\bigg]
      \langle...\Phi_{R_i,\bar{R}_i}^{s_i,\bar{s}_i}(x_i,t_i)...\Phi_{R_j,\bar{R}_j}^{s_j,\bar{s}_j}(x_j,t_j)...\rangle=0\nonumber
\end{align}
\begin{align}
       &\Rightarrow \bigg[\partial_{t_i}-\frac{1}{k_2}\sum_{j\neq i}\bigg\{\frac{\sum_a (t^a_{R_i,J}\otimes t^a_{R_j,K}+t^a_{R_i,K}\otimes t^a_{R_j,J})}{t_{ij}}\nonumber\\
      &\hspace{20pt} +\bigg(\frac{x_{ij}}{t_{ij}^2}-\frac{(k_1+2C_g)}{k_2t_{ij}}\bigg)\sum_a (t^a_{R_i,K}\otimes t^a_{R_j,K})\bigg\}\bigg]{\langle...\Phi_{R_i,\bar{R}_i}(x_i,t_i)...\Phi_{R_j,\bar{R_j}}(x_j,t_j)...\rangle}=0
\end{align}
which is in agreement with (\ref{nlkz1}).\\

Now,we consider another linear combination,
\begin{align}
    \epsilon\{(\ref{kz1}) \times\langle\bar{\phi}_{\bar{R}_1}^{\bar{r}_1}(\bar{w}_1)...\bar{\phi}_{\bar{R}_N}^{\bar{r}_N}(\bar{w}_N)\rangle-(\ref{kz2})\times\langle\phi_{R_1}^{r_1}(w_1)...\phi_{R_N}^{r_N}(w_N)\rangle\}=0
\end{align}
Again following similar steps as before,
\begin{align}
    &\Rightarrow \bigg( \partial_{x_i}-2\sum_{j\neq i}Y\bigg)\langle...\Phi_{R_i,\bar{R}_i}^{s_i,\bar{s}_i}(x_i,t_i)...\Phi_{R_j,\bar{R}_j}^{s_j,\bar{s}_j}(x_j,t_j)...\rangle=0\label{chh}
\end{align}
where,
\begin{align}
    Y=\epsilon\left\{\frac{\gamma}{t_{ij}+\epsilon x_{ij}}\sum_a (t^a_{R_i}\otimes' \bar{I})^{r_i,\bar{r}_i}_{s_i,\bar{s}_i}(t^a_{R_j}\otimes' \bar{I})^{r_j,\bar{r}_j}_{s_j,\bar{s}_j}-\frac{\bar{\gamma}}{t_{ij}-\epsilon x_{ij}}\sum_a (I\otimes'\bar{t}^a_{\bar{R}_i})^{r_i,\bar{r}_i}_{s_i,\bar{s}_i} (I\otimes'\bar{t}^a_{\bar{R}_j})^{r_j,\bar{r}_j}_{s_j,\bar{s}_j}\right\}
\end{align}
Similar to what we did for $X$, we get an equation analogous to (\ref{f3}) using (\ref{f4}) and (\ref{f5}),
\begin{align}
Y&=\frac{\epsilon}{4}\{\frac{\gamma}{t_{ij}}(1-\epsilon \frac{x_{ij}}{t_{ij}}+\epsilon^2 \frac{x^2_{ij}}{t^2_{ij}}+...) (t^{ij}_{JJ}-\frac{1}{\epsilon}(t^{ij}_{JK}+t^{ij}_{KJ})+\frac{1}{\epsilon^2}t^{ij}_{KK})\nonumber\\
&-\frac{\bar{\gamma}}{t_{ij}}(1+\epsilon \frac{x_{ij}}{t_{ij}}+\epsilon^2 \frac{x^2_{ij}}{t^2_{ij}}+...)(t^{ij}_{JJ}+\frac{1}{\epsilon}(t^{ij}_{JK}+t^{ij}_{KJ})+\frac{1}{\epsilon^2}t^{ij}_{KK})\}
\end{align}
Again using (\ref{gamrel}) and collecting the finite terms in the limit $\epsilon\rightarrow 0$, we get,
\begin{align}
 Y=-\frac{1}{2k_2} \frac{t^{ij}_{KK}}{t_{ij}}  
\end{align}
Putting back in \eqref{chh}, 
\begin{align}
       {\left( \partial_{x_i}+\frac{1}{k_2}\sum_{j\neq i} \frac{\sum_a (t^a_{R_i,K}\otimes t^a_{R_j,K})}{t_{ij}} \right)\langle...\Phi_{R_i,\bar{R}_i}(x_i,t_i)...\Phi_{R_j,\bar{R}_j}(x_j,t_j)...\rangle=0}
\end{align}
which is in agreement with (\ref{nlkz2}).

\newpage

\bibliographystyle{JHEP}
\bibliography{NLKM}

\end{document}